\documentclass[11pt,a4paper]{article}
\pdfoutput=1
\usepackage[english]{babel}
\usepackage[latin1]{inputenc}
\usepackage{amsfonts,amsbsy,bm,euscript,mathrsfs}
\usepackage{amssymb,stmaryrd,faktor,slashed,cancel}
\usepackage{color}
\usepackage[tbtags]{amsmath}
\usepackage[bookmarks=true,colorlinks=true,linkcolor=black,citecolor=black,urlcolor=black,bookmarksnumbered]{hyperref}

\numberwithin{equation}{section}

\makeatletter
\renewcommand\section{\@startsection {section}{1}{\z@}%
{-3.5ex \@plus -1ex \@minus -.2ex}%
{2.3ex \@plus.2ex}%
{\normalfont\large\bfseries}}
\renewcommand\subsection{\@startsection{subsection}{2}{\z@}%
{-3.25ex\@plus -1ex \@minus -.2ex}%
{1.5ex \@plus .2ex}%
{\normalfont\normalsize\bfseries}}

\def\maketag@@@#1{\hbox{\m@th\normalfont\normalsize#1}}
\makeatother

\def \be {\begin{equation}}
\def \ee {\end{equation}}
\def \bea {\begin{eqnarray}}
\def \eea {\end{eqnarray}}
\def \p {{\partial}}

\def \eps {{\epsilon}}
\def \Tr {{\rm Tr}}
\def \bi {\bibitem}

\def \ha {{1 \over 2}}
\def \td {\tilde}
\def \ci {\cite}

\def \const {{\rm const}}

\def \E {{\mathcal E}}

\def \z {\zeta}

\def \a {\alpha}
\def \b {\beta}
\def \e {\varepsilon}
\def \p {\phi}

\def \del {\partial}
\def \a {\alpha}

\def \g {\gamma}
\def \s {\sigma}
\def \z {\zeta}

\def \ov {\over}
\def \E {{\mathcal E}}
\def \w {\omega}
\def \b {\beta}
\def \l {\lambda}
\def \eps {\epsilon}

\def \det {\hbox{det}}
\def \ci {\cite}

\def \Tr {{\rm Tr}}

\def \l {\lambda}
\def \const {{\rm const}}

\def \td {\tilde}

\def \m {\mu}
\def \ie {i.e.}

\def \e {\epsilon}
\def \bi {\bibitem}
\def \la {\label}
\def \l {\lambda}

\def \adss {AdS$_5 \times S^5~$}
\def \ov {\over}

\def \F {{\cal F}}

\def \ha {{1\ov 2}}
\def \r {\rho}
\def \no {\nonumber}

\def \del {\partial}
\def \E {{\cal E}}

\def \bi {\bibitem}
\def \la {\label}

\def \adss {AdS$_5 \times S^5$\ }

\def \p {\phi}
\def \r {\rho}

\def \w {\omega}
\def \ov {\over}

\def \varpi {{\rm w}}

\def \ve {\varepsilon}
\def \pa {\partial}

\def \LL {{\cal L}}
\def \ep {\epsilon}

\def \Tr {{\rm Tr}}

\def \s {\sigma}
\def \eps {{\epsilon}}

\def \n {\nu}

\def \vp {\varphi}

\def \ed {\end{document}}
\def \te {\textstyle}

\def \ha {{{\textstyle{1 \ov2}}}}

\def \eqref {\rf}
\def \ha {{\textstyle {1 \ov 2}}}

\def \bi {\bibitem}
\def \be {\bea}
\def \ee {\eea}

\def \na {\nabla}

\def \iffa {\iffalse}

\def \vp {\varphi}

\def \p {\phi}

\def \iffa {\iffalse}

\def \ep {\epsilon}
\def \text { }
\def \adss {$AdS_5 \times S^5\ $}
\def \adst {$AdS_3 \times S^3\times T^4 \ $}
\def \adt {$AdS_2 \times S^2\times T^6\ $}

\def \rr {{\rm R}}
\def \te {\textstyle}

\def \ffb {F}
\def \ffg {H}
\def \goo {G_{\yy\yy}}
\def \goon {\hat G_{\ty\ty}}
\def \yy {{y}}
\def \ty {{\hat y}}
\def \by {K}
\def \my {A}
\def \otherc {\hat c}

\def \G {\Gamma}

\def \K {{\rm K}}

\def \K {{\rm K}}

\def \ie {\begin{equation}}
\def \fe {\end{equation}}

\def \te {\theta}
\def \edo {\end{document}}

\def \F {{\cal F}}
\def \T {{\cal T}}
\def \K {{\cal K}}
\def \te {\textstyle}
\def \ve {\varepsilon}
\def \ep {\epsilon}
\def \eps {\epsilon}
\def \vk {\varkappa}
\def \W {\wedge}
\def \rF {{\rm F}}
\def \emo {$\eta$-model}

\def \extraterm{{+\ 2I\cdot Z \F_{2n+1}}}

\def \ah {{\hat a}}
\def \hs {{\hat s}}
\def \hg {{\hat g}}

\def \hA {{\hat A}}
\def \hB {{\hat B}}
\def \hH {{\hat H}}
\def \hZ {{\hat Z}}
\def \hb {{\hat b}}
\def \hF {{\hat \F}}
\def \hf {{\hat f}}
\def \hp {{\hat \phi}}

\def \G {\Gamma}

\def \otherc {\hat c}
\def \by {K}
\def \my {A}
\def \ty {{\hat y}}
\def \yy {y}
\def \lmo {$\l$-model}

\def \bb {{\bar \beta}}
\def \tildec {\hat c}
\def \Ee {\Xi}

\newcommand{\rf}[1]{(\ref{#1})}

\newcommand{\foot}[1]{\footnote{#1\vspace{1pt}}}
\newcommand{\bei}{\begin{itemize}}
\newcommand{\eei}{\end{itemize}}
\newcommand{\bee}{\begin{enumerate}}
\newcommand{\eee}{\end{enumerate}}
\newcommand{\bal}{\begin{equation}\begin{aligned}[b]}
\newcommand{\eal}{\end{aligned}\end{equation}}
\newcommand{\arxivlink}[1]{\href{http://arxiv.org/abs/#1}{\texttt{arXiv:#1}}}

\begin{document}

\thispagestyle{empty}

\hfill\vspace{-2cm}
\rightline{\footnotesize ZMP-HH/15-27}
\rightline{\footnotesize TCDMATH 15-12}
\rightline{\footnotesize Imperial-TP-AT-2015-08}
\normalsize
\vspace{-0.2cm}

\begin{center}

{\Large\bf
Scale invariance of the $\eta$-deformed $AdS_5 \times S^5$ superstring,
\\\vspace{0.3cm}
T-duality and modified type II equations}

\vspace{0.8cm}{
G. Arutyunov$^{a,b,1}$,
S. Frolov$^{c,}$\foot{Correspondent fellow at Steklov Mathematical Institute, Moscow.},
B. Hoare$^{d}$,
R. Roiban$^{e}$
~and A.A.~Tseytlin$^{f,}$\foot{Also at Lebedev Institute, Moscow.
\\ gleb.arutyunov@desy.de; frolovs@maths.tcd.ie; bhoare@ethz.ch; radu@phys.psu.edu; tseytlin@imperial.ac.uk}

\vskip 0.25cm

{\em $^{a}$ II. Institut f\"ur Theoretische Physik, Universit\"at Hamburg, \\ Luruper Chaussee 149, 22761 Hamburg, Germany}
\vskip 0.05cm
{\em $^{b}$ Zentrum f\"ur Mathematische Physik, Universit\"at Hamburg, \\ Bundesstrasse 55, 20146 Hamburg, Germany}
\vskip 0.05cm
{\em $^{c}$ Hamilton Mathematics Institute and School of Mathematics, \\ ~~Trinity College, Dublin 2, Ireland}
\vskip 0.05cm
{\em $^{d}$ Institut f\"ur Theoretische Physik, ETH Z\"urich,\\ Wolfgang-Pauli-Strasse 27, 8093 Z\"urich, Switzerland}
\vskip 0.05cm
{\em $^{e}$ Department of Physics, The Pennsylvania State University, \\ University Park, PA 16802, USA}
\vskip 0.05cm
{\em $^{f}$ The Blackett Laboratory, Imperial College, London SW7 2AZ, U.K.}

\vskip 0.4cm}
\end{center}

\vskip 0.10cm
\begin{abstract}
We consider the ABF background underlying the $\eta$-deformed $AdS_5 \times
S^5$ sigma model. This background fails to satisfy the standard IIB
supergravity equations which indicates that the corresponding sigma model is
not Weyl invariant, i.e. does not define a critical string theory in the usual
sense. We argue that the ABF background should still define a UV finite theory
on a flat 2d world-sheet implying that the $\eta$-deformed model is scale
invariant. This property follows from the formal relation via T-duality
between the $\eta$-deformed model and the one defined by an exact type IIB
supergravity solution that has 6 isometries albeit broken by a linear dilaton.
We find that the ABF background satisfies candidate type IIB scale invariance
conditions which for the R-R field strengths are of the second order in
derivatives. Surprisingly, we also find that the ABF background obeys an
interesting modification of the standard IIB supergravity equations that are
first order in derivatives of R-R fields. These modified equations explicitly
depend on Killing vectors of the ABF background and, although not universal,
they imply the universal scale invariance conditions. Moreover, we show that it
is precisely the non-isometric dilaton of the T-dual solution that leads, after
T-duality, to modification of type II equations from their standard form. We
conjecture that the modified equations should follow from $\kappa$-symmetry of
the $\eta$-deformed model. All our observations apply also to
$\eta$-deformations of \adst and \adt models.
\end{abstract}

\newpage

\topmargin=0cm
\setcounter{equation}{0}
\setcounter{footnote}{0}

\tableofcontents

\section{Introduction}

The study of integrable deformations of the \adss superstring sigma model is an
important direction in the search for new solvable examples of AdS/CFT duality.
An interesting one-parameter integrable generalisation of the classical \adss
Green-Schwarz action related to the quantum deformation of the underlying
supergroup symmetry was found in \ci{dmv}. Just from the construction of this
``$\eta$-model" (based on a particular current-current deformation of the
supercoset action \ci{mt} generalising the bosonic model of \ci{kl}) there is
no a priori reason why it should define a scale invariant (UV finite) 2d theory
and, moreover, why it should preserve the conformal (Weyl) invariance and hence
still correspond to a consistent superstring theory as the undeformed \adss
model does.\foot{This is in contrast, e.g., to the integrable deformation
\ci{lm} based on TsT duality transformations, which preserve conformality. In
particular, the TsT deformed background is a solution of type IIB
supergravity.}

The only indication in this direction is that the \emo\ action, like the
original \adss action, is invariant under a version of fermionic $\kappa$-symmetry
\ci{dmv}, which reduces the number of fermions by half. However, the usual
claim that $\kappa$-symmetry implies the corresponding action can be interpreted
as that of a GS superstring propagating in a background that is a consistent
type II supergravity solution (and hence defines a consistent critical
superstring theory) assumes that the $\kappa$-symmetry is of the standard GS
``projector" form \ci{howe}. This is most probably not the case for the
$\eta$-model at higher orders in fermions. Indeed, it was found in
\ci{abf,abf2} that the target space background corresponding to the \emo\
action \ci{dmv}, interpreted as a GS action, does not represent a type IIB
supergravity solution.

Starting with the GS Lagrangian written in superspace form ($Z^M = (x^m,\theta^\a)$)
\be
L = (\sqrt h h^{ab} E_M^r E_N^s \eta_{rs} -\epsilon^{ab} B_{MN} ) \del_a Z^M \del_b Z^N\ ,
\ee
one can solve the standard type II superspace constraints and Bianchi
identities for $E(Z), B(Z)$ (which imply the supergravity equations) in order
to express the GS action in terms of component fields. One then observes that
the dilaton $\p$ (which is part of the dilaton superfield $\Phi(Z)$ that is
introduced in the process of solving the constraints) enters the world-sheet
action (i) in the combination $\F=e^\p F$ with the R-R field strengths starting
at order $\theta^2$ and (ii) via derivatives $\del_m \p$ starting at order
$\theta^4$ (see \ci{lin} and references therein). This action has classical
Weyl invariance and $\kappa$-symmetry, which will be broken, in general, by
quantum corrections. As for the bosonic string \ci{cal}, to cancel the 2d
stress tensor trace anomaly requires adding the familiar 1-loop dilaton
counterterm $ \sim \int d^2 z \sqrt h R^{(2)} \Phi(Z) $ (see
\ci{Grisaru:1988sa,bel} and references therein).\foot{This additional term is
certainly required to reproduce the standard 1-loop Weyl-invariance conditions
for the $G$ and $B$-field couplings or supergravity equations in NS-NS sector.
This term should also be required to cancel the quantum anomaly of
$\kappa$-symmetry.} 

The case relevant to our discussion below is a special isometric type II
solution for which the metric $G_{mn}$, $B$-field $B_{mn}$ and R-R fields $ \F_{m_1...m_n}$ are
invariant while $\p$ is linear in the isometric directions. In this case the GS
action will depend on the isometric coordinates only through their 2d
derivatives and can thus be T-dualised. As we shall see, in this case the
T-dual model will be scale invariant but may not be Weyl invariant 
(one may not be able to cancel the Weyl anomaly by a local counterterm), i.e. may
not correspond to a type II supergravity solution.

The ABF background \ci{abf,abf2}
includes the 10d metric $G$, the $B$-field and the R-R fields
$\F_n$ ($n=1,3,5$) that are extracted from the quadratic fermionic part of the action of \ci{dmv}
put into the usual GS form,
\be
&&{\cal A}= - T \int d^2 z \Big[\ \ha (\eta^{ab} G_{mn} - \eps^{ab} B_{mn}) \del_a x^m \del_b x^n
\no \\ && \qquad \qquad \qquad \qquad
+\, i \bar \theta \G_m D \theta\ \del x^m +
\bar \theta \G_m \ \F\cdot \G\ \G_n \theta\ \del x^m \del x^n + ... \Big] \ .\la{0}
\ee
For the standard GS action in a type IIB supergravity background $\F_n$ are
interpreted as the products of the dilaton and the R-R field strengths $\F_n =
e^\phi F_n$ but in the \emo\ case there is no independent information about the
dilaton, and there exists no dilaton field that completes $G,B, \F_n$ of
the ABF background to a type IIB solution \ci{abf2}.

While not solving the standard type IIB equations directly this ABF background
still turns out to be very special: it is related by T-duality to an
exact type IIB supergravity solution \ci{ht2,ht}. The latter HT background
involves a non-diagonal metric $\hat G$, an imaginary 5-form $\hat F_5$
and the dilaton $\hat \p$, and the T-duality applied in all 6 isometric
directions acts only on the fields $\hat G$ and $\hat \F_5 = e^{\hat \p} \hat
F_5$ entering the corresponding GS action \rf{0} on a flat 2d background.
The GS action for any type II solution (and thus for the HT background) should
be Weyl invariant and, in particular, scale invariant. As the T-duality
applied to the GS action \ci{gst} is a simple path integral transformation, the
T-duality relation between the ABF and HT backgrounds implies \ci{ht2} that the
\emo\ action should define a scale invariant 2d theory at least to 1-loop
order.

However, there may be a problem with Weyl invariance for the \emo\
action on a curved 2d background. The HT dilaton $\hat \p$ has a
term linearly depending on the isometric directions of $\hat G$ and $\hat \F_5$
and thus one cannot directly apply the standard T-duality transformation rules
\ci{bho} to the full HT background to get a full T-dual supergravity solution,
and thus the Weyl invariance of the T-dual sigma model requires further
investigation.\foot{The 1-loop Weyl invariance conditions of the NSR or GS type II
superstring sigma model are believed to be equivalent to the field equations of
type II supergravity. While this is a well-established fact in the NS-NS sector
\ci{cal,tse} this was never demonstrated directly with the R-R couplings included
(for some related work, mostly for the heterotic string, see
\ci{Grisaru:1988sa,bel,Berenstein:1999ip}). Given that the linearised equations
for all the supergravity fields follow from the condition of marginality of the
corresponding NSR vertex operators and that the type II action is a leading
term in the string effective action reconstructed from the superstring S-matrix
on flat space, it is usually assumed that the superstring sigma model defining
consistent critical string theories should correspond (to leading order in
$\a'$) to backgrounds that solve the 10d supergravity equations.} This is of
course consistent with the observation \ci{abf2} that the ABF background does
not satisfy the full set of type IIB supergravity equations.

\medskip

The aim of the present paper is to further clarify and extend these
observations. We shall demonstrate that the relation by formal T-duality between
the ABF and HT backgrounds implies that the former, while not a
supergravity solution, should satisfy the following two generalisations or
``modifications" of the type II supergravity equations:

(i) the scale invariance conditions for the type II superstring sigma model
(with equations on the R-R fields $\F$ being of 2nd order in derivatives)

(ii) a set of equations that are structurally similar to those of type II
supergravity (with 1st-order equations for the R-R fields $\F$)
but involving, instead of derivatives of the dilaton, a certain co-vector $Z_m$
playing now the role of the dilaton one-form and a Killing vector $I^m$
responsible for the ``modification" of the equations from their standard
form.\foot{In what follows we shall not distinguish between co-vectors and
vectors, referring to both $X^m$ and $X_m$ as vectors.}

While the scale invariance conditions are universal, the second set of
equations (which we shall refer to as ``$I$-modified" type II equations)
only apply to particular backgrounds with isometric $G,B$ and $\F$-fields,
which are related by formal T-duality to a type II solution ($\hat G,\hat B,
\hat \F, \hat \p$) with the dilaton $ \hat \p$ containing a term linear in
the isometric coordinates. Such a dilaton background, breaking isometries by a linear
term only, is special. As the type II supergravity equations written in terms of
the $\F$-fields only depend on the dilaton through its {\it derivatives},
they remain independent of the isometric directions. As a result, the
standard type II supergravity equations for the T-dual solution ($\hat G,\hat
B, \hat \F, \hat \p$) can be re-interpreted as certain modified
type II equations for the original fields $(G,B,\F)$, also depending on the
vectors $Z$ and $I$. The Killing vector $I^m$ dependence is fixed by the
term linear in isometric coordinates in the dilaton, while the vector $Z_m$ is
determined by applying the standard T-duality rules to the part of the dilaton
independent of the isometric coordinates.\foot{Let us stress that $\hat F$ and
$\hat \p$ explicitly depend on the isometric coordinates. It is $\hat G,\hat B,
\hat \F, d\hat \p$, and $G,B,\F, Z$ that are invariant under the isometries
generated by the Killing vectors $\hat I^m$ and $I^m$ respectively. That is,
Lie derivatives of the fields along the corresponding Killing vector are
zero.}

It is possible to express the modified equations for the NS-NS fields
in terms of just one single vector $X_m = Z_m +I_m$, which is the vector that
appears in the scale invariance conditions.
The superstring scale invariance conditions
generalise the familiar one-loop
scale invariance conditions for
the bosonic sigma model with couplings $G_{mn}$ and $B_{mn}$ (cf. \rf{0})
\be \la{1} \te
\beta^G_{mn} \equiv R_{mn} - {1\ov 4} H_{mkl} H_{n}^{\ kl} = - D_m X_n - D_n X_m \ , \\
\beta^B_{mn} \equiv \ha D^k H_{kmn} = X^k H_{kmn} + \del_m Y_n - \del_m Y_n \ .
\la{2}
\ee
Here the terms involving $X_m$ and $Y_m$ do not contribute to on-shell UV
divergences or, equivalently, reflect the freedom of renormalisation by
reparametrisations and $B$-field gauge transformations. The $X_m$ terms drop
out of the action if the sigma model field $x^m$ is subject to the classical
equations, or, equivalently, they can be absorbed in a field renormalisation,
$x^m \to x^m + X^m \log \eps$. The origin of the $X^k H_{kmn}$ term can be
understood either by starting with a counterterm proportional to $(D_m X_n +
D_n X_m )\del_a x^n \del_b x^n + ...$, integrating by parts and using the
equations of motion for $x^m$, or by observing that $B_{mn}$ transforms under a
combination of reparametrisations and gauge transformations as $ X^k \del_k
B_{mn} + \del_m X^k B_{kn} - \del_n X^k B_{km} + \del_m Y'_n -\del_n Y'_m = X^k 
H_{kmn} + \del_m Y_n - \del_n Y_m $ where $Y'_m$ or $Y_m$ drop out of the sigma
model action upon integration by parts.

The Weyl invariance conditions are equivalent to the
vanishing of the trace of the 2d stress tensor operator on a curved 2d background.
For the NSR type II superstring sigma model
they can be satisfied provided one adds
the dilaton term $\sim R^{(2)} \p(x)$
\ci{cal,Hull:1985rc,ts,Shore:1986hk,tse}: they are a
stronger form of the scale invariance conditions \rf{1},\rf{2} with $X_m$ and $Y_m$
no longer arbitrary, but given by
\be \la{2a} X_m = \del_m \p \ , \qquad \qquad Y_m =0 \ . \ee
The Weyl invariance equations \rf{1},\rf{2},\rf{2a} imply
the ``central charge" identity \ci{cal,Curci:1986hi}
\be \la{cp}
\del_m \bar \b^\p =0 \ , \ \ \ \ \ \ \
\bar \b^\p \equiv R - \te {1\ov 12} H^2_{mnk} + 4 D^2 \p - 4 \del^m\p \del_m \p \ ,
\ee
i.e. that the effective dilaton ``$\beta$-function" is a constant (which should be zero in critical string theory).
The full set of
Weyl invariance equations for $G,B$ and $\phi$ follows from the effective action
with the same form as the NS-NS sector of the type II supergravity action ($ \F \equiv e^\p F$)
\bea
S=\int d^d x\ \sqrt G\ \big(e^{-2\p}\ \bar \b^\p + \sum FF + ... \big)= \int d^d x\ \sqrt G\ e^{-2\p}\ \big( \bar \b^\p + \sum \F\F + ... \big) \ , \la{ef}
\eea
where we have indicated the presence of the R-R field strength terms for future reference.

The generalisation of the scale invariance conditions to the presence
of R-R fields is given by \rf{1},\rf{2} with
extra $\F\F$ terms, together with a set of {\it second}-derivative
equations for the R-R fields $\F$ that directly enter the GS action \rf{0},
$ \ha D^2 \F + ... = X \del \F + \F \del X $. Here the r.h.s. stands for reparametrisation (Lie derivative) terms
with the same $X$-vector as in \rf{1},\rf{2} and dots indicate non-linear terms.
In the special case when $X_m=\del_m \p$ these equations
are the consequence of the type IIB equations or Weyl invariance conditions,
which are 1st order in $F=e^{-\p} \F$, i.e. $d \star F + ... = 0$ and $d F+ ... = 0$.\foot{The relation between
the 1st-order and 2nd-order equations on $\F$ has the same spirit as the relation between
the Dirac and the Klein-Gordon (squared Dirac) equations for spinor fields.}
These universal scale invariance conditions will be satisfied by the ABF background
for a particular choice of the vectors $X_m$ and $Y_m$.

To explain the origin of the second ``$I$-modified" set of
equations let us
first ignore the R-R fields and
assume that there exists the following metric-dilaton background that solves
the Weyl invariance equations (i.e. $R_{mn} + 2 D_m D_n \p=0$, \ $ \bar \b^\p =$const)
\be\la{ba1}
\hat {ds}^2 = e^{2\hat a(x)} [d\hat y + \hat A_\mu(x) dx^\mu]^2 + g_{\mu\nu} (x) dx^\mu dx^\nu \ ,
\qquad
\hat \phi = - c\, \hat y + f(x) \ . \ee
Here the metric has an isometry which is broken by the linear term in the dilaton ($c=\const$).
Examples of such non-trivial solutions\foot{It is important that dilaton has a linear
term in a ``warped" isometric direction of the metric, i.e. $a(x),\, A_\m(x)$ are non-constant, otherwise
the effect of adding the linear dilaton would be trivial.}
can be found by taking special limits of gauged WZW
backgrounds \ci{ht}.
T-dualising this metric, we find a diagonal metric $G$ and $B$-field, i.e.
\be \la{bb1}
ds^2= e^{2a(x)} d y^2 + g_{\mu\nu} (x) dx^\mu dx^\nu \ , \qquad \qquad
\ B= \hat A_\mu(x) \, dy \wedge dx^\m \ , \ \ \ \ \ \ \ \ \ \ \ a= - \hat a \ . \ee
For $c=0$ (i.e. when $\hat \p$ is isometric)
these fields together with the T-duality transformed dilaton $\p = \hat \p - \hat a$ would
solve the standard Weyl invariance equations \rf{1},\rf{2} with $X_m= \del_m \phi$, $Y_m=0$. For non-zero $c$ the equation
$\hat R_{mn} + 2 \hat D_m \hat D_n \hat \p=0$
(for the original background \rf{ba1}) expressed in terms of the
dual fields $G,B$ will contain additional $c$-dependent
terms obstructing (for non-constant $a(x)$)
the introduction of a new dilaton scalar. Still, they can be put in a more general form
$R_{mn} + D_m X_n + D_n X_m =0$ with a special vector $X$ given by\foot{The need to introduce the
vector $X_m$, which is {\it not} simply a gradient of a scalar,
is therefore directly related to the feature $\del_{\hat y} \hat \p = - c \not = 0$.}
\be \la{ba2}
X_m dx^m \equiv I_m dx^m + Z_m dx^m =
c\, e^{-2 a} \, dy + \big[ \del_\m (\hat \p - \hat a) + c\, \hat A_\m\big] \, dx^\m \ .
\ee
The dilaton equation $\bar \b^\p=0$ for the original background \rf{ba1}
also can be rewritten as
the following generalised equation (cf. \rf{cp})\foot{Note that this equation is not present in the list of scale invariance
conditions, and Weyl invariance conditions require this relation to hold with
$X_m =\del_m \phi$ for some $\phi$.}
\be \la{xd}
\bar \b^X\equiv R -\te {1\ov 12} H^2_{mnk} + 4 D^m X_m - 4 X^m X_m =0 \ , \ee
that is satisfied for the T-dual background.

The T-dual background $(G,B)$ defines a sigma model that is scale invariant
on a flat 2d background (satisfying equations \rf{1},\rf{2} with $Y_m=X_m$) but which is not Weyl
invariant. The trace of stress tensor $T= \b^G_{mn} \del_a x^m \del^a x^n
+ \b^B_{mn} \ep^{ab} \del_a x^m \del_b x^n$ is a total derivative
$T=\nabla^a N_a, \ \ N_a = 2 ( X_m \del_a x^m + \ep_{a}^{\ b} Y_m \del_b x^m) $ (up to terms proportional to the $x^m$ equations of motion).
This cannot be cancelled
by a local counterterm (the classical dilaton term) unless \ $X_m =\del_m \phi, \ Y_m=0$ \
\ci{Hull:1985rc,ts}, which is not the case for the ABF background. The sigma
models based on \rf{bb1} (with explicit backgrounds given below) thus represent
particular examples of 2d scale invariant theories that avoid the
Zamolodchikov-Polchinski theorem \ci{pol} due to their non-compactness (and/or
non-unitarity related to the presence of time-like directions). It thus
remains unclear if such backgrounds related by formal T-duality to Weyl
invariant models \rf{ba1} can also be associated somehow with a consistent
critical string theory.

As we shall see below, a similar generalisation of the full set of the bosonic
type II supergravity equations also exists in the presence of R-R
fields $\F_n$ that have the same isometries as the metric (i.e. when \rf{ba1}
is extended to an analog of the HT solution \ci{ht2}). Thus in general, given a type II
solution with non-isometric linear dilaton there will be an associated
(``T-dual'' or ABF-like) background solving such a modified set of type II
equations.

\medskip

The rest of this paper is organised as follows. In section \ref{Sect:NSNS} we
shall present the general scale invariance conditions for the couplings $G,B$
of the sigma model \rf{0} that generalise \rf{1},\rf{2} to the presence of the
R-R fields $\F$ and show that there exist such vectors $X=Z+I$ and $Y$ that
these equations are satisfied by the ABF background. In section \ref{rreom} we
shall derive a modification of the standard 1st-order IIB supergravity
equations of the R-R fluxes that is ``driven'' by the special isometry vector
$I$ and which are satisfied by the ABF background. In section \ref{s4} we shall show
that combining these 1st-order equations one can find 2nd-order equations for
$\F$ that have the right structure (when generalised to arbitrary vector $X$)
to be interpreted as scale invariance conditions on the R-R couplings. In
section \ref{s5} we explain how the standard type II supergravity equations for a
solution with the dilaton linear along the isometric directions is mapped to
the modified equations for T-dual solution.

Our notation and some useful relations are summarised in Appendix \ref{A}. In
Appendix \ref{B} we present the explicit form of the ABF background and the
T-dual type IIB HT solution. Appendix \ref{C} contains the derivation of the
identity $\del_m \bb^X=0$ from the modified type II equations which is closely
related to the on-shell conservation of R-R stress tensor. In Appendix \ref{D},
starting with the modified type II equations, we derive the 2nd-order equations
for the R-R fields that are candidates for the corresponding scale invariance
conditions. In Appendix \ref{appderiv} we remark on an alternative derivation
of the relation \rf{bZ} for the vector $Z$, which plays the role of the dilaton
one-form in the modified equations. In Appendix \ref{F} we summarise the
analogs of the ABF and HT backgrounds in the \adt and \adst cases and give the
corresponding expressions for the vectors $X,Y$ and $I$ that solve the scale
invariance and modified type II equations. In Appendix \ref{R} we explain how
the 2nd-order equations for the R-R couplings $\F$ emerge as the one-loop
conditions of scale invariance for the GS sigma model \rf{0}.

\section{Scale invariance conditions and modified type II equations: NS-NS sector}\label{Sect:NSNS}

The scale invariance conditions for the bosonic sigma model \rf{1},\rf{2} have
a straightforward generalisation to the GS superstring case with non-zero R-R
couplings $\F=e^\p F$ (see Appendix \ref{R}). The $\bar \theta \F \theta \del
x \del x $ terms in the GS action \rf{0} should lead to one-loop diagrams (with
one bosonic and one fermionic line) contributing logarithmic UV divergences
$\sim \F \F \del x \del x $. These terms will produce extra $\F\F$ terms in
the $\beta$-functions in \rf{1} and \rf{2}. In particular, the analog of the
Einstein equation \rf{1} should pick up the R-R stress tensor term and the
$B$-field equation \rf{2}, the $\F \F$ term as in the II supergravity
equations.\foot{For an argument supporting this in the NSR formalism see
\ci{Berenstein:1999ip}.} This is expected as for $X_m=\del_m \phi, \ Y_m=0$ the
resulting equations are the Weyl invariance equations that
should be equivalent to the type II supergravity equations.

The scale invariance equations for the $\F$-fields (to be discussed in section \ref{s4})
will not, however, have the familiar supergravity form of 1st-order equations for $\F$
(these should follow from the Weyl invariance conditions). Instead
they will be of 2nd order, $D^2 \F + ...= X$-dependent terms, and for $X_m=\del_m \phi$ will be a consequence
of the 1st-order supergravity equations.

Explicitly, the scale invariance conditions \rf{1} and \rf{2} generalise to
\be \la{4} &
\b^G_{mn} \equiv \te R_{mn} - {1\ov 4} H_{mkl} H_{n}{}^{kl} - \T_{mn} = - D_m X_n - D_n X_m \ ,
\\
&\te \b^B_{mn} \equiv \ha D^k H_{kmn} + \K_{mn} = X^k H_{kmn} + \del_m Y_n -\del_n Y_m \ , \la{5}
\\
\la{444}
&\te \T_{mn} \equiv {1\ov 2} \F_m \F_n + {1\ov 4} \F_{mpq} \F_{n}{}^{pq} + {1\ov 4\times 4!} \F_{mpqrs} \F_{n}{}^{pqrs} - {1\ov 2} G_{mn} (
{1\ov 2} \F_k \F^k + {1\ov 12} \F_{kpq} \F^{kpq} )\ , \ \ \ \ \ \
\\\qquad
& \te \K_{mn} \equiv \ha \F^{k} \F_{kmn} + {1 \ov 12} \F_{mnklp} \F^{klp} \ .
\la{55}
\ee
Here $\F_m,\F_{mnk},\F_{mnklp}$ are R-R fields of type IIB supergravity (for notation see Appendix \ref{A}).
For $X_m=\del_m \p, \ Y_m=0$ these equations follow from type IIB supergravity action \rf{ef}.
$\T_{mn} $ is the familiar stress tensor that
follows from the type IIB action \rf{ef} upon variation over $G_{mn}$.\foot{Note that
in the first (NS-NS) term of \rf{ef} one does not need to vary the $\sqrt G$ factor
as its contribution vanishes after use of the dilaton equation
$\bar \beta^\p=0$ in \rf{cp}. This equation is not required for scale invariance.}

As was noted in \ci{ht2}, the existence of the HT solution related to the ABF background
by T-duality, suggests that the GS sigma model for the latter defined on a flat 2d background should be scale
invariant (at least to leading, 1-loop, order).
Our key observation is that indeed there exist vectors $X_m$ and $Y_m$ such that
eqs. \rf{4} and \rf{5}
are satisfied for the ABF background \rf{abf}.
The vector $X_m$ required to satisfy \rf{4} turns out to be
(see Appendix {B} for notation)
\begin{equation}\la{X}
\begin{split}
X \equiv X_m dx^m =\ & c_0 \, \frac{1+\rho^2}{1-\vk^2\rho^2} dt + c_1 \, \rho^2 \sin^2 \zeta \, d\psi_2 + c_2\, \frac{\rho^2\cos^2\zeta}{1+\vk^2\r^4 \sin^2\z} d\psi_1
\\ + &c_3 \, \frac{1-r^2}{1+\vk^2r^2} d\vp + c_4 \, r^2 \sin^2 \xi \, d\phi_2 + c_5 \, \frac{r^2\cos^2\xi}{1+\vk^2 r^4 \sin^2\xi} d\phi_1
\\ + &\frac{\vk^2\r^4 \sin2\z}{2(1+\vk^2\r^4 \sin^2\z)} d\z + \frac1\rho \big(1 -\frac{3}{1-\vk ^2 \rho ^2}+ \frac{2}{1+ \vk ^2 \rho ^4 \sin ^2\zeta}\big) d\r
\\ + &\frac{\vk^2r^4 \sin2\xi}{2(1+\vk^2r^4 \sin^2\xi)} d\xi + \frac1r \big(1 -\frac{3}{1+\vk ^2 r ^2}+ \frac{2}{1+ \vk ^2 r ^4 \sin ^2\xi}\big)d r \ .
\end{split}\end{equation}
$X_m$
can be split in the following way
\be
X_m = I_m + Z_m \ , \ \ \ \qquad D_m I_n + D_n I_m =0\ , \ \ \ \qquad D^m I_m =0\ , \la{7}\ee
where $I^m= \sum_{i=0}^5 c_i ( I^{(i)})^m $. The index $i$ labels the 6 
isometric directions $y^i=(t,\psi_2,\psi_1,\vp,\p_2,\p_1)$ of the 10d ABF metric and $c_i$ are arbitrary constant coefficients.
$( I^{(i)})^m $ are the 6 independent commuting Killing vectors of the ABF background: the
Lie derivatives of the $G,B$ and $\F$-fields in \ci{abf} along $I^m$ all vanish.
If we split the coordinates as $x^m= (y^i, x^\m)$ where $\mu=1,2,3,4$ labels
the non-isometric directions $x^\m=(\zeta,\rho,\xi, r)$, then
\be \la{8}
I_m = \sum_{i=0}^5 \delta_{m}^i c_i G_{ii} (x^\mu) \ , \ \qquad \ \ \ \ I^m = \delta_{m} ^i c_i =\const \ , \qquad \ \ \ \ 
Z_m = \delta_{m}^\m Z_\m( x^\n)
\ .
\la{9}\ee
The vector $Y_m$ required to satisfy \rf{5} on the ABF background is found to be\foot{$Y$ is of course defined modulo a total derivative.}
\begin{equation}\la{Y}
\begin{split}
Y \equiv Y_m dx^m =\ & 4\vk\, \frac{1+\rho^2}{1-\vk^2\rho^2} dt + 2\vk\, \frac{\rho^2\cos^2\zeta}{1+\vk^2\r^4 \sin^2\z} d\psi_1
\\+ & 4\vk \, \frac{1-r^2}{1+\vk^2r^2} d\vp - 2\vk \, \frac{r^2\cos^2\xi}{1+\vk^2 r^4 \sin^2\xi} d\phi_1
\\ + & \frac{\vk^2\r^4 \sin2\z}{2(1+\vk^2\r^4 \sin^2\z)} d\z + \frac{1}{\rho} \big(1 -\frac{3}{1-\vk ^2 \rho ^2}+ \frac{2(\vk^{-1}c_2-1)}{1+ \vk ^2 \rho ^4 \sin ^2\zeta}\big) d\r
\\ + &\frac{\vk^2r^4 \sin2\xi}{2(1+\vk^2r^4 \sin^2\xi)} d\xi + \frac{1}{ r} \big(1 -\frac{3}{1+\vk ^2 r ^2}
- \frac{2(\vk^{-1} c_5+1)}{1+ \vk ^2 r ^4 \sin ^2\xi}\big)d r \ .
\end{split}\end{equation}
We observe that if we fix $c_i$ in \rf{X} to the following specific values
\begin{equation}\la{c}
c_0 = c_3 = 4 \vk \ , \qquad\qquad
c_1 = c_4 = 0 \ , \qquad\qquad
c_2 = -c_5 = 2 \vk \ ,
\end{equation}
then $Y_m$ and $X_m$ coincide
\begin{equation}\la{xy}
Y_m
= X_m
\ .
\end{equation}
The next surprising observation is that for these specially chosen values of $c_i$ in \rf{c}
the vector $X_m$ satisfies also a direct generalisation \rf{xd}
of the dilaton equation \rf{cp} ($\del_m \p \to X_m$):\foot{Since $D_n X^n = D_\m Z^\m \ , \ \
X^m X_m = G^{ij} c_i c_j + G^{\m\n} Z_\m Z_\n$
this equation does not depend on signs of $c_i$.}
\be \la{10} \te
\bb^X\equiv R - {1\ov 12} H^2_{mnk} + 4 D_k X^k - 4 X_k X^k =0 \ . \ee
As we shall show in Appendix \ref{C} this $\bb^X$ satisfies the generalisation of the dilaton identity \rf{cp}
\be
\del_m \bb^X = 0 \ .
\ee

The reason for this particular choice of $c_i$ in \rf{c} can be traced to the form of the
linear terms in the dilaton $\hat \p$ of the T-dual HT solution \rf{hts}. That is the presence
of the $I$-term in $X_m$ in \rf{7} reflects the presence of the non-isometric linear terms in $\hat \p$.
Therefore, these terms drive the modification of the equations satisfied by the ABF background
from their standard type II form.
In this sense the $Z_m$ part of $X_m$ may be interpreted as the analog of $\del_m \phi$ in
the modified equations.
Indeed, one can check that for $I^m$ in \rf{9} with $c_i$ chosen as in \rf{c}
the following relation is satisfied
\bal\la{bZ}
\del_m Z_n - \del_n Z_m + I^k H_{kmn} =0\ .
\eal
This may be interpreted as a modified ``dilaton Bianchi identity":
if $I_m$ is formally set to zero then $Z_m$ becomes a derivative of a scalar, $\del_m \p$.
In general, assuming that $I_m$ represents an isometry of the $B$-field, i.e. the Lie derivative
($\LL_I B)_{mn} = I^k \del_k B_{mn} + B_{kn} \del_m I^k - B_{km} \del_n I^k$ vanishes (modulo a gauge transformation term
$ \del_m U_n - \del_n U_m$), we can solve \rf{bZ} as\foot{In general, we find
$ Z_m = \del_m \p + B_{km} I^k - U_m$.
Under gauge transformations of $B$ the vector $U_m$
transforms so that $\p$ may be assumed to be invariant.
In the particular case of the ABF background \rf{abf} with the $B$-field chosen in the manifestly symmetric form
we have $U_m=0$.}
\be
Z_m = \del_m \p + B_{km} I^k
\ , \la{uuu}\ee
where $\del_m \p$ term represents the trivial ``zero-mode" solution.
In the particular case of the ABF background with $Z_m$ and $I_m$ given by \rf{X},\rf{7},\rf{9} and $c_i$ fixed as in
\rf{c} we find
\be
&&X_m = Y_m = I_m + Z_m= \del_m \p + ( G_{km}+ B_{km} ) I^k \ , \label{vu1}\\
&&
\p= \ha \log{ (1-\kappa ^2 \rho ^2)^3 (1+\kappa ^2 r^2)^3 \ov
(1+\kappa ^2 \rho ^4 \sin ^2\zeta) (1+\kappa ^2 r^4 \sin ^2\xi
) } \label{sol}\ .
\ee
The scalar $\p$ in \rf{sol} is precisely the one that is found \ci{ht2}
by applying the standard T-duality transformation rule
to the isometric part of the dilaton $\hat \p$ of the HT solution in \rf{hts}
(cf. \rf{ba2}).

\section{Modified type II equations: first-order equations for R-R couplings}\label{rreom}

Let us now explore what modification of the type IIB equations for the R-R couplings is satisfied
by the ABF background.

The standard equations of type IIB supergravity \ci{schwarz} in the R-R sector
written in terms of the rescaled $\F=e^\p F$ field strengths are pairs
of dynamical equations and Bianchi identities (see Appendix \ref{A} for notation)\foot{Note that all equations including \rf{5} are invariant under the simultaneous change of sign of $H_3$ and $F_3$,
or of $H_3$, $F_1$ and $F_5$. The choice of sign of $H_3$ or $B$ can be changed by parity.}
\begin{align}
&\te D^m \F_m - Z^m\F_m - {1 \ov 6} H^{mnp} \F_{mnp} =0 \ , && d \F_1 - Z\W \F_1=0 \ , \la{1b} \\
&\te D^p \F_{pmn}- Z^p\F_{pmn} - {1 \ov 6} H^{pqr}\F_{mnpqr} =0 \ ,&& d \F_3- Z\W \F_3 + H_3 \wedge \F_1 =0 \ , \la{3b}\\
&\te D^r \F_{rmnpq}- Z^r\F_{mnpq} + {1\ov 36} \ve_{mnpqrstuvw}H^{rst}\F^{uvw} =0 \ , && \label{5b}
d \F_5- Z\W \F_5 + H_3 \wedge \F_3 =0 \ .
\end{align}
Here $Z = Z_m dx^m =d\phi $ is the dilaton one-form.
The five-form $\F_5$ is also required to satisfy
the self-duality equation $ \star \F_5 = \F_5$ which implies the equivalence of
the first and second equation in \rf{5b}.

An a priori surprising observation is that
there exist direct generalisations of the 1st-order equations
\rf{1b}--\rf{5b} involving $Z=Z_m dx^m$ and $I=I_m dx^m $ in \rf{X},\rf{7}, with fixed values of the
coefficients $c_i$ as given in \eqref{c}, which are solved by the ABF background \rf{abf}.
Explicitly, the equations for the one-form $\F_1$ in \rf{abf} are
\begin{align}
& D^m \F_m - Z^m \F_m - \te {1 \ov 6} H^{mnp} \F_{mnp} =0 \ , \qquad \qquad I^m \F_m = 0 \ , \label{ef1}\\
& (d \F_1 - Z \wedge \F_1)_{mn} - I^p \F_{mnp} = 0 \ . \label{bf1}
\end{align}
We have added the condition $ I^m \F_m = 0$ as an independent equation on $\F_1$.\foot{Alternatively, one
can derive this equation
from the Bianchi equation \eqref{bf1}, the invariance of $\F_1$ under the isometry, the orthogonality of $I$ and $Z$, and the condition that $Z$ is not an exact one-form. Indeed, multiplying \eqref{bf1} by $I^m$ one finds
$\pa_n(I^m \F_m) - Z_n I^m\F_{m} = 0$
Thus, if $I^m\F_{m} \neq 0$ then $Z = d\ln (I^m \F_m)$. We find, however, it more convenient to add $ I^m \F_m = 0$ as an independent equation, and infer from it the orthogonality of $I$ and $Z$.}

Similarly, the equations that generalise \rf{3b} and
are satisfied for the three-form $\F_3$ in \rf{abf} are found to be
\begin{align}
& D^p \F_{pmn} - Z^p \F_{pmn} - \te \frac{1}{6} H^{pqr}\F_{mnpqr} - (I\wedge \F_1)_{mn}= 0 \ , \label{ef3}
\\
& (d\F_3 - Z \wedge \F_3 + H_3 \wedge \F_1)_{mnpq} - I^r \F_{mnpqr} =0 \ . \label{bf3}
\end{align}
The equations satisfied by $\F_5$ of the ABF background are found to be
\begin{align}
& D^r \F_{rmnpq} - Z^r\F_{rmnpq} + \te {1 \ov 36} \ve_{mnpqrstuvw}H^{rst}\F^{uvw} - (I \wedge \F_3)_{mnpq} =0 \ , \label{ef5}
\\
& (d \F_5- Z \wedge \F_5 + H_3 \wedge \F_3)_{mnpqrs} +
\te \frac{1}{6}\ve_{mnpqrstuvw}I^t\F^{uvw} = 0 \ . \label{bf5}
\end{align}
These two are equivalent in view of the self-duality of $\F_5$.

These modified equations \rf{ef1}--\rf{bf5} reduce back to \rf{1b},\rf{3b},\rf{5b}
if we drop all terms with $I_m$ and assume that $dZ=0$, i.e. if we set
\be \la{back} Z_m \to \del_m\p \ , \qquad \qquad I_m \to 0 \ . \ee
The structure of \rf{ef1}--\rf{bf5} supports the interpretation of $Z$ as a generalised ``dilaton one-form",
while the isometry vector $I$ effectively drives the deformation of the standard type IIB equations.

An interesting observation is that there exist certain combinations of the equations
\eqref{ef1}--\eqref{bf5} that depend on $Z$ and $I$ only through
the combination $X = Z + I$, which entered the NS-NS equations of the previous section.
These are found by adding together equations of equal form degree,
for example, the equation of motion for the R-R three-form and the Bianchi identity for the
R-R one-form. The resulting $X$-dependent equations are given by
\begin{align}
&\te D^m \F_m - X^m \F_m - {1 \ov 6} H^{mnp} \F_{mnp} =0 \ ,\label{eqf0} \\
& \te D^p \F_{pmn} - X^p \F_{pmn} - \frac{1}{6} H^{pqr}\F_{mnpqr} \label{eqf2}
+ (d \F_1 - X \wedge \F_1)_{mn} = 0 \ , \\
&D^r \F_{rmnpq} - X^r\F_{rmnpq} +\te {1\ov 36} \ve_{mnpqrstuvw}H^{rst}\F^{uvw} +
(d\F_3 - X \wedge \F_3 + H_3 \wedge \F_1)_{mnpq} =0 \ . \label{eqf4}
\end{align}
Using the self-duality of $\F_5$ the last equation can be also written as
\bea
\la{BianF5}\te
(d\F_5-X\wedge \F_5+H_3\wedge \F_3)_{pqrlmn}-\frac{1}{6}\ve_{pqrlmn vstu}(D^v \F^{stu}-X^v\F^{stu}-\F^vH^{stu})=0\ .
\eea
As will be discussed below, these three equations are already
sufficient for deriving candidates for the scale invariance equations for the $\F$-fields, which are 2nd order in derivatives.

It is useful to rewrite \rf{1b}--\rf{5b} in the
notation of forms (see
Appendix \ref{A} for conventions).
To do so we
introduce
the dual forms defined by
\begin{equation}\la{dualF}
\F_1 = \star \F_9 \ , \qquad \F_3 = - \star \F_7 \ , \qquad \F_5 = \star \F_5 \ , \qquad \F_7 = - \star \F_3 \ , \qquad \F_9 = \star \F_1 \ .
\end{equation}
Then the complete set of the type II supergravity equations
for R-R strengths and Bianchi identities \rf{1b}--\rf{5b} is given by\foot{We assume that $\mathcal{F}_n = 0$ for $n< 0$ and $n>10$.}
\bal \label{feq}
& d\F_{2n+1} - Z\W \F_{2n+1} + H_3 \W \F_{2n-1} = 0 \ , \qquad\ \qquad n= 0, 1, ...\ ,
\\
& d \star \F_{2n+1} - Z \W \star \F_{2n+1} - H_3 \wedge \star \F_{2n+3} = 0 \ ,\ \ \qquad n=0,1,...\ ,
\eal
where $Z=d\phi$.

The ``$I$-modified'' equations \rf{ef1}--\rf{bf5} are given by\foot{Note that here we include
$n=-1$ as in
the deformed theory it is no longer trivial:
it gives the second equation in \rf{ef1}, i.e.
$\star (I\wedge \star \F_1) = I^m \F_m = 0$.}
\bal\label{defeq}
& d\F_{2n+1} - Z\W \F_{2n+1} + H_3 \W \F_{2n-1} - \star (I \wedge \star \F_{2n+3}) = 0 \ , \ \ \ \ \ \qquad n=-1, 0, ...\ ,
\\
& d \star \F_{2n+1} - Z \W \star \F_{2n+1} - H_3 \wedge \star \F_{2n+3} + \star (I \wedge \F_{2n-1}) = 0 \ , \ \ \ \qquad n=0,1,... \ .
\eal
Due to \eqref{dualF} the two equations in \rf{feq} are equivalent
and the same is true for \rf{defeq}.

Let us note that the deformed R-R equations \rf{defeq}
together with the relation \rf{bZ} or $dZ +\iota_I H_3=0$ imply the following relation
\begin{equation}\label{lif}
\LL_I \F_{2n+1} = (I\cdot Z) \F_{2n+1} \ .
\end{equation}
Thus the condition that the $\F$-fields are invariant under the isometry $I$
is equivalent to the condition $I\cdot Z = 0$, which is clearly satisfied
for the ABF background as is evident from \rf{X},\rf{9}.

\section{Second-order equations for R-R couplings as scale invariance conditions}\label{s4}

Let us return to the discussion of the scale invariance conditions for the couplings of the GS sigma model \rf{0}
in section \ref{Sect:NSNS} and consider the equations for the R-R couplings $\F$ that
should follow from the requirement of (1-loop) UV finiteness of the 2d model.
One can argue that the conditions analogous to eqs.~\rf{4},\rf{5} for the $G$ and $B$-field couplings should have the form
\be
\beta^{\F}_{k_1 ... k_s} \equiv \ha D^2 \F_{k_1 ... k_s} + ... =
X^m \del_m \F_{k_1 ... k_s} + \sum_i \F_{k_1 ...m ... k_s} \del_{k_i} X^m \ ,
\la{25a}
\ee
where we have omitted possible non-linear terms such as $R \F+ D H\F +... $ on the l.h.s.
The $X$-dependent Lie derivative term on the r.h.s.
reflects, as in \rf{4},\rf{5}, the reparametrisation (or off-shell $x^m$-renormalisation) freedom.
For example, starting with the linearized RG equation $ {d \F_n(x) \ov d t} =
\beta^{\F}_{n} = \ha \del^2 \F_n (x) , \ t=\log \epsilon$ and doing the
coordinate redefinition $x^m\to x^m + t X^m$, one ends up with $ {d \F_n(x) \ov
d t} = \ha \del^2 \F_n (x) - X^m \del_m \F_n - \F_m \del_n X^m$. 

We shall discuss the computation of 1-loop logarithmic UV divergences for the
GS action \rf{0} in Appendix \ref{R} clarifying the structure of $\beta^\F$.

For $X_m=\del_m \p$ the equations \rf{25a} should be a consequence of stronger Weyl invariance conditions,\foot{For
example, using the NSR approach on a flat background
we may consider the R-R vertex operators
built out of spin operators and consider the linearised conditions for conformal invariance (marginality).
Then $d F=0, \ d\star F =0$ will follow (see, e.g., \cite{Blumenhagen:2013fgp})
just like the usual transversality conditions
on the graviton operator follow from the marginality conditions
of the $h_{mn}(p) e^{ipx} \del x^m \del x ^n$ vertex.
On a curved 2d background these are equivalent to the decoupling of derivatives $\del_a \rho$
of the conformal factor of the 2d metric (see, e.g., \ci{Weinberg:1985tv,Callan:1986ja}).
These conditions are stronger than just scale invariance which requires only ``masslessness"
$p^2 F(p) =0$ or $\del^2 F=0$.}
which should be
equivalent to the type II supergravity equations \rf{1b}--\rf{5b} or \rf{feq} where $Z=X=d\phi$.
Indeed, combining (``squaring") the familiar $dF+...=0, \ d\star F + ...=0$ equations
leads to $ d\star d\star F + \star d\star d F + ...=0$ or $ D^2 F +...=0$, where the leading term is the Hodge-de Rham operator.

Moreover, the same equations should follow also
from the modified type II equations \rf{ef1}--\rf{bf5} or \rf{defeq}
(as, e.g., the ABF background that solves the modified equations should also be a solution of the scale invariance conditions).
This should provide a non-trivial consistency check: after properly ``squaring" \rf{ef1}--\rf{bf5} the dependence on the $Z$ and $I$ vectors
in any candidate scale invariance equations
should appear only through their sum $X= Z + I$ as in \rf{4},\rf{5}.

Starting from the modified type II equations \rf{ef1}--\rf{bf5} (which
include the standard type IIB supergravity equations
as a special case \rf{back}, $I_m=0$),
let us outline the derivation of the 2nd order equations for the R-R couplings
that should be equivalent to the scale invariance conditions for $\F_n$ of the GS sigma model \rf{0}.
To be a candidate for the scale invariance conditions these equations should have the following properties:

(i) vanish on the supergravity equations \eqref{4},\eqref{5},\eqref{10},\eqref{1b}--\eqref{5b} with $X = d\p$,\
\,$Y = 0$

(ii)
depend on $Z$ and $I$ through $X= Z + I$

(iii)
depend on $X$ through Lie derivatives.\foot{Moreover, since
the R-R fields $\F$ are invariant under the isometries generated by $I$, their
Lie derivatives along $I$ vanish, and therefore the scale invariance equations
in fact depend only on $Z$.}

Starting with the modified equations \eqref{defeq} and
acting with $\star d\star$ on the first equation and $d\star $ on the second
and then using the modified equations (as described in Appendix \ref{D})
we arrive at the following equation, which satisfies the above properties
\begin{equation}\begin{split}\label{candidate}
&d\star d\star \F_{2n+1} + \star d \star d \F_{2n+1} + \tfrac14 R\W \F_{2n+1} - \tfrac18\star (H_3 \W \star H_3) \W \F_{2n+1}
\\ &- H_3 \W \star (H_3 \W\star\F_{2n+1}) - \star (H_3 \W \star (H_3 \W \F_{2n+1}))
\\ & - d \star (H_3 \W \star \F_{2n+3}) - \star (H_3 \W \star d\F_{2n+3})
+ \star d \star (H_3 \W \F_{2n-1})+ H_3 \W \star d\star \F_{2n-1}
\\ & = \mathcal{L}_X \F_{2n+1} + \star \mathcal{L}_X \star \F_{2n+1} - (\star d \star X) \W \F_{2n+1} + \beta^B \W \F_{2n-1} -\star (\beta^B \W \star \F_{2n+3}) \ .
\end{split}\end{equation}
Here $\beta^B$ is the 2-form analog of \rf{5}, i.e.
\begin{equation}
\beta^B \equiv \tfrac12 \star d\star H_3 + \K = \star( X \W \star H_3) + d Y \ .
\end{equation}
This is then a candidate for the scale invariance equation for the R-R form $\F_{2n+1}$.

Using the identity
\bea
&& \la{ga}
\star \LL_X \star \F_{2n+1} = \LL_X \F_{2n+1} + \star (d\star X)\W \F_{2n+1} + \b^G \cdot \F_{2n+1}
\ ,\\
&& \no \ \ \ \ \ \ \
\b^G \cdot \F_{2n+1} \equiv \sum_i \beta^G_{m_i n} \F_{m_1 ...m_{i-1}}{}^n{}_{ m_{i+1} ... m_{2n+1}}\ ,
\eea
where $\beta^G_{mn}$ is defined in \rf{4},
we find that \eqref{candidate} becomes
\begin{equation}\begin{split}\label{candidate2}
&d\star d\star \F_{2n+1} + \star d \star d \F_{2n+1} + \tfrac14 R\W \F_{2n+1} - \tfrac18\star (H_3 \W \star H_3) \W \F_{2n+1} 
\\ &- H_3 \W \star (H_3 \W\star\F_{2n+1}) - \star (H_3 \W \star (H_3 \W \F_{2n+1}))
\\ & - d \star (H_3 \W \star \F_{2n+3}) - \star (H_3 \W \star d\F_{2n+3})
+ \star d \star (H_3 \W \F_{2n-1})+ H_3 \W \star d\star \F_{2n-1}
\\ & = 2\mathcal{L}_X \F_{2n+1} + \b^G \cdot \F_{2n+1}
+ \beta^B \W \F_{2n-1} -\star (\beta^B \W \star \F_{2n+3}) \ .
\end{split}\end{equation}
The dependence of these equations on $X$ rather than separately on $Z$ and $I$ can be related to their close connection to
the particular $X$-dependent combinations of the modified equations in \eqref{eqf0},\eqref{eqf2},\eqref{eqf4}, i.e. to
(here $n \in \mathbb{Z}$ as in \rf{defeq})
\begin{equation}\begin{split}\label{eeeeee}
\Ee_{2n}& \equiv d\F_{2n-1} - X \wedge \F_{2n-1} + H_3 \W \F_{2n-3}
\\ & \quad+ (-1)^n \star (d \F_{9-2n} - X \wedge \F_{9-2n} + H_3 \W \F_{7-2n}) = 0 \ .
\end{split}\end{equation}
We also define as in \rf{xd},\rf{5}
\begin{equation}\begin{split}\label{bbpp}
\bb^B & \equiv \tfrac12 \star d \star H_3 + \K - \star (X \W \star H_3) - dX = 0\ , \\
\bb^X & \equiv R-\tfrac12 \star (H_3 \W \star H_3) + 4 \star d\star X - 4 \star(X\W \star X) = 0 \ .
\end{split}\end{equation}
Deconstructing the derivation in Appendix \ref{D}, we find that the 2nd-order equation
for the R-R fluxes \eqref{candidate} can also be written as
\begin{equation}\begin{split}\la{432}
& d\Ee_{2n} - X\W \Ee_{2n} + H_3 \W \Ee_{2n-2} -\F_{2n-1} \W \bb^B
\\ & + (-1)^n \star (d \Ee_{8-2n} - X\W\Ee_{8-2n} + H_3 \W \Ee_{6-2n} - \F_{7-2n} \W \bb^B \big) + \tfrac14 \F_{2n+1} \W \bb^X = 0 \ .
\end{split}\end{equation}

Finally, let us present the explicit form of eq. \rf{candidate2} in components.
For $\F_1$ we find
\bal
& D^2 \F_m - R_{mn}\F^n + \tfrac14 (R - \tfrac3{4} H^2)\F_m
\\
& \hspace{20pt}+ \tfrac12 H^{pnk}H_{mpn}\F_k-\tfrac16 D_m H^{pnk} \F_{pnk} - \tfrac12 H^{pnk}D_p \F_{nkm} \\
& \hspace{20pt}= 2 (X^p D_p \F_m + D_m X^p\F_p ) + \beta^G_{mn} \F^n- \tfrac12 \beta^B_{nk} \F^{nk}{}_{m} \ .
\la{scinvF1}
\eal
Using the identity $D_{[m}H_{npk]}=0$ the term $\tfrac16 D_m H^{pnk} \F_{pnk}$ in \eqref{scinvF1} can be replaced
by $\tfrac{1}{2}D_p H_{mnk}\F^{pnk}$.
The equation for $\F_3$ may be written as
\bal
&D^2\F_{nkm}-R_{a[n}\F^a{}_{km]}+R_{ab[nk}\F^{ab}_{~~~m]} + \tfrac14 (R - \tfrac3{4} H^2)\F_{nkm}\\
&\quad +\tfrac12 H^{abc}H_{ab[n}\F_{km]c}-\tfrac12 H^{abc}H_{a[nk}\F_{m]bc} \\
&\quad +D^aH_{a[nk}\F_{m]} +H_{a[nk}D^a\F_{m]} -\F_aD^aH_{nkm}\\
& \quad -\tfrac16 D_{[n}H^{abc}\F_{km]abc} -\tfrac12 H^{abc}D_a\F_{bc nkm} \\
&\ \ \ ~~=2(X^aD_a\F_{nkm}+D_{[n}X^a\F_{km]a})+\b_{a[n}^G \F^{a}{}_{km]}+\b_{[nk}^B\F_{m]}-\tfrac12\b_{ab}^B\F^{ab}{}_{nkm}\ , \label{scinvF3}
\eal
while the equation for $\F_5$ can be put into the form
\bea
\la{scinvF5}
\begin{aligned}
&D^2\F_{ijklm}-R_{a[i}\F^a{}_{jklm]}+R_{ab[ij}\F^{ab}_{~~~klm]} +\tfrac{1}{4}(R-\tfrac{3}{4}H^2)\F_{ijklm} \\
&~~ +\tfrac{1}{2}H^{abc} H_{ab[i}\F_{jklm]c} - \tfrac{1}{2}H^{abc}H_{a[ij}\F_{klm]bc} \, \, \\
&~~+D^aH_{a[ij}\F_{klm]}+H_{a[ij}D^a\F_{klm]} -\F_{a[ij}D^a H_{klm]} \, \\
&~~+\tfrac{1}{12}\ve_{ijklm bdef}(D_aH^{abc}\F^{def}+H^{abc}D_a\F^{def}-\F^{abc} D_aH^{def} )=\\
&~~=2(X^aD_a\F_{ijklm}+ D_{[i} X^a\F_{jklm]a} ) +\beta^G_{a[i}\F^a{}_{jklm]} +\beta^B_{[ij}\F_{klm]} +\tfrac{1}{12}\ve_{ijklm abcde}(\beta^B)^{ab}\F^{cde}\ .
\end{aligned}
\eea
This expression is consistent with the self-duality of $\F_5$ (in particular, the third and forth lines are manifestly dual to each other).

These 2nd-order equations for $\F_1$, $\F_3$ and $\F_5$ exhibit obvious structural similarities.
In particular, they contain the expected Hodge-de Rham operator terms and the
vector $X$ only enters through the reparametrisation terms as in \rf{25a}. The
$\beta^G$ and $\beta^B$ terms in these equations are defined as in
\rf{4},\rf{5} but can also be replaced by expressions on the r.h.s. of
\rf{4},\rf{5}.

As we shall discuss in Appendix \ref{R}, similar equations come out of the
computation of the one-loop beta-functions for the R-R couplings in the GS
sigma model \rf{0}.

\section{Origin of modified equations: T-duality relation to type II equations \\ for backgrounds with non-isometric linear dilaton}\label{s5}

Given a scale invariant sigma model in flat 2d space T-duality in an
isometric direction should also produce a scale invariant sigma model.
Similarly, given a Weyl invariant sigma model on curved 2d space with all
couplings including the dilaton being isometric the T-dual background should
also be Weyl invariant (provided the dilaton transforms in the usual way
\ci{bush,swt}). As discussed in the introduction, in general this is not so
if the dilaton is not isometric: T-duality will still preserve scale invariance
but not Weyl invariance. Thus given a solution of type II supergravity
equations which has linear non-isometric term in the dilaton its T-duality
image will no longer solve the standard type II equations but will satisfy
instead a modified set of type II equations as discussed above.

\subsection{Simple examples}

Here we shall make the origin of the modified equations explicit by showing that they represent the original type II equations
for a solution with non-isometric linear dilaton, rewritten in terms of the fields of the T-dual background.
To explain how this happens in simple terms let us first start with a bosonic background \rf{ba1} with $A_\m=0$, i.e.
\be\la{ba15}
\hat {ds}^2 = e^{-2a(x)} d\hat y ^2 + g_{\mu\nu} (x) dx^\mu dx^\nu \ ,
\qquad\qquad \hat \phi = - c\, \hat y + \vp(x) - \ha a(x) \ . \ee
Then the corresponding Weyl anomaly coefficients
\bea \la{721}
\bar \beta^G_{mn} = \hat R_{mn} + 2 \hat D_m \hat D_n \hat \phi \ , \qquad \qquad
\bar \beta^\p= \hat R + 4 \hat D^2 \hat \p - 4 \hat G^{mn} \del_m \hat \p \del_n\hat \p \ ,
\eea
have the following components under the $\hat x^m = (\hat y, x^\m)$ splitting of coordinates\foot{Here $\rr_{\m\n}$ is the Ricci tensor of
$g_{\mu\nu} (x)$, see Appendix \ref{A}.}
\bea \la{77x} &&
\bar \beta^G_{\m\n } = \rr_{\m\n} - \del_\m a \del_\n a + 2 D_\m D_\n \vp \ , \qquad \quad \,
\bar \beta^G_{\hat y\hat y } = e^{-2a} ( D^\m D_\m a -2 \del^\m a \del_\m \vp) \ , \\ &&
\bar \beta^G_{\hat y \m } = -2 c\ \del_\m a \ , \qquad \qquad
\la{80x} \bar \beta^\p
= \rr - \del^\m a \del_\m a + 4 D^2 \vp - 4 \del^\m \vp \del_\m \vp - 4 c^2 e^{2a} \ .
\eea
We see that if $c=0$, i.e. the dilaton is isometric, then the Weyl invariance conditions $ \bar \beta^G_{\m\n } =0, \ \bar \beta^\p=0 $
are invariant under T-duality in $y$, i.e. under
$ \hat a=-a \to a, \ \hat \p \to \hat \p + a$ or $ \vp \to \vp $. The $c= - \del_{\hat y} \hat \p $ dependent terms in \rf{80x} thus
represent obstructions
to mapping one Weyl invariant model to another. The T-dual metric then solves weaker, modified, equations
\bea \la{81}
&& R_{mn} + D_m X_n + D_n X_m =0 \ , \ \ \ \ \ \ \ \qquad
\bar \beta^X= R + 4 D^m X_m - 4 X^m X_m =0 \ , \la{82}
\eea
with $X_m$ being (cf. \rf{ba2})
\be \la{83}
X_\m = Z_\m = \del_\m \phi = \del_\m (\vp +\ha a ) \ , \qquad \qquad
X_y = I_y = - G_{yy} \del_{\hat y} \hat \p = c e^{2a} \ .
\ee
Similar conclusions are reached in the case we have a non-diagonal metric
in \rf{ba1} (see the general discussion below). The presence of non-zero $\hat A_\m$ is in fact necessary to have a solution
of the Weyl invariance conditions when $c\not=0$ (cf. \rf{80x}) and the target space should thus be at least 3-dimensional.
An example of such a solution was found in \ci{ht}.
It represents a limit of the background associated with the $SO(4)/SO(3)$ gWZW model, which has the following metric and dilaton \ci{fl}
\bea
&&\hat ds^2 = dt^2 +{ \tan^2 t \ d p^2 + \cot^2 t \ dq^2 \ov 1 - p^2 - q^2 }
= dt^2 +{ \cot^2 t \ ( d \theta + \tan \psi \cot \theta d \psi)^2 + { \tan^2 t \ov \sin^2 \theta} d \psi^2 } \ , \ \quad
\la{d1} \\
&&\hat \phi = - \ln \big( \sqrt{ 1 - p^2 - q^2} \sin 2t \big) 
= - \ln \big( \sin \theta \ \cos\psi\ \sin 2t \big) \la{d2} \ , \ \ \ \eea
where $p= \sin \psi , \ q= \cos \theta \cos\psi $ and $t,\psi,\theta$ are angles of the coset parametrisation of the $SO(4)$ group element.
This background (which solves the Weyl invariance condition $\bar \beta^G=0$ with $\bar\beta^\p =\const$)
has no isometries.
One option to generate an isometry is to set $t=iz$ and then shift $z$ by an infinite constant.
Doing so we get linear dilaton in $z$, but the $z$ direction decouples in the metric.
A non-trivial alternative is to set $\psi = i \hat y$ and to shift $\hat y$ by an infinite constant
(which corresponds to infinite rescaling of $p,q$ generating a scaling isometry in the metric \rf{d1}).
The resulting background (we drop an infinite constant in the dilaton)
\bea
&&\hat ds^2 = dt^2 - { \tan^2 t \ d p^2 + \cot^2 t \ dq^2 \ov p^2 + q^2 } =
dt^2 + \cot^2 t \ ( d \theta + \cot \theta\, d \hat y)^2 - { \tan^2 t \ov \sin^2 \theta} \, d\hat y^2 \ ,
\la{d3} \\
&&\hat \phi= - \ln \big( \sqrt{ p^2 + q^2} \sin 2t \big)
= - \hat y - \ln \big( \sin \theta \ \sin 2t \big) \la{d4} \ , \ \ \ \eea
is therefore of the same type as in \rf{ba1} and defines a conformal sigma
model\foot{The central charge for this $d=3$ conformal 
model is given by $c= d - \tfrac32 \a' (R - \tfrac 1{12} H^2 + 4 D^2 \phi - 4 D_m \phi D^m \phi) + ... = 3 - \tfrac 32 \a' \times
12 + ...$. Here the scale of the space was set to one, 
so that $\a'$ is then the inverse of the WZW level $k$. This is in
agreement with the usual count of the central charge for the $SO(4)/SO(3)$
gWZW model $c= 6 k /(k+4) - 3k/(k+2) = 3 - 18/k + ...$, 
which should be unchanged in the coordinate limit leading from \eqref{d1} to \eqref{d3}.}
Similar higher dimensional backgrounds can be constructed starting from
$SO(n)/SO(n-1)$ gWZW models with $n >4$ \ci{ht}.

T-dualising the metric \rf{d3} along $\hat y$ we get a $(G,B)$ background that will solve the
modified $(G,B)$ equations \rf{4},\rf{5} with non-trivial $X_m = I_m + Z_m$,
where $I^y = - \del_{\hat y} \hat \p$ and $Z_m$ is given by \rf{uuu}, with
$\p = \hat \p - \ha \log G_{\hat y \hat y} $. These modified equations will be the original
Weyl invariance conditions rewritten in terms of the dual $G$ and $B$-fields.

Given the 2d CFT in \eqref{d3} with 2d stress-tensor defined taking into account the dilaton
in \eqref{d4}, one may formally compactify $\hat y$ and ask if this CFT has T-duality as
a symmetry of its spectrum. The answer appears to be no as the 2d stress-tensor will not
be invariant under T-duality (i.e. mapping momentum into winding modes). \foot{Given 
a free compact scalar CFT $L= r^2 (\del \phi)^2$ with $\phi \equiv \phi + 2
\pi$ the spectrum of dimensions of primary operators (like $e^{i n \phi + i m
\tilde \phi}$, etc.) is T-duality symmetric. If one formally adds a linear
dilaton term $q \int d^2 z \sqrt h\, R^{(2)} \phi$, or equivalently modifies
the 2d stress tensor by $q\del^2\phi$ terms (which are invariant under shifts
of $\phi$ and thus defined for a compact boson) then the T-duality symmetry of
the spectrum is broken by extra terms $\sim q n$. The formal symmetry would
be restored in the ``doubled" formulation if the linear dilaton
 term were given by $q \phi +\tilde q \tilde \phi$ where $\tilde \phi$ is the dual
field (with ${1\ov \sqrt r } q \leftrightarrow \sqrt { {r\ov \sqrt {\a'}}} \td q $ under T-duality). } This is compatible with our expectation that formally T-dualising the
metric \eqref{d3} will not lead to a consistent CFT.

The same conclusions are reached for type II solutions with a linear dilaton and non-zero R-R fluxes
(with isometric $G,B$ and $\F_n= e^\p F_n$), such as the HT solution dual to ABF background in the \adss case and its counterparts
in the \adt and \adst cases discussed in Appendix \ref{F}.
Explicitly, in the case of a solution ($\hat G,\hat B, \hat \F_n, \hat \p)$ with several isometries broken only by the linear dilaton term
\begin{equation}\la{sev}
\hat \phi = \phi_0 - c_i \hat y^i + f(x^\m) \ ,
\end{equation}
we will get a
generalisation of the type II supergravity equations,
depending on the following two vectors $Z$ and $I$
(cf.\rf{7},\rf{9},\rf{uuu})
\begin{equation}\la{iz}
I = c_i G_{y_i y_i} dy^i \ , \qquad Z = d\p + \iota_I B \ , \qquad \p = f - \ha \sum_i \log \hat G_{\hat y_i \hat y_i} \ .
\end{equation}
Here $G$ and $B$ are the background fields of the T-dual background.

Below we shall illustrate these relations in the general case with one isometry.

\subsection{NS-NS sector}

We consider the following two $d$-dimensional backgrounds (here we use $K_\m$ instead of $\hat A_\m$ in \rf{ba1})
\begin{equation}\begin{split}\label{back1}
ds^2 & = e^{2a} (d \yy + \my_\mu dx^\mu)^2 + g_{\mu\nu} dx^\mu dx^\nu \ ,
\\
B & = \by_\nu (d\yy + \tfrac12 \my_\mu dx^\mu) \wedge dx^\nu + \tfrac12 b_{\mu\nu} dx^\mu \wedge dx^\nu \ ,
\\
\phi & = - c\, \yy + \varphi + \tfrac12 a \ , \qquad
I^{\yy} = \otherc\, \ , \qquad I^\mu = 0 \ ,
\end{split}\end{equation}
\begin{equation}\begin{split}\label{back2}
\hat{ds}^2 & = e^{-2a} (d\ty + \by_\mu dx^\mu)^2 + g_{\mu\nu} dx^\mu dx^\nu \ ,
\\
\hat{B} & = \my_\nu (d\ty + \tfrac12 \by_\mu dx^\mu) \wedge dx^\nu + \tfrac12 b_{\mu\nu} dx^\mu \wedge dx^\nu \ ,
\\
\hat{\phi} & = - \otherc\, \ty + \varphi -\tfrac12 a \ , \qquad
\hat I^\ty = c \ , \qquad \hat{I}^\mu = 0 \ .
\end{split}\end{equation}
Here $\yy$ and $\ty$ are the directions that are assumed to be
(shift) isometries of their respective metrics and $B$-fields.
We use the indices $\mu,\nu,...=1, ...,d-1$ and $m,n,...=1, ..., d$.
 For $c=\hat c =0$ \eqref{back1} and \eqref{back2} are
related by standard T-duality, such that $\vp$ is the analog of the duality-invariant dilaton field.
Let us also define
\begin{equation}\begin{split}\label{756}
Z & = d \phi + \iota_{I} B = -c\, d\yy + d \varphi + \tfrac12 da + \otherc\, \by_\mu dx^\mu \ ,
\\ X & = Z + I =(-c+\otherc\, e^{2a}) d\yy + d \varphi + \tfrac12 da + (\otherc\, \by_\mu +\otherc\, e^{2a} \my_\mu) dx^\mu \ ,
\\ \hat Z & = d \hat\phi + \iota_{\hat I}\hat B = -\otherc\, d \ty + d \varphi - \tfrac12 da + c\, \my_\mu dx^\mu \ ,
\\ \hat{X} & = \hat Z + \hat I =(-\otherc\, + c\, e^{-2a}) d \ty + d \varphi - \tfrac12 da + ( c \, \my_\mu + c\, e^{-2a} \by_\mu) dx^\mu \ ,
\end{split}\end{equation}
where
\begin{equation}
Z \cdot I = \hat Z \cdot \hat I = -c \, \otherc\, \ .\la{559}
\end{equation}
The two $(G,B)$ backgrounds in \eqref{back1} and \eqref{back2} are T-dual to each other
and thus for $c=\hat c=0$ solve the equivalent Weyl invariance equations (see, e.g., \ci{Haagensen:1996np}
and references therein).
We will now show how this relation also extends to the more general case with linear dilatons.

Let us first consider the generalised dilaton equation
\begin{equation}\label{check}
R - \tfrac1{12} {H}^2 + 4 {D}^m {X}_m -4 {X}^m {X}_m = 0 \ .
\end{equation}
The question we want to address is: if the background \eqref{back1} satisfies \eqref{check}
does that imply that \eqref{back2} satisfies \eqref{check}.
As
the two backgrounds \eqref{back1} and \eqref{back2} are related by the obvious
symmetry
\begin{equation}\label{ssyymm}
a \to - a \ , \qquad \my_\mu \leftrightarrow \by_\mu \ , \qquad c \leftrightarrow \otherc\, \ , \qquad \yy \leftrightarrow \ty \ ,
\end{equation}
it is sufficient to compute the left-hand side of \eqref{check} for \eqref{back1} and
check that it is invariant (or at least covariant) under \eqref{ssyymm}.

For \eqref{back1} we have
\begin{equation}\label{xx}
X_{\yy} = - c + \otherc\, e^{2a} \ , \qquad \qquad X_\mu = \partial_\mu \varphi + \tfrac12\partial_\mu a
+ \otherc\, \by_\mu + \otherc\, e^{2a} \my_\mu \ .
\end{equation}
It will also be useful to define the following objects
\begin{equation}\label{fbfgh}
\ffb_{\mu\nu} \equiv \partial_\mu \my_\nu - \partial_\nu \my_\mu \ , \qquad
\ffg_{\mu\nu} \equiv \partial_\mu \by_\nu - \partial_\nu \by_\mu \ , \qquad
h_{\mu\nu\rho} \equiv (db + \tfrac12 \my \wedge d \by + \tfrac12 \by\wedge d\my)_{\mu\nu\rho} \ ,
\end{equation}
where we observe that $h$ is invariant under \eqref{ssyymm}.
Now using the dimensional reduction formulae in Appendix \ref{A} we find
\begin{equation}\begin{split}\label{outhere1}
R & - \tfrac1{12} {H}^2 + 4 {D}^m {X}_m -4 {X}^m {X}_m
\\ & = \rr
- \partial^\mu a \partial_\mu a
- \tfrac1{12} h^2
- \tfrac 14 e^{2a} \ffb^{\mu\nu}\ffb_{\mu\nu}
- \tfrac14 e^{-2a} \ffg^{\mu\nu}\ffg_{\mu\nu} + 4\nabla^\mu \partial_\mu \vp - 4\partial^\mu \vp \partial_\mu \vp
\\ & \qquad + 4c\, \nabla^\mu \my_\mu + 4 \otherc\, \nabla^\mu \by_\mu- 8(c\, \my^\mu + \otherc\, \by^\mu) \partial_\mu \vp
\\ & \qquad- 4c^2 (\my^\mu \my_\mu + e^{-2a}) - 4 \otherc^2( \by^\mu \by_\mu +e^{2a}) + 8 c \, \otherc\, (1 - \my^\mu \by_\mu)\ ,
\end{split}\end{equation}
which is indeed invariant under \eqref{ssyymm}.
Therefore, if \eqref{check} is satisfied for
background \eqref{back1} it is satisfied for background \eqref{back2}
and vice versa.\foot{This generalises the usual ($c= \otherc= 0$) discussion of the T-duality invariance of
the string effective
action \rf{ef} with $\sqrt G \, e^{-2\p}=\sqrt g\, e^{-2\vp}$.}

Let us now turn to the modified metric and $B$-field equations to
show that the two combinations appearing in \rf{1} and \rf{2}
\begin{align}\label{tcc1}
& {R}_{mn} -\tfrac14 {H}_{mpq}{H}_n{}^{pq} + {D}_m {X}_n + {D}_n {X}_m \ ,
\\ \label{tcc2}
& \tfrac12 {D}^p H_{mnp} - X^p {H}_{mnp} - D_m X_n + D_n X_m \ ,
\end{align}
are covariant under the symmetry \eqref{ssyymm}.\foot{Here we assume $Y_m$ in \rf{2},\rf{5} is equal to $X_m$ in \rf{1},\rf{4}
as is the case for the $I$-modified equations satisfied by
the ABF background. More generally, given a scale invariant sigma model with an isometry and the $G$ and $B$-field couplings
satisfying \rf{1},\rf{2}, its T-dual counterpart will also satisfy \rf{1},\rf{2} with the roles of $X_m$ and $Y_m$ interchanged.}
Then if they vanish for the background \eqref{back1} this implies that they vanish for \eqref{back2} and vice versa.
As \eqref{tcc1} is symmetric and \eqref{tcc2} is antisymmetric, we may just consider their difference
\begin{equation}\label{cccc}
C_{mn} \equiv {R}_{mn} -\tfrac14 {H}_{mpq}{H}_n{}^{pq} - \tfrac12 {D}^p H_{mnp} + 2 {D}_m {X}_n + X^p {H}_{mnp} \ ,
\end{equation}
which we can decompose into a part independent of $c$ and $\otherc$ ($C^{(0)}_{mn}$) and a part linear
in $c$ and $\otherc$ ($C^{(1)}_{mn}$) as
\begin{align}
C_{mn} & = C^{(0)}_{mn} + 2C^{(1)}_{mn}\ ,
\\\label{c00}
C^{(0)}_{mn} & = {R}_{mn} -\tfrac14 {H}_{mpq}{H}_n{}^{pq} - \tfrac12 {D}^p H_{mnp} + 2 {D}_m {X}^{(0)}_n + X^{(0)}{}^p {H}_{mnp} \ ,
\\\label{c11}
C^{(1)}_{mn} & = {D}_m {X}^{(1)}_n +\tfrac12 X^{(1)}{}^p {H}_{mnp} \ .
\end{align}
Here we have used the fact that all the $c$ and $\otherc$ dependence is contained in $ X = X^{(0)} +X^{(1)} $, where
$ X^{(0)}$ is the $c$- and $\otherc$-independent and $ X^{(1)}$ is the $c$- and $\otherc$-dependent part.
Using the specific form of ${X}$ for the background \eqref{back1}, as given in \eqref{xx},
we have
\begin{align}
X^{(0)}_{\yy} = 0 \ , \quad X^{(0)}_\mu = \partial_\mu\varphi + \tfrac12 \partial_\mu a \ , \quad X^{(1)}_{\yy} = - c + \otherc\, e^{2a} \ , \quad
X^{(1)}_\mu = \otherc\, \by_\mu + \otherc\, e^{2a} \my_\mu \ . \label{xx2}
\end{align}
Using the formulae in Appendix \ref{A} we find the following relations for $C^{(0)}_{mn}$ and $C^{(1)}_{mn}$
evaluated on the background \eqref{back1}
{\small
\begin{align}\nonumber
& C^{(0)}_{\yy \yy} = 2 e^{2a} \partial^\mu \vp \partial_\mu a - e^{2a} \nabla^\mu \partial_\mu a + \tfrac14 e^{4a} \ffb_{\mu\nu}\ffb^{\mu\nu} - \tfrac14 \ffg_{\mu\nu}\ffg^{\mu\nu} \ ,
\\\nonumber
& C^{(0)}_{\yy \mu} - \frac{{G}_{\yy \mu}}{\goo} C^{(0)}_{\yy \yy} = e^{2a}(\tfrac12 \nabla^\nu + \partial^\nu a - \partial^\nu \vp) \ffb_{\mu\nu} + \tfrac14 h_{\mu\nu\rho}\ffg^{\nu\rho}
+ (\tfrac12 \nabla^\nu - \partial^\nu a - \partial^\nu \vp) \ffg_{\mu\nu} + \tfrac14 e^{2a} h_{\mu\nu\rho}\ffb^{\nu\rho} \ ,
\\\nonumber
& C^{(0)}_{\mu\yy} - \frac{{G}_{\mu\yy }}{\goo} C^{(0)}_{\yy \yy} = e^{2a}(\tfrac12 \nabla^\nu + \partial^\nu a - \partial^\nu \vp) \ffb_{\mu\nu} + \tfrac14 h_{\mu\nu\rho}\ffg^{\nu\rho}
- (\tfrac12 \nabla^\nu - \partial^\nu a - \partial^\nu \vp) \ffg_{\mu\nu} - \tfrac14 e^{2a} h_{\mu\nu\rho}\ffb^{\nu\rho} \ ,
\\\nonumber
& C^{(0)}_{\mu\nu} - \frac{{G}_{\mu\yy }}{\goo} C^{(0)}_{\yy \nu} - \frac{{G}_{\yy\nu}}{\goo} C^{(0)}_{\mu\yy} + \frac{ G_{\mu\yy}}{\goo}\frac{{G}_{\yy \nu}}{\goo} C^{(0)}_{\yy \yy}
= \rr_{\mu\nu} - \partial_\mu a\partial_\nu a + 2 \nabla_\mu \partial_\nu \vp
- \tfrac12 e^{2a} \ffb_{\mu\nu}\ffb_\nu{}^\rho - \tfrac12 e^{-2a} \ffg_{\mu\rho} \ffg_\nu{}^\rho
\\ & \hspace{220pt}- \tfrac12 \nabla^\rho h_{\mu\nu\rho} - \tfrac14 h_{\mu\rho\sigma}h_\nu{}^{\rho\sigma} + h_{\mu\nu\rho}\partial^\rho \vp \ ,
\end{align}
\begin{align}\nonumber
& C^{(1)}_{\yy \yy}
=e^{2a}(c \my^\mu + \otherc\, \by^\mu ) \partial_\mu a \ ,
\\\nonumber
& C^{(1)}_{\yy \mu} - \frac{{G}_{\yy \mu}}{\goo} C^{(1)}_{\yy \yy} = - \tfrac12 (e^{2a} \ffb_{\mu\nu} + \ffg_{\mu\nu})(c\my^\nu + \otherc\, \by^\nu) + (c-\otherc\, e^{2a}) \partial_\mu a \ ,
\\\nonumber
& C^{(1)}_{\mu\yy} - \frac{{G}_{\mu\yy }}{\goo} C^{(1)}_{\yy \yy} = - \tfrac12 (e^{2a} \ffb_{\mu\nu} - \ffg_{\mu\nu})(c \my^\nu \otherc\, \by^\nu) + (c+\otherc\, e^{2a}) \partial_\mu a \ ,
\\\nonumber
& C^{(1)}_{\mu\nu} - \frac{{G}_{\mu\yy }}{\goo} C^{(1)}_{\yy \nu} - \frac{{G}_{\yy\nu}}{\goo} C^{(1)}_{\mu\yy} + \frac{ G_{\mu\yy}}{\goo}\frac{{G}_{\yy \nu}}{\goo} C^{(1)}_{\yy \yy}
= \tfrac12 c (\nabla_\mu \my_\nu + \nabla_\nu \my_\mu)
+ \tfrac12 \otherc\, e^{2a} (\nabla_\mu \my_\nu - \nabla_\nu \my_\mu)
\\ & \hspace{190pt} \nonumber
+ \tfrac12 \otherc(\nabla_\mu \by_\nu + \nabla_\nu \by_\mu)
+ \tfrac12 c e^{-2a} (\nabla_\mu \by_\nu - \nabla_\nu \by_\mu)
\\ & \hspace{190pt}
+ \tfrac12 h_{\mu\nu\rho} (c \my^\rho + \otherc\, \by^\rho) \ .\vphantom{\int}
\end{align}}
Then using the map \eqref{ssyymm}
between the backgrounds \eqref{back1} and \eqref{back2}, we find the following relations
for $C_{mn}$
\begin{equation}\begin{split}\label{ttraas}
&\te \frac{C_{\yy \yy}}{\goo} = - \frac{\hat C_{ \ty \ty}}{\hat G_{\ty\ty}}
\ , \qquad \qquad
{\te \frac{1}{\sqrt{\goo}} }\big[C_{\yy \mu} - \frac{{G}_{\yy \mu}}{\goo} C_{\yy \yy}\big]
={\te \frac{1}{\sqrt{\goon}}}\big[\hat C_{ \ty \mu} -\frac{ \hat{G}_{ \ty \mu}}{\goon} \hat C_{ \ty \ty}\big]\ ,
\\
& \hspace{110pt}\te {\te \frac{1}{\sqrt{\goo}}} \big[C_{\mu \yy} - \frac{{G}_{\mu\yy }}{\goo} C_{\yy \yy}\big]
= - {\te \frac{1}{\sqrt{\goon}}}\big[\hat C_{\mu \ty} - \frac{\hat{G}_{\mu \ty }}{\goon} \hat C_{ \ty \ty} \big]\ ,
\\
&\te C_{\mu\nu} - \frac{{G}_{\mu\yy} }{\goo} C_{\yy \nu} - \frac{{G}_{\yy\nu}}{\goo} C_{\mu\yy} + \frac{ G_{\mu\yy}}{\goo}\frac{{G}_{\yy \nu}}{\goo} C_{\yy \yy} 
=
\hat C_{\mu\nu} - \frac{\hat{G}_{\mu \ty} }{\goon} \hat C_{ \ty \nu} - \frac{\hat{G}_{ \ty\nu}}{\goon} \hat C_{\mu \ty} + \frac{\hat G_{\mu \ty}}{\goon}\frac{\hat{G}_{ \ty \nu}}{\goon}\hat C_{ \ty \ty}\ ,
\end{split}\end{equation}
where the left-hand side is evaluated on \eqref{back1} and the right-hand side on \eqref{back2}.
From these equalities it follows that the vanishing of the tensors \eqref{tcc1},\eqref{tcc2} on the background \eqref{back1}
implies their vanishing also on the background \eqref{back2}, and in this
sense are covariant under T-duality.

Finally, let us note that the same analysis can be carried out with $I^\mu$ and
$\hat I^\mu$ in the backgrounds \eqref{back1} and \eqref{back2}
non-vanishing (i.e. when there are extra isometries in the $x^\m$ directions).
The T-duality relation between the equations for \eqref{back1} and
\eqref{back2} still holds with $I^\mu = \hat I^\mu$.
The same is also true for
the modified equations of motion for the R-R fields discussed in the following
section.
Furthermore, as $I^\mu = \hat I^\mu$ should also be a Killing vector
we still have $Z \cdot I = \hat Z \cdot \hat I = - c \hat c$.

Therefore, we can now use these relations to perform the sequence of
T-dualities required to transform from the
HT supergravity solution of \cite{ht2} to the ABF background \eqref{abf} 
and its $AdS_3 \times S^3$ and $AdS_2 \times S^2$ counterparts \eqref{ads3},\eqref{ads2}.

\subsection{R-R sector}

Let us now consider the case of non-zero isometric R-R fields $\F_n$.
The contribution of the R-R fields to the metric and $B$-field equations appears in
the usual unmodified form and hence we can focus our attention on the modified
equations of motion for the R-R fields \rf{defeq}. Written in terms of the
forms $f_k\equiv e^{-a/2}\F_k$ these take the following form (dropping the distinction between $y$ and $\ty$)
\bal
\label{eqfk2}
&\E_k\equiv df_{k} - Z'\W f_{k} + H_3 \W f_{k-2} -\tildec\, \iota_{y}f_{k+2}= 0 \ ,
\eal
\bal
\label{eqfk3}
&\hat{\E}_k\equiv d\hf_{k} - \hZ'\W \hf_{k} + \hH_3 \W \hf_{k-2} - c\, \iota_{y}\hf_{k+2}= 0 \ ,
\eal
where we have introduced ($K= K_\m dx^\m, \ A= A_\m dx^\m$)
\bal
Z' = Z - \ha da =-c\, dy +d\vp + \tildec\, K \ ,\qquad\quad \hZ' \equiv \hat{Z}-\ha d\ah=-\tildec\, dy +d\vp + c\, A \ ,
\eal
which are related to each other under T-duality as
\bal
\hZ' = Z' +c(dy + A) -\tildec (dy+K)\equiv Z' + \delta Z\ .
\eal
Recall that the invariance of the R-R forms under the isometry along $y$ requires
\be
c\, \tildec=0 \ . \la{cec} \ee
This follows from the condition $I\cdot Z = 0$, which is implied by
the invariance of R-R fields as $\LL_I \F_k = (I\cdot Z) \F_k$.

We want to show that if $f_k$ satisfies the equation $\E_k=0$, then $\hf_k$ satisfies $\hat{\E}_k=0$.
Taking into account the T-duality relations in Appendix \ref{A}
one finds
\bal
-\hZ'\W \hf_{k} =& -(dy+K)\W Z'\W f_{k-1}+(dy+K )\W(dy+A )\W \iota_{y}(Z'\W f_{k-1}) - \iota_{y}(Z'\W f_{k+1})\\
&+c (dy+K)\W(dy+A )\W f_{k-1} - c f_{k+1} -\delta Z\W (-(dy+K )\W f_{k-1} - \iota_{y}f_{k+1})\ ,\no
\eal
\bal
\hH_3\W \hf_{k-2} =& (dy+K)\W H_3\W f_{k-3}-(dy+K)\W(dy+A)\W \iota_{y}(H_3\W f_{k-3}) + \iota_{y}(H_3\W f_{k-1})\\
&+(dy+K)\W(dy+A)\W \iota_{y}H_3\W f_{k-3} - \iota_{y}H_3\W f_{k-1} \\
&+((dy+A)\W H_2 - (dy+K)\W F_2)\W (-(dy+K)\W f_{k-3}- \iota_{y}f_{k-1})\ ,\no
\eal
\bal
-c\, \iota_{y}\hf_{k+2}=&-\tildec(dy + K)\W \iota_{y}f_{k+1}+\tildec(dy + K)\W \iota_{y}f_{k+1}+cf_{k+1}-c(dy + A)\W \iota_{y}f_{k+1} \ .
\eal
Here we used that
$\delta Z\W (dy+K)\W(dy+A)\W \iota_{y}f_{k-1} =0$ and $((dy+A)\W H_2 - (dy+K)\W F_2)\W (dy+K)\W(dy+A)\W \iota_{y}f_{k-1} =0$.
Further using that $\iota_{y}H_3=-H_2$, one finds
\bal
\hat{\E}_k =& (dy+K)\W \E_{k-1}-(dy+K)\W(dy+A)\W \iota_{y}\E_{k-1} + \iota_{y}\E_{k+1}
\\
&-H_2\W f_{k-1}+H_2\W(dy+A)\W \iota_{y}f_{k-1}-(dy+K)\W F_2\W \iota_{y}f_{k-1}\\
&+c (dy+K)\W(dy+A)\W f_{k-1} - c f_{k+1} -\delta Z\W (-(dy+K)\W f_{k-1}- \iota_{y}f_{k+1})\\
&-(dy+K)\W(dy+A)\W H_2\W f_{k-3} +H_2\W f_{k-1}\\
&+((dy+A)\W H_2 - (dy+K)\W F_2)\W (-(dy+K)\W f_{k-3} - \iota_{y}f_{k-1})\\
&+\tildec(dy + K)\W \iota_{y}f_{k+1}+cf_{k+1}-c(dy + A)\W \iota_{y}f_{k+1}\ .
\eal
If we set $\E_{k}=0$ and $c=\tildec=0$ we get
\bal
&H_2\W(dy+A)\W \iota_{y}f_{k-1}-(dy+K)\W F_2\W \iota_{y}f_{k-1}\\
&-(dy+K)\W(dy+A)\W H_2\W f_{k-3} \\
&+((dy+A)\W H_2 - (dy+K)\W F_2)\W (-(dy+K)\W f_{k-3} - \iota_{y}f_{k-1})=0\ ,
\eal
as expected. It remains to consider the $c$ and $\tildec$ dependent terms only
\bal
\hat{\E}_k =& (dy+K)\W \E_{k-1}-(dy+K)\W(dy+A)\W \iota_{y}\E_{k-1} + \iota_{y}\E_{k+1}\\
+
&c (dy+K)\W(dy+A)\W f_{k-1} - c f_{k+1} \\
&-(c(dy + A) -\tildec (dy+K))\W (-(dy+K)\W f_{k-1} - \iota_{y}f_{k+1})\\
&+\tildec(dy + K)\W \iota_{y}f_{k+1}+cf_{k+1}-c(dy + A)\W \iota_{y}f_{k+1}\\
=& (dy+K)\W \E_{k-1}-(dy+K)\W(dy+A)\W \iota_{y}\E_{k-1} + \iota_{y}\E_{k+1}\ .
\eal
Thus, if ${\E}_k=0$ then $\hat{\E}_k=0$, i.e. the backgrounds \eqref{back1} and \eqref{back2}
supplemented by R-R fields have their corresponding modified equations mapped into each other
by this generalised T-duality.

\section{Concluding remarks}

There are several open problems and puzzling questions.
First, it remains unclear if the scale invariant but
arguably not Weyl invariant $\eta$-model can still be used to define a critical
superstring theory. This might be possible in view of the existence of the
$\lambda$-model \ci{hms} which is classically related to the $\eta$-model by
the Poisson-Lie duality combined with an analytic continuation of the
deformation parameter, and for which there is a candidate supergravity solution
\ci{st} (i.e. it should represent a Weyl invariant sigma model). In fact, a
special limit \ci{ht} of this solution should be essentially equivalent to the
HT solution \ci{ht2}.\foot{The need for T-duality in order to relate the HT
solution to the ABF background can be understood from the two facts: that the
$\lambda$-model is a deformation of the non-abelian T-dual of the \adss sigma
model and that in the limit of \ci{ht}, which enhances the Cartan directions
making them the isometries, the non-abelian T-duality along these isometric
directions turns into the standard abelian one.} Thus if the classical
Poisson-Lie duality relation \ci{kli2} between the \emo\ and \lmo\ \ci{vic,ht2}
extends to the full quantum level there may be a way to associate a string
theory to the ABF background. This might also require increasing the number of
2d fields (such as in a doubled or phase space formulation). Indeed, already at the
classical level, establishing the connection between the two models calls for
the use of the phase space formalism.
The quantum \emo\ defined in terms of an extended number of fields
(including, e.g., analogs of 2d gauge fields of the gWZW part of $\l$-model) may
then be Weyl invariant, and integrating out extra fields might produce the
GS action corresponding to the ABF background plus extra non-local terms
required for restoring its Weyl invariance.

As we have seen above, the fact that the HT background solves the type IIB
equations implies that the T-dual ABF background should satisfy the
$I$-modified type II equations. These explicitly depend on the isometry vector $I$,
whose origin can be traced to the presence of the linear term in the dilaton of
the HT solution. One can ask whether these $I$-modified equations are
Lagrangian, i.e. if they can be derived from the action principle. Answering
this question may require the introduction of R-R potentials and
understanding whether one should treat the vector $I$ as an external source or
as an auxiliary field with no physical degrees of freedom. In view of our
analysis of T-duality in section \ref{s5}, it would be interesting to know if there
exist more general $I$-modified equations that are compatible with T-duality
and have $c\, \hat c\neq 0$ in \rf{559}. One would also like to understand how
the usual action of the T-duality group $O(d,d)$ is modified.

In the present work we discussed only the $I$-modified equations for bosonic
fields. It is an interesting question how the equations for the fermionic fields
of type II theory are modified. Furthermore, if the $I$-modification destroys
the local supersymmetry of type II theory one may ask if there is
still any (hidden) symmetry of the $I$-modified equations for bosonic and fermionic
fields.

To better understand the nature of the ABF background it would be important to
derive the quartic fermionic action for the $\eta$-model of \ci{dmv} and to
show that the $I$-modified equations indeed follow from the $\kappa$-symmetry
\ci{dmv} of this action.
Starting with the standard GS action for the HT solution \ci{ht2} (which, as
was mentioned in the Introduction, is invariant under shifts of the 6 isometric
coordinates) and performing the T-dualities one will get $\theta^4$ and higher
terms in the $\eta$-model GS action depending on the vectors $I$ and $Z$.
These will originate from the dilaton, $\theta^4\del \p$, etc., terms in the HT
GS action. The resulting $\eta$-model action should still be invariant under the
$\kappa$-symmetry defined in \cite{dmv}, however it is then probable that the
structure of these transformations will deviate from those of the usual GS
action.

The knowledge of the quartic fermionic action should
also enable one to perform the full computation of one-loop divergences of the
$\eta$-model in the R-R sector (completing our discussion in Appendix \ref{R})
and hence check the agreement between the 2nd-order equations for R-R fields
derived from the modified type II equations with the scale invariance
beta-functions for $\F_n$.

It would also be important to attempt a direct analysis of the Weyl invariance
conditions, which should lead to 1st-order conditions for R-R strengths
equivalent to type II supergravity equations. More generally, one may study the
one-loop renormalisation of a generic $\kappa$-symmetric sigma model with 8
bosonic and 8 fermionic degrees of freedom, and classify interaction terms for
which the corresponding model is either conformal or scale invariant only. It
is possible that the class of conformally invariant models may be bigger than
just the usual type II GS superstring sigma models.

Finally, it would be interesting to perform a similar analysis for the
deformations of $AdS_n \times S^n$ backgrounds constructed from other solutions
of the modified classical YB equation \cite{dmv,hh}, or solutions of the
classical (non-modified) YB equation, see, e.g.,
\cite{Kawaguchi:2014qwa,Tongeren}. In the latter case many of the resulting
metrics and $B$-fields can be completed to full type II supergravity solutions,
however it remains to verify that these completions are indeed realised by the
supercoset action. Indeed, the analysis of \ci{abf2} has shown that the large
$\vk$-limit of the $\eta$-model does not coincide with the \adss mirror sigma
model \cite{ArT2} even though the bosonic part of the model does.

\section*{Acknowledgments}

We thank R. Borsato, O. Lunin, S. J. van Tongeren and L. Wulff for useful discussions and O. Lunin and L. Wulff for comments on the draft.
The work of G.A. is supported by the German Science Foundation (DFG) under the Collaborative Research Center (SFB) 676 Particles, Strings and the Early Universe.
The work of B.H. is partially supported by grant no. 615203 from the European Research Council under the FP7.
The work of R.R. is supported in part by the US Department of Energy under DOE Grants No: de-sc0013699.
The work of A.A.T. is supported by the ERC Advanced grant No.290456, the STFC grant ST/J0003533/1 and by the Russian Science Foundation grant 14-42-00047 associated with Lebedev Institute.

\appendix

\section{Conventions and some standard relations}\label{A}
\def\theequation{A.\arabic{equation}}
\setcounter{equation}{0}

\subsection*{Conventions for forms}\la{appforms}

We have for any $m$-form $Y$ and $n$-form $Z$ on a manifold of dimension $d$
\bal
&Z=\te {1\ov n!} Z_{i_1\cdots i_n} dx^{i_1}\wedge\cdots\wedge d x^{i_n}\ ,\qquad (\star Z)_{i_1\cdots i_{d-n}} = \frac{1}{n!}\ve_{i_1\cdots i_{d-n}j_1\cdots j_{n}} Z^{j_1\cdots j_n}\ ,\\
& \iota_I Z =\te {1\ov (n-1)!} I^pZ_{p i_2\cdots i_n} dx^{i_2}\wedge\cdots\wedge dx^{i_n}\ ,\qquad
(Y\wedge Z)_{i_1\cdots i_mj_1\cdots j_n} =Y_{[i_1\cdots i_m} Z_{j_1\cdots j_n]} \ , \\
& Y\wedge Z =\te {1\ov m!n!} Y_{i_1\cdots i_m} Z_{j_1\cdots j_n}dx^{i_1}\wedge\cdots\wedge dx^{i_m}\wedge dx^{j_1}\wedge\cdots\wedge d x^{j_n}
\ ,
\eal
where the antisymmetrisation is understood as
\bal\te
Y_{[i_1\cdots i_m} Z_{i_{m+1}\cdots i_{m+n}]}dx^{i_1}\wedge\cdots\wedge dx^{i_{m+n}}= {(m+n)!\ov m!n!} Y_{i_1\cdots i_m} Z_{i_{m+1}\cdots i_{m+n}}dx^{i_1}\wedge\cdots\wedge dx^{i_{m+n}}\ .
\eal
In $d$ dimensions with Lorentzian signature we have
\be
\star^2 Z_n = (-1)^{dn + n+1} Z_n\ ,
\quad \
\left[\star(Y_m \W \star Z_n)\right]_{i_1\cdots i_{n-m}} = {(-1)^{nd+n+1}\ov m!} Y^{j_1\cdots j_{m}}Z_{i_1\cdots i_{n-m}j_1\cdots j_{m} }
\ .
\ee
In particular for $m=1$ and even $d$ one has
\bal
\star(I \W \star Z_n) &= \iota_I Z_n
\ .
\eal

\subsection*{Dimensional reduction formulae}

Let us take the metric and $B$-field as in \eqref{back1}
\begin{equation}\begin{split}\label{backred}
ds^2 & = e^{2a} (d \yy + \my_\mu dx^\mu)^2 + g_{\mu\nu} dx^\mu dx^\nu \ ,
\\
B & = \by_\nu (d\yy + \tfrac12 \my_\mu dx^\mu) \wedge dx^\nu + \tfrac12 b_{\mu\nu} dx^\mu \wedge dx^\nu \ ,
\end{split} \end{equation}
where $y$ is an isometric direction.
It is useful to define
\begin{equation}\la{usd}
\ffb_{\mu\nu} \equiv \partial_\mu \my_\nu - \partial_\nu \my_\mu \ , \qquad
\ffg_{\mu\nu} \equiv \partial_\mu \by_\nu - \partial_\nu \by_\mu \ . \qquad
h_{\mu\nu\rho} \equiv (db + \tfrac12 \my \wedge d \by + \tfrac12 \by \wedge d\my)_{\mu\nu\rho} \ .
\end{equation}
We can now write the various $d$-dimensional quantities appearing in the modified type II
equations in terms of $(d-1)$-dimensional ones as follows
{\small
\begin{equation}\begin{split}
{G}^{\yy \yy} & = e^{-2a} + \my^2 \ ,
\qquad G^{\yy \mu} = - \my^\mu \ , \qquad G^{\mu\nu} = g^{\mu\nu} \ ,
\\
{H}_{\yy \mu\nu} & = - \ffg_{\mu\nu} \ , \quad \qquad \ H_{\mu\nu\rho} = h_{\mu\nu\rho} - (\my\wedge d\by)_{\mu\nu\rho} \ , \qquad
\end{split} \end{equation}
\bal
\G_{yy}^y &\te= e^{2a}A^\mu\pa_\mu a\ ,\qquad \G_{yy}^\mu = -e^{2a}\pa^\mu a\ ,\qquad \G_{y\mu}^y = \pa_\mu a + e^{2a}A_\mu A^\nu\pa_\nu a
+{1\ov 2}e^{2a}A^\nu F_{\nu\mu}\ , \\
\G_{\mu\nu}^y &\te = {1\ov 2}(\na_\mu A_\nu+\na_\nu A_\mu)+A_\nu\pa_\mu a +A_\mu\pa_\nu a+ e^{2a}A_\mu A_\nu A^\rho\pa_\rho a+{1\ov 2} e^{2a}A^\rho(A_\mu F_{\rho\nu}+A_\nu F_{\rho\mu})\ ,
\\
\G_{y\nu}^\mu &=\te -e^{2a}A_\nu\pa^\mu a-{1\ov 2} e^{2a}F^{\mu}{}_{\nu}\ , \qquad
\G_{ym}^m = 0\ ,\qquad \G_{\mu m}^m = \pa_\mu a + \g_{\mu\nu}^\nu\ , 
\\
\G_{\mu\nu}^\rho &\te = \g_{\mu\nu}^\rho- e^{2a}A_\mu A_\nu \pa^\rho a-{1\ov 2} e^{2a}(A_\mu F^{\rho}{}_{\nu}+A_\nu F^\rho{}_{\mu})= \g_{\mu\nu}^\rho+\delta\G_{\mu\nu}^\rho\ ,
\eal
\bal
R_{yy}&=\te -e^{2a}\na^\mu\pa_\mu a-e^{2a}\pa^\mu a\pa_\mu a +{1\ov4}e^{4a}F_{\mu\nu}F^{\mu\nu}\ ,\\
R_{y\mu}&=\te R_{yy}A_\mu-{3\ov2}e^{2a}\pa^\nu aF_{\nu\mu}-{1\ov2}e^{2a}\na^\nu F_{\nu\mu}\ , \\
R_{\mu\nu}&=\te \rr_{\mu\nu} -\na_\mu\pa_\nu a -\pa_\mu a\pa_\nu a + A_\mu R_{y\nu}+A_\nu R_{y\mu}
-A_\mu A_\nu R_{yy}-{1\ov2}e^{2a}F_{\mu}{}^\rho F_{\nu\rho}\ ,\\
R & =\te \rr - \tfrac 14 e^{2a} \ffb^{\mu\nu}\ffb_{\mu\nu} - 2 \partial^\mu a \partial_\mu a - 2 D^\mu \partial_\mu a\ ,
\eal
\bal
H_y{}^{kl} H_{ykl}&= H^{\mu\nu}H_{\mu\nu} \ ,\qquad H_y{}^{kl} H_{\mu kl} = -H^{\nu\rho}h_{\mu\nu\rho}+A_\mu H^{\nu\rho}H_{\nu\rho} \ ,\\
H_\mu{}^{kl} H_{\nu kl} &=h_{\mu}{}^{\rho\s}h_{\nu \rho\s}-h_{\mu}{}^{\rho\s}A_\nu H_{\r\s}-h_{\nu}{}^{\r\s}A_\mu H_{\r\s}+A_\mu A_\nu H^{\r\s}H_{\r\s}+2e^{-2a}H_{\mu\r}H_\nu{}^\r
\ ,\\
H^{\mu\nu\r} H_{\mu\nu\r} &=h^{\mu\nu\r}h_{\mu\nu\r}+3e^{-2a}H^{\mu\nu}H_{\mu\nu}\ ,
\eal
\bal
D^kH_{\mu yk}&\te =\na^\nu H_{\mu \nu} -H_{\mu\nu}\pa^\nu a+{1\ov2}e^{2a}F^{\nu\r}h_{\mu\nu\r}\ ,\\
D^kH_{\mu\nu k}&\te =\na^\r h_{\mu\nu\r} +(h_{\mu\nu\r}+A_\mu H_{\nu\r}-A_\nu H_{\mu\r})\pa^\r a - e^{2a}F^{\r\s}A_{[\mu}h_{\nu]\r\s}- 2A_{[\mu}\na^\r H_{\nu]\r}\ .
\eal
}
Also, for a vector $X = X_{\yy} d\yy + X_\mu dx^\mu$ we have
{\small
\begin{equation}\begin{split}\label{inhere}
D^m X_m & = \nabla^\mu X_\mu - X_{\yy} \nabla^\mu \my_\mu - X_{\yy} \my^\mu \partial_\mu a + X^\mu \partial_\mu a -\my^\mu \partial_\mu X_{\yy} \ ,
\\
X^m X_m & = e^{-2a} X_{\yy}^2 + X_{\yy}^2 \my^\mu\my_\mu - 2 X_{\yy} \my^\mu X_\mu + X^\mu X_\mu \ ,
\end{split}\end{equation}
\begin{equation}\begin{split}
D_{\yy}X_{\yy} & = e^{2a} X^\mu \partial_\mu a - e^{2a} X_{\yy} \my^\mu \partial_\mu a \ ,
\\
D_{\yy} X_{\mu} & = \tfrac12 e^{2a} (-\ffb_{\mu\nu} + 2 \partial_\nu a \my_\mu) (X^\nu - X_{\yy} \my^\nu) - X_{\yy} \partial_\mu a \ ,
\\
D_{\mu} X_\yy & = \tfrac12 e^{2a} (-\ffb_{\mu\nu} + 2 \partial_\nu a \my_\mu) (X^\nu - X_{\yy} \my^\nu) - X_{\yy} \partial_\mu a + \partial_\mu X_{\yy} \ ,
\\
D_{\mu} X_\nu & = \nabla_\mu X_\nu - X_{\yy} \my_\mu \partial_\nu a - X_{\yy} \my_\nu \partial_\mu a -\tfrac 12 X_{\yy} \nabla_\mu\my_\nu - \tfrac12 X_{\yy} \nabla_\nu\my_\mu
\\ &\qquad\qquad - \tfrac12 e^{2a} (\my_\mu \ffb_{\nu\rho} + \my_\nu \ffb_{\mu\rho} - 2 \partial_\rho a \my_\mu\my_\nu )(X^\rho - X_{\yy} \my^\rho) \ .
\end{split}\end{equation}
}
Here $Q_{[\mu\nu]}\equiv \ha ( Q_{\mu\nu}-Q_{\nu\mu})$, $\nabla_\m$ is the
covariant derivative with respect to the $(d-1)$-dimensional metric $g_{\m\n}$
with connection $\g_{\mu\nu}^\l$, and $\rr_{\mu\nu}$ and $\rr$ are the
$(d-1)$-dimensional Ricci tensor and scalar respectively.

\subsection*{T-duality rules}

Let us consider two isometric backgrounds related by T-duality, with the fields of the dual background denoted with hats.
The metric and $B$-field will be
taken in the form of \eqref{back1}, and we will also consider the isometric
dilaton $\p$ and the R-R field strengths $\F_{k}\equiv e^\p F_{k}$ of type II theory
\begin{equation}\begin{split}\label{backgr1}
ds^2 & = e^{2a} (d \yy + \my_\mu dx^\mu)^2 + g_{\mu\nu} dx^\mu dx^\nu\ ,\qquad\qquad \p\ ,\qquad \F_{k} \ ,
\\
B & = \by_\nu (d\yy + \tfrac12 \my_\mu dx^\mu) \wedge dx^\nu + \tfrac12 b_{\mu\nu} dx^\mu \wedge dx^\nu \ ,
\end{split} \end{equation}
\begin{equation}\begin{split}\label{backgr2}
d\hs^2 & = e^{2\ah} (d\hat \yy + \hat\my_\mu dx^\mu)^2 + \hg_{\mu\nu} dx^\mu dx^\nu\ ,\qquad\qquad \hp\ ,\qquad \hF_{k} \ ,
\\
\hB & = \hat\by_\nu (d\hat \yy + \tfrac12 \hat\my_\mu dx^\mu) \wedge dx^\nu + \tfrac12 \hb_{\mu\nu} dx^\mu \wedge dx^\nu \ .
\end{split} \end{equation}
The T-duality rules for the NS-NS fields are (see \rf{usd})
\bal\la{TrulesNS}
a=&-\ah\ ,\quad A_{\mu} = {\hat K_{\mu}} \ ,\quad g_{\mu\nu}=\hg_{\mu\nu}\ ,
\quad b_{\mu\nu}=\hb_{\mu\nu}\ ,\quad \p=\hp - \ah =\hp + a\ ,\\
K_{\mu}= & \hA_\mu\ ,\quad
F_{\mu\nu} = {\hat H_{\mu\nu}} \ ,\quad H_{\mu\nu} = {\hat F_{\mu\nu}} \ ,\quad h_{\mu\nu\r} = {\hat h_{\mu\nu\r}} \ .\quad
\eal
In terms of the forms $A=A_\mu dx^\mu$, $ K= K_\mu dx^\mu$, $H_2=dK$, $H_3=dB$, and the corresponding hatted ones, one
has\foot{Here for notational simplicity we use the same $y$ for the isometric direction and its dual -- whether it is
$y$ or $\hat y$ is clear from context.}
\bal\la{TruleH}
H_3 & = \hH_3 + (dy+\hA)\W \hH_2 - (dy+\hat K)\W \hat F_2\ ,
\\ \hH_3 & = H_3 + (dy+A)\W H_2 - (dy+K)\W F_2\ .
\eal
To write the T-duality rules for the R-R fields it is convenient to introduce
\bal
f_{k}\equiv e^{-a/2}\F_{k}= e^{\p -a/2}\, F_{k} \ ,\qquad\qquad \hat f_{k}\equiv e^{-\ah/2}\hF_{k}= e^{\hp -\ah/2}\, \hat F_{k}\ .
\eal
Then
\bal\la{TrulesRR}
\hf_{k}=&-(dy+K)\W f_{k-1}+(dy+K)\W(dy+A)\W \iota_{y}f_{k-1} - \iota_{y}f_{k+1}\ ,
\eal
where $ \iota_{y}f_{k}\equiv \iota_{I_y}f_{k}$, $I_y^m=\delta^m_y$. Also using
the assumption of invariance of the R-R forms under the isometry, $\LL_{I_y}f^{(k)}=0$, one has
\bal\la{TrulesRR2}
\iota_{y}\hf_k =& -f_{k-1}+(dy + A)\W \iota_{y}f_{k-1} \ ,\\
d\hf_{k}
=&(dy+K)\W df_{k-1}-(dy+K)\W(dy+A)\W \iota_{y}df_{k-1} + \iota_{y}df_{k+1} \\
&-H_2\W f_{k-1}+H_2\W(dy+A)\W \iota_{y}f_{k-1}-(dy+K)\W F_2\W \iota_{y}f_{k-1} \ .
\eal

\section{ABF background and T-dual HT solution}\label{B}
\def\theequation{B.\arabic{equation}}
\setcounter{equation}{0}

The ABF background \ci{abf,abf2} represents the couplings in the
$\eta$-deformed \adss action \ci{dmv} expanded to quadratic order in fermions
and formally identified with a GS action. This background for the type IIB
fields $(G_{},B_{}, \F_1, \F_{3}, \F_{5})$ (but not the dilaton which cannot be
extracted from the DMV action, and, in fact, does not exist) is given by

\footnotesize
{\allowdisplaybreaks
\begin{align}\nonumber
ds^2 & = -\frac{1+\rho^2}{1-\varkappa^2\rho^2} dt^2 + \frac{d\rho^2}{(1-\varkappa^2\rho^2)(1+\rho^2)} + \frac{\rho^2\cos^2\zeta}{1+\varkappa^2 \rho^4\sin^2\zeta}d\psi_1^2 + \frac{\rho^2d\zeta^2}{1+\varkappa^2 \rho^4 \sin^2 \zeta} +\rho^2 \sin^2\zeta d\psi_2^2
\\\nonumber & \hspace{12pt} +\frac{1-r^2}{1+\varkappa^2r^2}d\varphi^2 + \frac{dr^2}{(1+\varkappa^2r^2)(1-r^2)} + \frac{r^2 \cos^2\xi}{1+\varkappa^2 r^4 \sin^2 \xi} d\phi_1^2 + \frac{r^2d\xi^2}{1+\varkappa^2 r^4 \sin^2 \xi} + r^2 \sin^2\xi d\phi_2^2\ ,
\\\nonumber
B & = \frac{\varkappa \rho^4 \sin \zeta\cos\zeta}{1+\varkappa^2 \rho^4 \sin^2\zeta} d\psi_1 \W d\zeta - \frac{\varkappa r^4 \sin\xi\cos\xi}{1+\varkappa^2 r^4 \sin^2\xi} d\phi_1 \W d \xi \ ,
\\\nonumber
\F_1& = \vk^2 \rF\,
\Big[\rho^4 \sin^2\zeta \,d \psi_2 - r^4 \sin^2 \xi\, d \phi_2\Big] \ ,
\\\nonumber
\F_3& =\vk\, \rF\,
\Big[\frac{\rho^3\sin^2\zeta}{1-\vk^2\r^2} dt\wedge d\psi_2 \wedge d\r + \frac{r^3 \sin^2\xi}{1+\vk^2 r^2} d\varphi\wedge d\phi_2\wedge dr
\\\nonumber
& \hspace{30pt}
+\frac{\r^4\sin\z\cos\z}{1+\vk^2\r^4 \sin^2\z}d\psi_2\W d\psi_1 \W d\zeta
+\frac{r^4\sin\xi\cos\xi}{1+\vk^2 r^4 \sin^2\xi}d\phi_2\W d\phi_1 \W d\xi
\\\nonumber
& \hspace{30pt} + \frac{\vk^2\r r^4\sin^2\xi}{1-\vk^2\r^2}dt \W d\r \W d\phi_2 - \frac{\vk^2 \r^4 r \sin^2\z}{1+\vk^2 r^2} d\psi_2\W d \varphi \W d r
\\\nonumber
& \hspace{30pt}+\frac{\vk^2\r^4 r^4\sin\z\cos\z\sin^2\xi}{1+\vk^2 \rho^4 \sin^2 \z} d\psi_1 \W d\z \W d \phi_2
+\frac{\vk^2\r^4 r^4\sin^2\z\sin\xi\cos\xi}{1+\vk^2 r^4 \sin^2 \xi} d\psi_2 \W d \phi_1\W d \xi \Big]\ ,
\\\nonumber
\F_5 & = \rF\,
\Big[
\frac{\r^3\sin\z\cos\z}{(1-\vk^2\r^2)(1+\vk^2\r^4\sin^2\z)}dt \W d\psi_2 \W d\psi_1 \W d\z \W d \r
\\\nonumber
&\hspace{30pt}
- \frac{r^3\sin\xi\cos\xi}{(1+\vk^2 r^2)(1+\vk^2 r^4 \sin^2 \xi)} d \varphi \W d \phi_2 \W d \phi_1 \W d\xi \W dr
\\\nonumber
&\hspace{30pt}
-\frac{\vk^2\r r}{(1-\vk^2\r^2)(1+\vk^2 r^2)} (\r^2 \sin^2 \z\, dt \W d \psi_2 \W d\r \W d\varphi \W dr
+ r^2 \sin^2 \xi \, dt \W d\r \W d\varphi \W d\phi_2 \W dr)
\\\nonumber
&\hspace{30pt}
+ \frac{\vk^2\r^4 r^4\sin\z\cos\z \sin\xi \cos\xi}{(1+\vk^2\r^4\sin^2\z)(1+\vk^2 r^4 \sin^2 \xi)}
(d\psi_2 \W d\psi_1 \W d\z \W d\phi_1 \W d\xi -
d\psi_1 \W d\z \W d\phi_2 \W d\phi_1 \W d\xi)
\\\nonumber
&\hspace{30pt}
+ \frac{\vk^2\r r^4 \sin\xi\cos\xi}{(1-\vk^2\r^2)(1+\vk^2 r^4 \sin^2\xi)}(\r^2\sin^2\z\,dt \W d\psi_2 \W d\r \W d\phi_1 \W d\xi
- dt \W d\r \W d\phi_2 \W d\phi_1 \W d\xi)
\\\nonumber
&\hspace{30pt}
-\frac{\vk^2 \r^4 r\sin\z\cos\z}{(1+\vk^2 r^2)(1+\vk^2 \r^4 \sin^2 \z)} (r^2\sin^2\xi\, d\psi_1 \W d\z \W d \varphi \W d\phi_2 \W dr
+ d\psi_2 \W d\psi_1 \W d\z \W d \varphi \W dr)
\\\nonumber
& \hspace{30pt} -\frac{\vk^4 \r^5 r^4 \sin\z\cos\z\sin^2\xi}{(1-\vk^2\r^2)(1+\vk^2\r^4\sin^2\z)} dt \W d\psi_1 \W d\z \W d\r\W d\phi_2
\\\nonumber
& \hspace{30pt}
- \frac{\vk^4 \r^4 r^5 \sin^2\z \sin\xi\cos\xi}{(1+\vk^2 r^2)(1+\vk^2 r^4 \sin^2 \xi)} d\psi_2 \W d\varphi \W d\phi_1 \W d\xi \W dr\Big] \ , \\
& \rF \equiv \frac{4 \sqrt{1+\vk^2}}{\sqrt{1-\vk^2\r^2}\sqrt{1+\vk^2\r^4\sin^2\z}\sqrt{1+\vk^2r^2}\sqrt{1+\vk^2r^4\sin^2\xi}} \ . \label{abf}
\end{align}}\normalsize
Here $\vk= { 2 \eta \ov 1 - \eta^2}$ is a continuous deformation parameter
of the \emo: $\vk=0$ corresponds to the standard \adss solution \ci{schwarz}.

$ds^2 \equiv G_{mn} (x) dx^m dx^n$ defines the 10d metric $G$ and the sign of $B$-field is chosen as in \ci{abf},
i.e. it corresponds to the sign in \rf{0}.
$\F_k\equiv e^\p F_k$ are effective R-R $k$-form strengths of type IIB theory that appear in the GS action.
The self-duality equation satisfied by the R-R 5-form is
\begin{equation}
{F}_{mnpqr} =\te \frac{1}{5!}
\ve_{mnpqrstuvw}{F}^{stuvw} \ , \quad \ve_{mnpqrstuvw} \equiv \sqrt G\ \e_{mnpqrstuvw}\ , \quad
G=|\det\ G_{mn}| \ , \la{f5f}
\end{equation}
where we order the coordinates as $x^m = ( t,\psi_2,\psi_1,\z,\r,\vp,\phi_2,\phi_1,\xi, r)$ and
take $\epsilon_{t\psi_2\psi_1\z\r\vp\phi_2\phi_1\xi r} = -1$.

As found in \ci{ht2}, there exists an exact solution
of the standard type IIB supergravity equations that
is T-dual to the ABF background (provided we ignore the dilaton transformation).
This HT background has the following explicit form\foot{Here we have redefined $\hat \phi_2 \to - \hat \phi_2$ compared to \ci{ht2} to account for
the opposite definition we use for the Hodge dual. Also, recall that to perform the T-duality in $t$ we first analytically continue to Euclidean time, then T-dualise
and finally continue back.}

\footnotesize
{\allowdisplaybreaks
\begin{align}
\hat{ds}^2= &
- { 1-\varkappa^2\rho^2 \ov 1+\rho^2}d\hat t^2
+\frac{d\rho^2}{(1+\rho^2) (1-\varkappa^2\rho^2) }
+{d\hat \psi_1^2 \ov \rho^2\cos^2\zeta } \no
+ (\r\, d\z + \vk \r \tan\z\, d\hat \psi_1)^2
+{d\hat \psi_2^2 \ov \rho^2\sin^2\zeta } \\
&+ { 1+\varkappa^2 r^2 \ov 1- r^2}d\hat \vp ^2
+\frac{d r^2}{(1-r^2) (1+\varkappa^2 r^2) }
+{d\hat \phi_1^2 \ov r^2\cos^2\xi }
+ (r\, d\xi - \vk r \tan\xi\, d\hat \phi_1)^2
+{d\hat \phi_2^2 \ov r^2\sin^2\xi } \ ,\no
\\ \hat B & = 0 \ , \qquad \hat \F_1 = \hat \F_3 = 0 \ ,\no
\\ \no
\hat \F_5 & = \frac{4i\sqrt{1+\vk^2}}{\sqrt{1+\r^2}\sqrt{1-r^2}}\Big[
(d\hat t + \frac{\vk \r d\r}{1-\vk^2\r^2})\W \frac{d\hat \psi_2}{\r \sin \z} \W \frac{d\hat \psi_1}{\r \cos \z} \W (r d\xi - \vk r\tan \xi \, d\hat \phi_1) \W (\frac{dr}{1+\vk^2 r^2} + \vk r d\hat \vp)
\\ \no & \hspace{80pt}
- (d\hat \vp - \frac{\vk r dr}{1+\vk^2 r^2}) \W \frac{d \hat \phi_2}{r \sin\xi} \W \frac{d\hat \phi_1}{r \cos \xi} \W (\r d\z +\vk \r\tan\z \, d\hat\psi_1)) \W (\frac{d\r}{1-\vk^2r^2} + \vk \r d\hat t )
\Big]
\\ \la{hts}
{\hat\phi} & =
{\phi_0 -4\vk(\hat t + \hat\vp) -2\vk (\hat\psi_1-\hat \phi_1)} + \log\frac{(1-\vk^2\r^2)^2 (1+\vk^2r^2)^2}{\r^2r^2\sqrt{1+\r^2}\sqrt{1-r^2}\sin 2\z \sin 2\xi} \ . 
\end{align}}\normalsize
When written in terms of the following ``boosted''/``rotated'' vielbein basis
\begin{equation}\label{boandro}
\begin{split}
e^0 & \te = \frac{1}{\sqrt{1+\r^2}} \big(d\hat t + \frac{\vk \r d\r}{1-\vk^2 \r^2}\big)
\ , \qquad
e^1\te = \frac{d \hat \psi_2 }{\r \sin\zeta}
\ , \qquad
e^2 = \frac{d\hat \psi_1}{\r \cos\zeta} \ ,
\\
e^3 & \te = \r\, d\z + \vk \r \tan \z \, d\hat \psi_1 \ , \qquad e^4 = \frac{1}{\sqrt{1+\r^2}} \big(\frac{d\r}{1-\vk^2\r^2} + \vk \r d\hat t\big) \ ,
\\
e^5 &\te = \frac{1}{\sqrt{1-r^2}} \big(d\hat \vp - \frac{\vk r dr}{1+\vk^2 r^2}\big)
\ , \qquad
e^6 = \frac{d \hat \phi_2 }{r \sin\xi}
\ , \qquad
e^7 = \frac{d\hat \phi_1}{r \cos\xi} \ ,
\\
e^8 &\te = r\, d\xi - \vk r \tan \xi \, d\hat \phi_1 \ , \qquad e^9 = \frac{1}{\sqrt{1-r^2}} \big(\frac{dr}{1+\vk^2r^2} + \vk r d\hat \vp\big) \ ,
\end{split}
\end{equation}
the metric and $\hat\F_5$ in \rf{hts} take the following remarkably simple form \cite{ht2}
\begin{equation}\la{gf5}
\hat{ds}^2 = \eta_{MN}e^M e^N \ , \qquad \hat \F_5 = 4i\sqrt{1+\vk^2}\big(e^0 \W e^1 \W e^2 \W e^8 \W e^9 - e^3 \W e^4 \W e^5 \W e^6 \W e^7\big) \ ,
\end{equation}
where $M,N = 0,...,9$ are flat tangent-space indices, and $\eta_{MN}$ is the Minkowski metric. 

\section{Conservation of R-R stress tensor and dilaton beta function identity}\label{C}
\def\theequation{C.\arabic{equation}}
\setcounter{equation}{0}

Given a Weyl invariant sigma model the dilaton beta function $\bar
\beta^{\phi}$ in \rf{cp} represents a natural definition of the central charge:
it appears as the coefficient of the $R^{(2)}$-term in the expectation value of the trace
of the stress tensor on a curved 2d background \ci{ts,Shore:1986hk}, and for
this reason must be a constant \cite{cal}.\foot{The one-loop equation $\del_m
\bar \beta^{\phi}=0$ is a special case of the Curci-Paffuti identity
\cite{Curci:1986hi} that extends to higher loops.}

In the case of the ABF background we found an analog of the dilaton beta-function
\be
\label{ccq1}
\bb^X\equiv R-\tfrac{1}{12}H_{nkl}H^{nkl}+4D_nX^n-4X^2 \ ,
\ee
and the equation $\bb^X=0$ was used in section \ref{Sect:NSNS} to determine the
isometric part $I$ of the diffeomorphism vector $X$. In this Appendix we
reverse the logic and show that the modified type II
equations for the NS-NS and R-R fields with the same vector $X$
implies the constancy of $\bb^X$. In other words, on the equations of motion
(\ref{4})-(\ref{55}), (\ref{ef1})-(\ref{bf5}) governed by the vector (\ref{c})
we have the dilaton beta-function identity $\del_m \bb^X=0$.

To proceed, we first need to derive the conservation law for the R-R stress tensor $\T_{mn}$ in \rf{444}
that should hold on the R-R equations of motion.
First, consider
\bea
(D^n-2Z^n)(\F_m\F_n)
=-(d\F_1)_{mn}\F^n +\tfrac{1}{2}D_m(\F_n\F^n) +\F_m(D^n-Z^n)\F_n - \F_m Z^n\F_n \ .
\eea
Now using (\ref{ef1}) and (\ref{bf1}), we find
\bea
&&(D^n-2Z^n)(\F_m\F_n)=\tfrac{1}{2}(D_m-2Z_m)(\F_n\F^n)-I^p(\F^n\F_{mnp}) +\tfrac{1}{6}\F_m H^{abc}\F_{abc}\ .
\label{EQ1}
\eea
Next, we have
\bal
(D^n-2Z^n)(\F_{mpq}\F_n{}^{pq})=&-\tfrac{1}{3}(d\F_3)_{mnpq}\F^{npq}+\tfrac{1}{6}D_m(\F_{npq}\F^{npq}) \\
&+\F_{m}{}^{pq}(D^n-Z^n)\F_{npq}-\F_{m}{}^{pq}Z^n\F_{npq}\ ,
\eal
such that using (\ref{bf3}) and (\ref{ef3}), we obtain
\bal
(D^n-2Z^n)(\F_{mpq}\F_n{}^{pq})=&\tfrac{1}{6}(D_m-2Z_m)(\F_{npq}\F^{npq})-\tfrac{1}{3}\F_m H^{abc}\F_{abc}+H_{mpq}\F_n\F^{npq} \\
&+\tfrac{1}{6}H^{abc}\F_m{}^{pq}\F_{pqabc}-2I^p(\F^n\F_{mnp})-\tfrac{1}{3}I^p\F^{abc}\F_{mabcp}\ . \label{EQ2}
\eal
Finally, we need
\bal
(D^n-2Z^n)(\F_{mpqrs}\F_n{}^{pqrs})=&-\tfrac{1}{5}(d\F_5)_{mnpqrs}\F^{npqrs} \\
&+ \F_m{}^{pqrs}(D^n-Z^n)\F_{npqrs}-Z^n\F_n{}^{pqrs}\F_{mpqrs}\, \\
=&\tfrac{1}{5}(H_3\wedge \F_3)_{mnpqrs}\F^{npqrs} +\tfrac{1}{30}\ve_{mabcdnpqrs}I^a\F^{bcd}\F^{npqrs} \\
&-\tfrac{1}{36}\F_m{}^{pqrs}\ve_{pqrsabcde}H^{abc}\F^{def}-4I^p\F^{abc}\F_{mabcp}\ ,
\eal
where we have used (\ref{ef5}) and (\ref{bf5}) and that $\F_5^2 = 0$.
Taking into account the self-duality of $\F_5$, which also implies that
$\F_m{}^{pqrs}\ve_{pqrsabcde}=-24g_{m[a}\F_{bcdef]}$,
we find
\bea
(D^n-2Z^n)(\F_{mpqrs}\F_n{}^{pqrs})=4H_m{}^{np}\F^{abc}\F_{npabc}-4\F_m{}^{np}H^{abc}\F_{npabc}-8I^p\F^{abc}\F_{mabcp}\ . \label{EQ3}
\eea
Combining (\ref{EQ1}),(\ref{EQ2}),(\ref{EQ3}) we find the following conservation law for the stress tensor $\T_{mn}$
\bea\label{conservT1}
(D^n-2Z^n)\T_{mn}=2\K_{mn}I^n+\tfrac{1}{2}H_{mkn}\K^{kn}\ ,
\eea
where $\K_{mn}$ is defined in \rf{55}.
We would like to rewrite this formula in terms of $X=Z+I$. We have
\bea
(D^n-2X^n)\T_{mn}=2(\K_{mn}-\T_{mn})I^n+\tfrac{1}{2}H_{mkn}\K^{kn}\ .
\eea
Further, we use (\ref{4}) and (\ref{5}) (with $Y=X$) to write
\bal
& (\K_{mn}-\T_{mn})I^n=-\tfrac{1}{2}D^kH_{kmn}I^n+Z^kH_{kmn}I^n \\
& \qquad + (D_mI_n-D_nI_m)I^n-R_{mn}I^n+\tfrac{1}{4}H_m{}^{kl}H_{nkl}I^n-(D_mZ_n+D_nZ_m)I^n\ .
\eal
Notice that due to the properties of $I_m$ in \rf{7}
one has $[D_n,D_m]I^n=R_{mn}I^n=-D^nD_nI_m$, which implies the following identity
\bea\label{idRI}
R_{mn}I^n=\tfrac{1}{2}D^n(D_mI_n-D_nI_m)\ .
\eea
Then
\bal
(\K_{mn}-\T_{mn})I^n=&-\tfrac{1}{2}D^k(H_{kmn}I^n)-\tfrac{1}{2}H_{mkn}D^kI^n+Z^kH_{kmn}I^n+\tfrac{1}{4}H_m{}^{kl}H_{kln}I^n \\
& +(D_mI_n-D_nI_m)I^n-\tfrac{1}{2}D^n(D_mI_n-D_nI_m)-(D_mZ_n+D_nZ_m)I^n\ .
\eal
Now using (\ref{bZ}), we obtain
\bal
(\K_{mn}-\T_{mn})I^n=&-\tfrac{1}{2}D^n(D_mZ_n-D_nZ_m)-\tfrac{1}{2}D^n(D_mI_n-D_nI_m) \\
&-\tfrac{1}{2}H_{mkn}D^kI^n-\tfrac{1}{2}H_{mkn}D^kZ^n \\
&+Z^n(D_mZ_n-D_nZ_m)+I^n(D_mI_n-D_nI_m)-(D_mZ_n+D_nZ_m)I^n \ .\eal
Taking into account that
\bal
-(D_mZ_n+D_nZ_m)I^n& =(D_mZ_n-D_nZ_m)I^n-2D_mZ_nI^n \\
&=(D_mZ_n-D_nZ_m)I^n+2Z^nD_mI_n
\\ &=I^n(D_mZ_n-D_nZ_m)+Z^n(D_mI_n-D_nI_m)\ , 
\eal
we find
\bea
&&(\K_{mn}-\T_{mn})I^n=-\tfrac{1}{2}(D^n-2X^n)(D_mX_n-D_nX_m)-\tfrac{1}{2}H_{mkn}D^kX^n\ .\eea
Thus, the conservation law \eqref{conservT1} acquires the following form depending only on the vector $X$
\bea\label{conservT2}
(D^n-2X^n)\T_{mn}=\tfrac{1}{2}H_{mkn}\K^{kn}-(D^n-2X^n)(D_mX_n-D_nX_m)-H_{mkn}D^kX^n\ .
\eea
Here, using \eqref{5}, the tensor $\K^{kn}$ can be eliminated such that the r.h.s.
of \rf{conservT2} is written solely in terms of $H_3$ and $X$.

Now we ready to show the constancy of $\bb^X$. We have from \rf{ccq1}
\bea
\pa_m\bb^X=2D^nR_{mn}-\tfrac{1}{6}H^{nkl}D_{[m}H_{nkl]}-\tfrac{1}{2}H^{nkl}D_nH_{mkl}+4D_mD_nX^n-8X^nD_mX_n\ .
\eea
Since $D_{[m}H_{nkl]}=0$ this can be rewritten as
\bal\label{ddee}
\pa_m\bb^X= &\, 2D^n\big(R_{mn}-\tfrac{1}{4}H_{mkl}H_n{}^{kl}\big)-4X^nR_{mn}\\
&+\tfrac{1}{2}D_nH^{nkl}H_{mkl}+4D^nD_mX_n-8X^nD_mX_n\ .
\eal
Furthermore, using
\bal
4D^nD_mX_n & = 2D^n(D_mX_n+D_mX_n)-2D^n(D_nX_m-D_mX_n) \ ,\\
 -8X^nD_mX_n & = -4X^n(D_mX_n+D_nX_m)+4X^n(D_nX_m-D_mX_n)\ ,
\eal
we may combine the terms in \eqref{ddee} as
\bal
\pa_m\bb^X=\,&2(D^n-2X^n)\big(R_{mn}-\tfrac{1}{4}H_{mkl}H_n{}^{kl}+D_mX_n+D_nX_m\big) \\
&+\big(\tfrac{1}{2}D_nH^{nkl}-X_nH^{nkl}\big)H_{mkl}-2(D^n-2X^n)(D_nX_m-D_mX_n)\ .
\eal
Finally, using eq.(\ref{4}) we have
\bal
\pa_m\bb^X=\,&2(D^n-2X^n)\T_{mn}\\
&-H_{mkn}\K^{kn}+H_{mkn}D^kX^n+2(D^n-2X^n)(D_mX_n-D_nX_m)=0\ ,
\eal
where the r.h.s. vanishes due to the conservation law (\ref{conservT2}).
This proves that $\bb^X$ is a constant (actually zero)
on the modified equations of motion.
The same is then true also in the spacial case of the standard type IIB supergravity equations (i.e. in the limit \rf{back})
with the R-R strengths non-zero.\foot{While expected, this was not explicitly shown before in the literature.
This provides a consistency check of the equivalence of the supergravity equations of motion with the sigma model Weyl invariance conditions.}

\section{Derivation of second-order equations for R-R strengths \\ from modified type II equations}\label{D}
\def\theequation{D.\arabic{equation}}
\setcounter{equation}{0}

Here we present the derivation of the 2nd-order equations for the R-R field strengths, which, as discussed in section \ref{s4},
are candidates for the scale invariance conditions of the GS sigma model,
starting with the modified type II equations \rf{ef1}--\rf{bf5} or \rf{defeq}, i.e. ($n \in\mathbb{Z} $)
\bal
\label{defeq1}
&d\F_{2n+1} - Z\W \F_{2n+1} + H_3 \W \F_{2n-1} - \star (I \wedge \star \F_{2n+3}) = 0 \ ,
\\
&d\star\F_{2n+1} - Z\W \star\F_{2n+1}-H_3 \W \star\F_{2n+3} + \star (I \wedge \F_{2n-1}) = 0 \ .
\eal
Our aim is to derive \eqref{candidate}.
Acting on the first equation by $\star d\star$ and on the second equation by $d\star$ we get
\bal
\label{defeq2}
&\star d\star d\F_{2n+1} - \star d\star(Z\W \F_{2n+1}) + \star d\star (H_3 \W \F_{2n-1}) + \star d(I \wedge \star \F_{2n+3}) = 0 \ ,\\
&d\star d\star\F_{2n+1} - d\star(Z\W \star\F_{2n+1})-d\star(H_3 \W \star\F_{2n+3}) - d(I \wedge \F_{2n-1}) = 0 \ .
\eal
Taking the sum of these equations and using $\star(I \W \star Z_n) = \iota_I Z_n$, we find
\bal
\label{defeq3}
&
\star d\star d\F_{2n+1}+d\star d\star\F_{2n+1} - \star\,\LL_X\star \F_{2n+1} - \LL_X\F_{2n+1}\\
&\qquad + \star d\star (H_3 \W \F_{2n-1})-d\star(H_3 \W \star\F_{2n+3}) \\
&\qquad + \star\,\iota_Z\,d(\star \F_{2n+1}) +\iota_Z\,d\F_{2n+1}+ \star d(I \wedge \star \F_{2n+3}) - d(I \wedge \F_{2n-1}) \extraterm = 0 \ ,
\eal
where we have used \eqref{lif}: ${\LL_Z \F_{2n+1}=\LL_X \F_{2n+1} -\LL_I \F_{2n+1} = \LL_X \F_{2n+1} - (I\cdot Z) \F_{2n+1}}$.

The terms on the first line are the same as in \eqref{candidate}, so we consider the last line in \eqref{defeq3}
\bal
\label{defeq4}
& \star\,\iota_Z\,d\star \F_{2n+1} +\iota_Z\,d\F_{2n+1}+ \star d(I \wedge \star \F_{2n+3})- d(I \wedge \F_{2n-1}) \extraterm \\
=& \star\,\iota_Z( Z\W \star\F_{2n+1}+H_3 \W \star\F_{2n+3} - \iota_I \star \F_{2n-1} ) +\iota_Z(Z\W \F_{2n+1} - H_3 \W \F_{2n-1} + \iota_I \F_{2n+3}) \\
&+ \star( dI \wedge \star \F_{2n+3})- dI \wedge \F_{2n-1}- \star (I \wedge d\star \F_{2n+3})+ I \wedge d\F_{2n-1} \extraterm \\
=& \star( dI \wedge \star \F_{2n+3})+\star(\iota_Z(H_3) \W \star\F_{2n+3})- dI \wedge \F_{2n-1}-\iota_Z(H_3) \W \F_{2n-1}\\
&- \star(H_3 \W\iota_Z( \star\F_{2n+3}))+ \star\,\iota_Z( Z\W \star\F_{2n+1} - \iota_I \star \F_{2n-1} )\\
& +H_3 \W \iota_Z\F_{2n-1}+\iota_Z(Z\W \F_{2n+1} + \iota_I \F_{2n+3}) - \star (I \wedge d\star \F_{2n+3})+ I \wedge d\F_{2n-1} \extraterm\ .
\eal
Now we use \rf{5} with $Y=X$ or
\bal
dI +\iota_ZH_3 = \b^B \ ,
\eal
to get
\bal
\label{defeq5}
&\star( \b^B \wedge \star \F_{2n+3})-\b^B \wedge \F_{2n-1}\\
&- \star(H_3 \W\iota_Z( \star\F_{2n+3}))+ \star\,\iota_Z( Z\W \star\F_{2n+1} - \iota_I \star \F_{2n-1} )\\
& +H_3 \W \iota_Z\F_{2n-1}+\iota_Z(Z\W \F_{2n+1} + \iota_I \F_{2n+3}) - \star (I \wedge d\star \F_{2n+3})+ I \wedge d\F_{2n-1} \extraterm\ .
\eal
The two terms on the first line are the same as in \eqref{candidate}.
To derive the remaining terms of \eqref{candidate}, we use the relations
\bal
- \star (I \wedge d\star \F_{2n+3}) =& - \star (H_3 \W \star d\F_{2n+3}) - \star (H_3 \W \star (H_3 \W \F_{2n+1})) \\
&+ \star (H_3 \W \star (Z \W \F_{2n+3})) - \star (I \W Z\W \star \F_{2n+3})+\star(I\W \star(I\W F_{2n+1}))\ ,\\
I \wedge d\F_{2n-1} =\,& H_3 \W \star d\star \F_{2n-1} - H_3 \W \star (H_3 \W\star\F_{2n+1}) \\
&- H_3 \W \star (Z \W\star\F_{2n-1}) +I\W Z\W \F_{2n-1}+I\W\star(I\W\star \F_{2n+1})\ ,
\eal
which transform the last two lines of \eqref{defeq5} into
\bal
\label{defeq7}
& - \star (H_3 \W \star d\F_{2n+3}) - \star (H_3 \W \star (H_3 \W \F_{2n+1})) + H_3 \W \star d\star \F_{2n-1} - H_3 \W \star (H_3 \W\star\F_{2n+1}) \\
& +\star (I \W Z\W \star \F_{2n+3})-\star(I\W \star(I\W F_{2n+1})) +I\W Z\W \F_{2n-1}+I\W\star(I\W\star \F_{2n+1})\\
&+ \star\,\iota_Z( Z\W \star\F_{2n+1} - \iota_I \star \F_{2n-1} )+\iota_Z(Z\W \F_{2n+1} + \iota_I \F_{2n+3}) \extraterm
\ .
\eal
Now using the identities
\bal
& \star (I \W Z\W \star \F_{2n+3}) = \iota_Z \iota_I \F_{2n+3}\ ,\quad \qquad \! \star\,\iota_Z \iota_I \star \F_{2n-1}=I\W Z\W \F_{2n-1}\ ,\\
&\star(I\W \star(I\W F_{2n+1}))= \iota_I(I\W \F_{2n+1})\ ,\quad \star\,\iota_I( I\W \star\F_{2n+1} )=I\W\star(I\W\star \F_{2n+1})=I\W \iota_IF_{2n+1} \ ,
\eal
one finds
\bal
\label{defeq8}
& - \star (H_3 \W \star d\F_{2n+3}) - \star (H_3 \W \star (H_3 \W \F_{2n+1})) + H_3 \W \star d\star \F_{2n-1} - H_3 \W \star (H_3 \W\star\F_{2n+1})\\
& +\iota_X(X) \F_{2n+1}\\
=&- \star (H_3 \W \star d\F_{2n+3}) - \star (H_3 \W \star (H_3 \W \F_{2n+1})) + H_3 \W \star d\star \F_{2n-1} - H_3 \W \star (H_3 \W\star\F_{2n+1})
\\
&\te + ({1\ov4}R-{1\ov8}\star(H_3\W\star H_3) )\F_{2n+1} + (\star d\star X)\F_{2n+1}\ .
\eal
This leads precisely to \eqref{candidate}.

\section{Derivation of ``Bianchi identity" for $Z$}\label{appderiv}
\def\theequation{E.\arabic{equation}}
\setcounter{equation}{0}

Here we observe that the
modified ``Bianchi identity for the dilaton" \eqref{bZ} that holds for the ABF background
may be derived more generally
from the Bianchi equations for $\F_k$, the invariance of
the R-R fields under the isometry $\mathcal{L}_I \F_{k} = 0$,
the conditions $\F_1 \wedge \F_3 \neq 0$, $\F_1 \wedge \F_5 \neq 0$ or $\F_3 \wedge \F_5 \neq 0$ and
the condition $\iota_I \F_1 = 0$.
Starting from (see \rf{ef1}--\rf{bf5})
\begin{align}
& d \F_1 - Z \wedge \F_1 - \iota_I \F_{3} = 0 \ , \label{bf1b}\\
& d\F_3 - Z \wedge \F_3 + H_3 \wedge \F_1 - \iota_I \F_{5} =0 \ , \label{bf3b}
\end{align}
we take the differential of \eqref{bf1b} and use \eqref{bf3b} to give\foot{Note
that here we use the condition $\iota_I \F_{1} = 0$, which if $dZ \neq 0$ follows from \eqref{bf1b} after acting on it with $\iota_I$
\be
\no \iota_I d \F_1 + Z \iota_I\F_1= 0\quad \Rightarrow\quad d \, \iota_I \F_1 - Z \, \iota_I\F_1= 0\label{bf1bb}\ ,
\ee
where we have used $\iota_I Z=0$ and $\mathcal{L}_I \F_1=0$. We see that if $\iota_I \F_1\neq 0$ then $Z=d\log \iota_I \F_1$,
which contradicts our assumption.}
\bal
&- dZ \wedge \F_1+Z \wedge d\F_1 - d(\iota_I \F_{3}) = - dZ \wedge \F_1 + Z \wedge \iota_I \F_{3}+ \iota_I d\F_{3}\\
 = &- dZ \wedge \F_1 + Z \wedge \iota_I \F_{3}+ \iota_I(Z \wedge\F_{3}) - \iota_I(H \wedge\F_{1})\\
 = &- dZ \wedge \F_1 + \iota_IZ \wedge\F_{3} - \iota_IH \wedge\F_{1}= - (dZ + \iota_I H )\wedge\F_{1}=0\ .
\eal
Thus
\bal
dZ + \iota_IH \sim \F_1\ .
\eal
A similar analysis of the Bianchi equations for $\F_3$ and $\F_5$ gives
\bal\la{conbf3}
(dZ + \iota_IH )\wedge\F_{3}=0\ ,\qquad\qquad (dZ+\iota_I H_3)\wedge \F_5 =0\ .
\eal
Thus if $\F_1\wedge \F_3\neq 0$, $\F_1\wedge \F_5\neq 0$ or $\F_3 \wedge \F_5 \neq 0$ then $dZ + \iota_IH=0$.

\section{Deformed $AdS_3 \times S^3$ and $AdS_2 \times S^2$ cases }\label{F}
\def\theequation{F.\arabic{equation}}
\setcounter{equation}{0}

In the deformed $AdS_3 \times S^3$ case
the (complete) T-dual HT background \cite{ht2} consists of
just the metric, dilaton and a single R-R 3-form flux,
and therefore has a simple embedding into Type IIB supergravity -- one just needs to add 4 extra toroidal dimensions.
Explicitly, this background which is T-dual to the
$\eta$-deformed $AdS_3 \times S^3$ background (cf. \ci{hrt,lrt})
is given by

\footnotesize
{\allowdisplaybreaks
\begin{align}\nonumber
\hat{ds}^2 & = - \frac{1-\vk^2 \r^2}{1+\r^2}d\hat t^2 + \frac{d\r^2}{(1-\vk^2\r^2)(1+\r^2)} + \frac{d\hat \psi_1^2}{\r^2}
+ \frac{1+\vk^2r^2}{1-r^2}d\hat\vp^2 + \frac{dr^2}{(1+\vk^2 r^2)(1-r^2)} + \frac{d\hat \phi_1^2}{r^2} + dx_adx_a \ ,
\\ \nonumber
\hat B & = 0 \ , \qquad \hat\F_1 = \hat\F_5 = 0 \ ,
\\ \nonumber
\hat \F_3 & = \frac{2i \sqrt{1+\vk^2}}{\sqrt{1+\r^2}\sqrt{1-r^2}} \Big[(d\hat t +\frac{\vk \r d\r}{1-\vk^2\r^2}) \W \frac{d\hat\psi_1}{\r} \W (\frac{dr}{1+\vk^2r^2}+\vk rd\hat \vp)
\\ & \nonumber
\hspace{100pt} + (d\hat \vp - \frac{\vk r dr}{1+\vk^2 r^2})\W \frac{d\hat \phi_1}{r} \W (\frac{d\r}{1-\vk^2\r^2} + \vk \r d\hat t) \Big]\ ,
\\
{\hat \phi} & ={\phi_0-2\vk(\hat t+\hat \vp)} + \log \frac{(1-\vk^2\r^2)(1+\vk^2r^2)}{\r r\sqrt{1+\r^2}\sqrt{1-r^2}} \ .
\label{ads3tdual}
\end{align}
}\normalsize
When written in terms of the ``boosted''/``rotated'' vielbein basis \cite{ht2}
\begin{equation}\label{boandro3}
\begin{split}
e^0 & =\te \frac{1}{\sqrt{1+\r^2}} \big(d\hat t + \frac{\vk \r d\r}{1-\vk^2 \r^2}\big)
\ , \qquad
e^1 =\te \frac{d\hat \psi_1}{\r} \ ,
\qquad e^2 = \te \frac{1}{\sqrt{1+\r^2}} \big(\frac{d\r}{1-\vk^2\r^2} + \vk \r d\hat t\big) \ ,
\\
e^3 & =\te \frac{1}{\sqrt{1-r^2}} \big(d\hat \vp - \frac{\vk r dr}{1+\vk^2 r^2}\big)
\ , \qquad
e^4 = \te \frac{d\hat \phi_1}{r} \ ,
\qquad e^5 =\te \frac{1}{\sqrt{1-r^2}} \big(\frac{dr}{1+\vk^2r^2} + \vk r d\hat \vp\big) \ ,
\end{split}
\end{equation}
the metric and $\hat \F_3$ take the following simple form (cf. \rf{gf5})
\begin{equation}
\hat{ds}^2 = \eta_{MN}\, e^M e^N + dx_a dx_a \ , \qquad\qquad \hat \F_3 = 2i\sqrt{1+\vk^2}\big(e^0 \W e^1 \W e^5 + e^2 \W e^3 \W e^4\big) \ .
\end{equation}
As in the $AdS_5 \times S^5$ case, the dilaton and the R-R flux $F_3$ depend on the isometric directions of the metric, but
this dependence is such that $e^\phi F = \F$ is invariant under the isometries.
Therefore, we can formally T-dualise the metric and $\hat\F$ to find the following analog of the ABF background (cf. \rf{abf})

\footnotesize
{\allowdisplaybreaks
\begin{align}\nonumber
ds^2 & = -\frac{1+\rho^2}{1-\varkappa^2\rho^2} dt^2 + \frac{d\rho^2}{(1-\varkappa^2\rho^2)(1+\rho^2)} + \r^2 d\psi_1^2
+\frac{1-r^2}{1+\varkappa^2r^2}d\varphi^2 + \frac{dr^2}{(1+\varkappa^2r^2)(1-r^2)} + r^2 d\phi_1^2 + dx_a dx_a \ ,
\\\nonumber
B & = 0 \ ,
\\\nonumber
\F_1& = \vk \,\rF\,
\Big[\rho^2 d \psi_1 + r^2 d \phi_1 \Big] \ ,
\\\nonumber
\F_3& =\rF\,
\Big[\frac\r{1-\vk^2\r^2} (dt \W d\psi_1 \W d\r + \vk^2 r^2 \, dt \W d\phi_1 \W d\r) - \frac r{1+\vk^2 r^2} (d\vp \W d\phi_1 \W dr -\vk^2\r^2 \, d\vp \W d\psi_1 \W dr) \Big] \ ,
\\\nonumber
\F_5 & = \vk\, \rF\,
\Big[
\frac{\r r}{(1-\vk^2\r^2)(1+\vk^2 r^2)}(dt \W d\r \W d\vp \W d\phi_1 \W dr - dt \W d\psi_1 \W d\r \W d\vp \W dr)\\ \nonumber & \hspace{232pt} - (\r^2 d \psi_1 + r^2 d\phi_1) \W dx_1 \W dx_2 \W dx_3 \W dx_4 \Big] \ ,
\\
& \rF \equiv \frac{2 \sqrt{1+\vk^2}}{\sqrt{1-\vk^2\r^2}\sqrt{1+\vk^2r^2}} \ . \label{ads3}
\end{align}}\normalsize
As in the $AdS_5 \times S^5$ case, it turns out that
there exist vectors $X$ and $Y$ such that the scale invariance conditions for the
metric and $B$-field \eqref{4},\eqref{5} are satisfied (cf. \rf{X},\rf{Y})
\begin{align}
X = X_m dx^m = & c_0 \frac{1+\r^2}{1-\vk^2\r^2} dt + c_1 \r^2 d\psi_1 - \frac{\vk^2 \r}{1-\vk^2 \r^2}d\r \nonumber
\\ & + c_2 \frac{1-r^2}{1+\vk^2r^2} d\vp + c_3 r^2 d\phi_1 + \frac{\vk^2 r}{1+\vk^2 r^2}dr + k_a dx^a \ ,\label{ads3x}
\\ Y = Y_m dx^m = & 2\vk \frac{1+\r^2}{1-\vk^2\r^2} dt - \frac{\vk^2 \r}{1-\vk^2 \r^2}d\r + 2\vk \frac{1-r^2}{1+\vk^2r^2} d\vp + \frac{\vk^2 r}{1+\vk^2 r^2}dr \ .\label{ads3y}
\end{align}
The parameters $c_i$ and $k_a$ are eight arbitrary constants parametrising the Killing vector part of $X_m$, while
$Y$ is defined up to a total derivative.
As in the $AdS_5 \times S^5$ case, we may split the vector $X$
into two parts: $I$, containing the $8$ commuting Killing vectors, and $Z$, which contains the rest.
If we fix the constants $c_i$ and $k_a$ as
\begin{equation}
c_0 = c_2 = 2 \vk \ , \qquad c_1 = c_3 = k_a = 0 \ ,\label{ads3coeff}
\end{equation}
so that $Y_m=X_m$
then the equations \eqref{xy},\eqref{10},\eqref{bZ},\eqref{ef1}--\eqref{bf5} are
all satisfied, and hence the background \eqref{ads3} solves the same system of equations as the ABF background \eqref{abf}.
Finally, we find $\phi$ in \eqref{uuu} is given by
\begin{equation}
\phi = \tfrac12 \log (1-\vk^2\r^2)(1+\vk^2 r^2) \ . \label{ads3ps}
\end{equation}

For the deformed $AdS_2 \times S^2$ case, the T-dual HT background \cite{ht2} consists of
just the metric, dilaton and a single R-R 2-form flux.
It can be again embedded into Type IIB supergravity by adding 6-torus
$T^6$ and combining the 2-form with the holomorphic 3-form on $T^6$ to give
a self-dual 5-form:

\footnotesize
{\allowdisplaybreaks
\begin{align}\nonumber
\hat{ds}^2 & = - \frac{1-\vk^2 \r^2}{1+\r^2}d\hat t^2 + \frac{d\r^2}{(1-\vk^2\r^2)(1+\r^2)}
+ \frac{1+\vk^2r^2}{1-r^2}d\hat\vp^2 + \frac{dr^2}{(1+\vk^2 r^2)(1-r^2)} + dx_adx_a \ ,
\\ \nonumber
\hat B & = 0 \ , \qquad \hat\F_1 = \hat \F_3 = 0 \ ,
\\ \nonumber
\hat \F_5 & = \frac{i \sqrt{1+\vk^2}}{\sqrt{2}\sqrt{1+\r^2}\sqrt{1-r^2}}
\Big[(d\hat t +\frac{\vk \r d\r}{1-\vk^2\r^2}) \W (\frac{dr}{1+\vk^2r^2}+\vk rd\hat \vp)\W(\w_r +\w_i)
\\ & \nonumber
\hspace{100pt}
+ (d\hat \vp - \frac{\vk r dr}{1+\vk^2 r^2})\W (\frac{d\r}{1-\vk^2\r^2} + \vk \r d\hat t) \W (\w_r - \w_i)\Big] \ ,
\\
{\hat \phi} & ={\phi_0-\vk(\hat t+\hat \vp)} + \log \frac{(1-\vk^2\r^2)(1+\vk^2r^2)}{\sqrt{1+\r^2}\sqrt{1-r^2}} \ ,
\label{ads2tdual}
\end{align}
}\normalsize
where $\w_r$ and $\w_i$ are the real and imaginary parts of the holomorphic 3-form on $T^6$, e.g.,
\begin{equation}\begin{split}
& \omega_r = dx^1 \W dx^3 \W dx^5 - dx^1 \W dx^4 \W dx^6 - dx^2 \W dx^3 \W dx^6 - dx^2 \W dx^4 \W dx^5 \ ,
\\
& \omega_i = dx^2 \W dx^4 \W dx^6 - dx^2 \W dx^3 \W dx^5 - dx^1 \W dx^4 \W dx^5 - dx^1 \W dx^3 \W dx^6 \ .
\end{split}\end{equation}
As in the $AdS_5\times S^5$ and $AdS_3 \times S^3$ cases, when written in terms of the ``boosted''/``rotated'' vielbein basis \cite{ht2}
\begin{equation}\label{boandro2}
\begin{split}
e^0 & =\te \frac{1}{\sqrt{1+\r^2}} \big(d\hat t + \frac{\vk \r d\r}{1-\vk^2 \r^2}\big) \ ,
\qquad e^1 =\te \frac{1}{\sqrt{1+\r^2}} \big(\frac{d\r}{1-\vk^2\r^2} + \vk \r d\hat t\big) \ ,
\\
e^2 & =\te \frac{1}{\sqrt{1-r^2}} \big(d\hat \vp - \frac{\vk r dr}{1+\vk^2 r^2}\big)
\ ,
\qquad e^3 =\te \frac{1}{\sqrt{1-r^2}} \big(\frac{dr}{1+\vk^2r^2} + \vk r d\hat \vp\big) \ ,
\end{split}
\end{equation}
the metric and $\hat \F_5$ have take the following simple form
\begin{equation}
\hat{ds}^2 = \eta_{MN}e^M e^N + dx_a dx_a \ , \qquad \hat \F_5 = \te \frac{i}{\sqrt 2}\sqrt{1+\vk^2}\big[e^0 \W e^3 \W (\w_r +\w_i) - e^1 \W e^2 \W (\w_r - \w_i)\big] \ .
\end{equation}
Applying T-duality to the metric and $\hat \F$ gives the analog of ABF background for the $AdS_2 \times S^2$ \emo

\footnotesize
{\allowdisplaybreaks
\begin{align}\nonumber
ds^2 & = -\frac{1+\rho^2}{1-\varkappa^2\rho^2} dt^2 + \frac{d\rho^2}{(1-\varkappa^2\rho^2)(1+\rho^2)}
+\frac{1-r^2}{1+\varkappa^2r^2}d\varphi^2 + \frac{dr^2}{(1+\varkappa^2r^2)(1-r^2)} + dx_a dx_a \ ,
\\\nonumber
B & = 0 \ , \qquad
\F_1 = 0 \ ,
\\\nonumber
\F_3& =\tfrac12 \vk\,\rF\,
\Big[(-\r+r) \w_r + (\r+r) \w_i\Big] \ ,
\\\nonumber
\F_5 & =\tfrac12 \rF\,
\Big[\frac{1-\vk^2 \r r}{1-\vk^2 \r^2}dt \W d\r \W \w_r - \frac{1+\vk^2 \r r}{1-\vk^2 \r^2} dt \W d\r \W \w_i
+ \frac{1+\vk^2 \r r}{1+\vk^2 r^2} d\vp \W dr \W \w_r +
\frac{1-\vk^2 \r r}{1+\vk^2 r^2}d \vp \W dr \W \w_i\Big] \ ,
\\
& \rF \equiv \frac{\sqrt{2} \sqrt{1+\vk^2}}{\sqrt{1-\vk^2\r^2}\sqrt{1+\vk^2r^2}} \ .
\label{ads2}
\end{align}}\normalsize
Here again the scale invariance conditions for the
metric and $B$-field \eqref{4},\eqref{5} are satisfied provided
\begin{align}
X = X_m dx^m = & c_0 \frac{1+\r^2}{1-\vk^2\r^2} dt + c_1 \frac{1-r^2}{1+\vk^2r^2} d\vp + k_a dx^a \ ,\label{ads2x}
\\ Y = Y_m dx^m = & \vk \frac{1+\r^2}{1-\vk^2\r^2} dt + \vk \frac{1-r^2}{1+\vk^2r^2} d\vp \ .\label{ads2y}
\end{align}
The parameters $c_i$ and $k_a$ are eight arbitrary constants parametrising the Killing vector part of $X_m$, while
$Y$ is defined up to a total derivative.
Here $X_m$ is given just by $I_m$ (i.e. the sum of commuting Killing vectors) and thus $Z_m=0$.
If we fix the constants $c_i$ and $k_a$ as
\begin{equation}
c_0 = c_1 = \vk \ , \qquad k_a = 0 \ ,\label{ads2coeff}
\end{equation}
so that $Y_m=X_m$
then the equations \eqref{xy},\eqref{10},\eqref{bZ},\eqref{ef1}--\eqref{bf5} are
all satisfied, i.e. the background \eqref{ads2} solves the same system of equations as the ABF background \eqref{abf} in the \adss case.
Finally, here we find that $\phi$ in \eqref{uuu}
is given by
\begin{equation}
\phi = 0 \ .\label{ads2ps}
\end{equation}

As in the $AdS_5 \times S^5$ case, the coefficients in \eqref{ads3coeff},\eqref{ads2coeff}
are equal to (minus) the corresponding coefficients of the isometric coordinates in the linear terms of the
dual dilatons $\hat \p$ of the T-dual HT backgrounds \cite{ht2}. Furthermore, the ``dilatons''
$\phi$ in \eqref{ads3ps},\eqref{ads2ps} are again
found by applying the standard T-duality rules to the remaining parts (depending only on non-isometric coordinates)
of the dilatons $\hat \p$ of the T-dual solutions.
Therefore, these examples also fit into the general picture described in section the main text.

\section{Second-order equations for R-R fields from scale invariance conditions for type II GS sigma model}\label{R}
\def\theequation{G.\arabic{equation}}
\setcounter{equation}{0}

In this Appendix we shall expand on the discussion in section \ref{s4} and explain
how the 2nd-order equations for the R-R couplings $\F$ such as \rf{scinvF1}--\rf{scinvF5}
can emerge as the one-loop conditions of scale invariance (UV finiteness) of the GS sigma model \rf{0}.
While we will not compute the beta-functions for R-R couplings in full, our aim will be to illustrate
how the relevant structures come out of logarithmically divergent parts of the corresponding one-loop
Feynman graphs.\foot{Previous studies of the UV finiteness conditions of the GS string \ci{Grisaru:1988sa,bel}
did not include R-R couplings,
but special cases of $AdS_5 \times S^5$ \ci{Drukker:2000ep}
and pp-wave backgrounds \ci{Russo:2002qj}
were explicitly discussed.
The vanishing of the beta-functions for the R-R couplings was not checked as the fermionic
coordinate was assumed to have trivial background.}

We shall consider the type IIB GS sigma model \ci{howe}
with couplings representing a generic type IIB superspace background subject to constraints
required for $\kappa$-symmetry: we will assume $\kappa$-symmetry to be able to gauge fix it
but otherwise will keep the R-R fields unconstrained.
The GS sigma model action expanded in powers of fermions may be written as
(see, e.g., \ci{gst,lin}, cf. \rf{0})\foot{In this Appendix we use
$\a,\b,\g, ...$ for 2d indices, with $\gamma^{\alpha\beta} \equiv \sqrt h\, h^{\a\b}$.
$\m,\n, ...$ are 10d coordinate indices, and $a,b,c, ...$ are tangent space indices with $G_{\m\n} = e^a_\m e^b_\m \eta_{ab} $.
The indices $I,J,K= 1,2$ label two MW spinors of type IIB action. 
}
\be\la{r1}
L_{GS} &=& L_b+L_{2f}+L_{4f}+\dots \ ,
\\
L_b&=&\te \frac{1}{2}\gamma^{\alpha\beta} \partial_\alpha x^\mu \partial_\beta x^\nu G_{\mu\nu}
- \frac{1}{2}\epsilon^{\alpha\beta} \partial_\alpha x^\mu \partial_\beta x^\nu B_{\mu\nu}\ ,
\\ \la{r2}
L_{2f}&=& i(\gamma^{\alpha\beta}\delta^{IJ} - \epsilon^{\alpha\beta} s^{IJ}){\bar\theta}^I e_\alpha^a\Gamma_a
{\rm D}_\beta^{JK}\theta^K \ , \qquad \qquad \qquad \ \ e_\alpha^a = e^a_\m (x) \del_\a x^\m \ , \
\\\te
{\rm D}_\m &=& \te (\partial_\m + \frac{1}{4} \omega_{\m}{}^{ab} \Gamma_{ab}) - \frac{1}{8} s_3 H_{ab\m} \Gamma^{ab}
+ \frac{1}{8} e^{\phi} \big[{\rm F}\llap/{}_{(1)} s_0 + {\rm F}\llap/{}_{(3)} s_1
+ \frac{1}{2} {\rm F}\llap/{}_{(5)} s_0 \big] \Gamma_\m
\cr
&=& \te {\cal D}_\m
+ \frac{1}{8} e^{\phi} \big[{\rm F}\llap/{}_{(1)} s_0 + {\rm F}\llap/{}_{(3)} s_1
+ \frac{1}{2} {\rm F}\llap/{}_{(5)} s_0 \big] \Gamma_\m \ , \qquad \qquad \quad {\rm D}_\a =\del_\a x^\m {\rm D}_\m \ ,
\la{r3} \\ \la{r4}
L_{4f}&=& K^{\alpha\beta}_{IJKL X Y} {\bar \theta}^IM^X_\alpha \theta^J{\bar\theta^K}N^Y_\beta \theta^L \ .
\ee
In \rf{r4} the indices $X$ and $Y$ stand for multi-indices of the same type as the one carried by the fermions.
The $2 \times 2$ matrices appearing in $L_{2f}$ are $s\equiv s_3=\sigma_3$, $s_1 = \sigma_1$, $s_0 = i \sigma_2$.
The R-R couplings are
\be\la{r5} \te
e^\phi{\rm F}\llap/{}_{n}=\frac{1}{n!}e^\phi F_{a_1\dots a_n}\Gamma^{a_1\dots a_n} \equiv \frac{1}{n!}{\cal F}_{a_1\dots a_n}\Gamma^{a_1\dots a_n} \ ,
\ee
where $F_{n}= e^{-\phi} {\cal F}_n $ are not required a priori to be field strengths.\foot{Let us
  make a comment on 
different normalizations of RR field strength $F_5$  used in the literature.
In this paper we use the now standard normalization  in which  the type IIB   Lagrangian  is 
$L= e^{-\phi} R - {1\ov4\cdot 5!} F_5^2 + ...$.
With  the selfduality  condition on $F_5$ imposed at the level of equations of motion the analog of the Einstein equation  becomes as in (2.1),(2.3): $R_{mn} = {1\ov 4 \cdot 4!} e^{2\p} F_{mpqrs} F_n^{\ pqrs} + ...$.  In particular, in the case of unit-radius $AdS_5 \times S^5$   solution  with  $R_{mn}\big|_{AdS}  = -4 g_{mn}\big|_{AdS} $    one has 
$e^\phi F_5= 4( \epsilon_5 +  \epsilon_5^*)$. 
The generalized covariant derivative  in the gravitino variation or in the GS action in \rf{r4}  then contains 
${\rm D}_\mu =\del_\mu + {1\ov 16\cdot 5!} F_{abcde} \G^{abcde} s_0 \G_\m + ...$.
To  compare, in ref.\ci{schwarz} (or section 13 in \ci{Green:2012pqa}) with  the constant $\kappa$ fixed to be 1 
one had $(F_5)_{\rm schwarz}=- {1\ov 4} F_5$. 
In refs. \ci{Drukker:2000ep,Metsaev:2002re,Frolov:2002av} it  was assumed that 
$(F_5)_{\rm there}= -2(F_5)_{\rm schwarz}=   {1\ov 2} F_5$.
In this case for  the $AdS_5 \times S^5$    background   one has 
$e^\phi (F_5)_{\rm there}= 2( \epsilon_5 +  \epsilon_5^*)$
(note that ref. \ci{Frolov:2002av} contained a misprint in this expression given there above eq.(2.5)). 
}

We shall first
fix the $\kappa$-symmetry gauge $\theta^1 = \theta^2$ and
also consider flat 2d space (or, equivalently, fix the conformal gauge for 2d diffeomorphisms) and then
expand $(x, \theta)$ near some background values $(\bar x, \Theta)$.
The aim will then be to compute the one-loop UV divergences that renormalise the R-R couplings in
the quadratic fermionic term \rf{r2}, i.e. ($\bar \F = \F ( \bar x))$
\be
{\bar L}_{2f} &=&\te\frac{1}{4}\epsilon^{\alpha\beta}{\bar\Theta} {\bar e}_\alpha^a\Gamma_a {\bar H}_{abc} {\bar e}_\beta^c \Gamma^{ab}\Theta
+ {\bar L}^{\cal F}_{2f} 
\\
\la{r6}
{\bar L}^{\cal F}_{2f} &=&\te \frac{1}{4}\eta^{\alpha\beta} {\bar\Theta} {\bar e}_\alpha^a\Gamma_a \Sigma_{e} {\bar e}_\beta^b\Gamma_b\Theta
+ \frac{1}{4}\epsilon^{\alpha\beta} {\bar\Theta} {\bar e}_\alpha^a\Gamma_a \Sigma_{o} {\bar e}_\beta^b\Gamma_b\Theta \ ,
\\
\Sigma_{e} &=& {\bar{\F}}\llap/{}_{(3)}
~,\qquad\qquad
\Sigma_{o} =\te {\bar {\F}}\llap/{}_{(1)} +\frac{1}{2}{\bar {\F}}\llap/{}_{(5)} \ , \la{r7}
\\
\delta { L}_{2f} &=&\te \epsilon^{\alpha\beta} {\bar\Theta} O^H_{\alpha\beta} \Theta + \delta { L}^{\cal F}_{2f} 
\label{noF}
\\
\delta { L}^{\cal F}_{2f} &=&\te \eta^{\alpha\beta} {\bar\Theta} E_{\alpha\beta} \Theta + \epsilon^{\alpha\beta} {\bar\Theta} O_{\alpha\beta} \Theta \ .
\label{r8}
\ee
Here the classical term ${\bar L}^{\cal F}_{2f} $ and the
expected divergent term $\delta {\bar L}^{\cal F}_{2f} $ are decomposed into
parity-even and parity-odd
parts containing the linearly-independent combinations
of antisymmetrised products of Dirac matrices.
The combinations $E$ and $O$ should then represent the R-R beta-functions
that should be set to zero modulo use of equations of motion on $(\bar x, \Theta)$ or
modulo target space (super)reparametrisations. A further contribution to the two-fermion divergence is the first term in $\delta { L}_{2f}$; it should
be proportional to the NS-NS fields (vielbein and $H$) beta-functions and thus should contain terms independent of the R-R fields.

Introducing the fluctuations $(\xi^\m, \theta)$ around $({\bar x^\m}, \Theta)$ as
\be\la{r9}
x^\m \rightarrow \bar x^\m + \pi^\m (\xi) ~,\qquad\qquad
\theta \rightarrow \Theta + \theta \ ,
\ee
the standard relations of the bosonic normal coordinate expansion are
\be\la{r10} \no
\te \partial_\alpha (\bar x^\mu + \pi^\mu) &=& \te \partial_\alpha \bar x^\mu +\nabla_\alpha \xi^\mu
+\frac{1}{3} R^\mu{}_{\lambda\sigma\nu}\partial_\alpha \bar x^\nu \xi^\lambda\xi^\sigma+{\cal O}(\xi^3) \ ,
\\ \te
\partial_\alpha (\bar x^\mu + \pi^\mu) e_\mu^a &=&\te \zeta_\alpha^a +\nabla_\alpha \xi^a
+\frac{1}{2} R^a{}_{bcd} \zeta_\alpha^b \xi^c \xi^d+{\cal O}(\xi^3) \ , \qquad
\zeta_\alpha^a \equiv \partial_\alpha {\bar x}^\mu {\bar e}_\mu^a \ ,
\\
g_{\mu\nu} &=& \te {\bar g}_{\mu\nu} +\frac{1}{3} R_{\mu\lambda\sigma\nu}\xi^\lambda\xi^\sigma+{\cal O}(\xi^3)\ , \no
\\
e_{\mu}^a &=& {\bar e}_{\mu}^a +\frac{1}{6} R^a{}_{\lambda\sigma\mu}\xi^\lambda\xi^\sigma+{\cal O}(\xi^3) \ , \no
\\
\omega_\mu{}^a{}_b&=&\te {\bar \omega}_\mu{}^a{}_b+\frac{1}{2}\xi^\nu R^a{}_{b\nu\mu} + \frac{1}{3} \xi^\nu \xi^\rho \nabla_\rho R^a{}_{b\nu\mu} +{\cal O}(\xi^3) \ . \la{r11}
\ee
The normal coordinate expansion of the R-R tensor fields ($\bar \F \equiv \F ( \bar x)$)
\be\te
\F_{\mu_1\dots \mu_n} = \bar \F_{\mu_1\dots \mu_n} + \xi^\nu\nabla_\nu\bar \F_{\mu_1\dots \mu_n}
+ \frac{1}{2}\xi^\mu\xi^\nu \big( \nabla_\mu\nabla_\nu \bar \F_{\mu_1\dots \mu_n} + \frac{1}{3} \sum_{j=1}^n \bar \F_{\mu_1\dots \sigma_j \dots\mu_n} R^{\sigma_j}{}_{\mu\nu\mu_j}\big)
+{\cal O}(\xi^3)\no
\ee
takes a simpler form using tangent space indices:
\be
\F_{a_1\dots a_n} = \bar \F_{a_1\dots a_n} + \xi^\nu\nabla_\nu \bar \F_{a_1\dots a_n} + \te \frac{1}{2}\xi^\mu\xi^\nu \nabla_\mu\nabla_\nu \bar \F_{a_1\dots a_n}
+{\cal O}(\xi^3) \ . \la{r12}
\ee
Note that the beta-functions for the couplings $\F_{a_1\dots a_n}$ and $\F_{\mu_1\dots \mu_n}$ are related
by extra terms involving beta-functions of vielbein $e^a_\m$ or the metric $G_{\m\n}$; this is related to the presence of $\beta^G$ terms in
\rf{candidate2} or \rf{scinvF1}--\rf{scinvF5}.

The expanded Lagrangian \rf{r1} has the following structure:
\be
\label{r13}
L &=& L_b + L^{\xi\xi}_f + L^{\xi\theta}_f + L^{\theta\theta}_f + ... \ ,
\\
L_b &=&\te \frac{1}{2}\eta^{\alpha\beta}\nabla_\alpha\xi^a \nabla_\beta\xi^b \eta_{ab} + \nabla_\alpha\xi^a \xi^b U_{ab}^\alpha + \frac{1}{2} \xi^a \xi^b X_{ab}\ , \la{r113}
\\ \la{r14}
L^{\xi\xi}_f &=&\te \frac{1}{2}\nabla_\alpha\xi^a \nabla_\beta\xi^b C^{\alpha\beta}_{ab} + \nabla_\alpha\xi^a \xi^b C_{ab}^\alpha + \frac{1}{2} \xi^a \xi^b C_{ab}\ ,
\\ \la{r155}
L^{\xi\theta}_f &=& \te \nabla_\alpha\xi^a {\bar \Psi}^{\alpha\beta}_a{\cal D}_\beta\theta
+ \nabla_\alpha \xi^a {\bar\Psi}_a^\alpha \theta+ \xi^a {\bar\Psi}_a\theta\ ,
\\ \la{r15}
L^{\theta\theta}_f &=& i {\bar\theta}\rho^\alpha {\cal D}_\alpha \theta + {\bar\theta} Y_{0F} \theta + {\bar\theta} Y_{2f} \theta \ ,
\ee
where $Y_{0F} $ and $Y_{2f}$
contain zero and two background fermions $\Theta$, respectively.
We have also defined $\rho_\alpha \equiv \zeta_\alpha^a\Gamma_a$.
It will be sufficient to further assume that the induced metric is trivial, i.e.
$G_{\m\n} (\bar x) \del_\a \bar x^\m \del_\b \bar x^\n=\eta_{\a\b}$ and
$\rho_{(\alpha} \rho_{\beta )} = \eta_{\alpha\beta}$.

The explicit form of the quadratic terms in \rf{r13} is ($L_f = L_{2f} + L_{4f} + ...$, see \rf{r2},\rf{r4})
\be
L_b &=& \te
\frac{1}{2}\eta^{\alpha\beta} \left[\nabla_\alpha\xi^a \nabla_\beta\xi^b\eta_{ab}
+ R_{a c d b} \zeta_\alpha^a \zeta_\beta^b \xi^c \xi^d\right]
+ \frac{1}{2}\epsilon^{\alpha\beta} \left[\zeta_\alpha^a \nabla_\beta\xi^b \xi^c H_{abc}
+ \frac{1}{2} \zeta_\alpha^a \zeta_\beta^b \xi^d \xi^c \; \nabla_d H_{ab c} \right]\no
\\
-i L_{2f}^{\xi\xi}&=&\te
\frac{1}{12}\xi^c \xi^d (R^a{}_{cde}\zeta_\alpha^e\zeta_\beta^b
+ R^b{}_{cde}\zeta_\alpha^a \zeta_\beta^e)({\eta^{\alpha\beta}}{ }{\bar\Theta}^I \Gamma_a \Sigma_e\Gamma_b\Theta
-{\epsilon^{\alpha\beta}}{ }{\bar\Theta}^I \Gamma_a \Sigma_o\Gamma_b\Theta)
\cr
&&\te
+{ 1 \ov 4} \nabla_\alpha\xi^a\nabla_\beta\xi^b ({\eta^{\alpha\beta}}{}{\bar\Theta}^I \Gamma_a \Sigma_e\Gamma_b\Theta
-{\epsilon^{\alpha\beta}}{}{\bar\Theta}^I \Gamma_a \Sigma_o\Gamma_b\Theta)
\cr
&&\te
+{ 1 \ov 4} (\nabla_\alpha\xi^a\zeta_\beta^b + \zeta_\alpha^a \nabla_\beta\xi^b)\xi^d ({\eta^{\alpha\beta}}{}{\bar\Theta}^I \Gamma_a \nabla_d \Sigma_e\Gamma_b\Theta
-{\epsilon^{\alpha\beta}}{}{\bar\Theta}^I \Gamma_a \nabla_d \Sigma_o\Gamma_b\Theta)
\cr
&&\te
+\frac{1}{24}\,\xi^c\xi^d (\zeta_\alpha^f \zeta_\beta^b R_{f cd}{}^a +\zeta_\alpha^f \zeta_\beta^a R_{f c d}{}^b ) ({\eta^{\alpha\beta}}{}{\bar\Theta}^I \Gamma_a \Sigma_e\Gamma_b\Theta
-{\epsilon^{\alpha\beta}}{}{\bar\Theta}^I \Gamma_a \Sigma_o\Gamma_b\Theta) \,
\cr
&&\te
+\frac{1}{8}\zeta_\alpha^a \zeta_\beta^b \,\xi^c\xi^d
({\eta^{\alpha\beta}}{}{\bar\Theta}^I \Gamma_a \nabla_c\nabla_d \Sigma_e\Gamma_b\Theta
-{\epsilon^{\alpha\beta}}{}{\bar\Theta}^I \Gamma_a \nabla_c\nabla_d \Sigma_o\Gamma_b\Theta) \ , \no
\\
-i L_{2f}^{\xi\theta}&=&\te 4 \eta^{\alpha\beta} \nabla_\alpha \xi^a {\bar\Theta}\Gamma_a {\cal D}_\beta \theta
+ \frac{1}{2} \eta^{\alpha\beta} \zeta_\alpha^a \zeta_\beta^c \xi^b R^{de}{}_{b c}{\bar\Theta}\Gamma_a \Gamma_{de} \theta
\cr
&&\te +\frac{1}{4}\epsilon^{\alpha\beta}(\zeta_\alpha^a \nabla_\beta \xi^c + \nabla_\alpha \xi^a \zeta_\beta^c)(\bar\Theta \Gamma_a H_{cde}\Gamma^{de}\theta +{\bar\theta} \Gamma_a H_{cde}\Gamma^{de}\Theta )
\cr
&&+\eta^{\alpha\beta}(\zeta_\alpha^a \nabla_\beta \xi^c + \nabla_\alpha \xi^a \zeta_\beta^c)(\bar\Theta \Gamma_a \Sigma_e \Gamma_c \theta +{\bar\theta}\Gamma_a \Sigma_e \Gamma_c \Theta )
\cr
&&\te -\epsilon^{\alpha\beta}(\zeta_\alpha^a \nabla_\beta \xi^c + \nabla_\alpha \xi^a \zeta_\beta^c)(\bar\Theta \Gamma_a \Sigma_o \Gamma_c \theta +{\bar\theta}\Gamma_a \Sigma_o \Gamma_c \Theta )
\cr
&&+ \te \eta^{\alpha\beta} \zeta_\alpha^a \zeta_\beta^c \xi^d (\bar\Theta \Gamma_a \nabla_d \Sigma_e \Gamma_c \theta +{\bar\theta}\Gamma_a \nabla_d \Sigma_e \Gamma_c \Theta )
\cr
&&-\epsilon^{\alpha\beta} \zeta_\alpha^a \zeta_\beta^c \xi^d (\bar\Theta \Gamma_a \nabla_d \Sigma_o \Gamma_c \theta +{\bar\theta}\Gamma_a \nabla_d\Sigma_o \Gamma_c \Theta )\ ,\no
\\
-i L_{2f}^{\theta\theta}&=&\te 2\eta^{\alpha\beta}\zeta_\alpha^a\zeta_\beta^b {\bar\theta}\Gamma_a {\bar {\cal D}}_b\theta
+\frac{1}{4}\epsilon^{\alpha\beta}\zeta_\alpha^a\zeta_\beta^b {\bar\theta}\Gamma_aH_{bcd}\Gamma^{cd}\theta
\cr
&&\te +\frac{1}{4}\eta^{\alpha\beta}\zeta_\alpha^a\zeta_\beta^b {\bar\theta}\Gamma_a\Sigma_e \Gamma_b \theta
-\frac{1}{4}\epsilon^{\alpha\beta}\zeta_\alpha^a\zeta_\beta^b {\bar\theta}\Gamma_a\Sigma_o \Gamma_b \theta \la{r200}
\\
L_{4f}^{\theta\theta}&=&{\rm K}^{\alpha\beta}_{X Y} ({\bar \Theta} M^X_\alpha \Theta {\bar\theta}N^Y_\beta \theta
+{\bar \Theta} M^X_\alpha \theta {\bar\Theta}N^Y_\beta \theta+{\bar \Theta} M^X_\alpha \theta {\bar\theta}N^Y_\beta \Theta
\cr
&&
+{\bar \theta} M^X_\alpha \Theta {\bar\Theta}N^Y_\beta \theta+{\bar \theta} M^X_\alpha \Theta {\bar\theta}N^Y_\beta \Theta
+{\bar \theta} M^X_\alpha \theta {\bar\Theta}N^Y_\beta \Theta)\ , \qquad {\rm K}^{\alpha\beta}_{X Y} = \sum_{IJKL} {K}^{\alpha\beta}_{IJKL X Y} \no
\ee
Thus the matrix coefficients appearing in \rf{r13}-\rf{r15} are
\be
U^\alpha_{ab} &=&\te \frac{1}{2}\epsilon^{\alpha\beta} \zeta_\beta^c H_{abc}\ , \no
\\
X_{ab} &=&\te \eta^{\alpha\beta} \zeta_\alpha^c\zeta_\beta^d R_{cabd} + \frac{1}{4}\epsilon^{\alpha\beta} \zeta_\alpha^c \zeta_\beta^d
( \nabla_a H_{bcd}+\nabla_b H_{acd}) \ ,\no
\\
C^{\alpha\beta}_{ab} &=&\te { 1 \ov 4} i ({\eta^{\alpha\beta}}{}{\bar\Theta}^I \Gamma_a \Sigma_e\Gamma_b\Theta
-{\epsilon^{\alpha\beta}}{}{\bar\Theta}^I \Gamma_a \Sigma_o\Gamma_b\Theta) \ ,
\qquad \qquad \textrm{ etc.} \la{r25}
\ee
It is straightforward to find the UV-divergent term \rf{r8}
for the general Lagrangian \rf{r13}.
It receives contributions from Feynman graphs with one $C$ or one $Y_{2f}$ vertex
(each containing two background fermions) and from
Feynman graphs with two vertices of the type $\Psi$ (each containing a single background fermion).
The result has the form
\be\la{r16}
\delta L ^{{\cal F}}_{2f} = \delta L_1 + \delta L_2+\delta L_3 +\delta L_4 +\delta L_5 \ ,
\ee
where $\delta L_1$ contains one vertex from $L_f^{\xi\xi}$ and any number of vertices from $L_B$,
$\delta L_5$ contains $Y_{2f}$, $\delta L_2$ contains two vertices from $L^{\xi\theta}_f$,
$\delta L_3$ contains two vertices from $L^{\xi\theta}_f$ and one from $L_B$,
$\delta L_4$ contains two vertices from $L^{\xi\theta}_f$ and more than one from $L_B$.
Explicitly,
\be
\delta L_1 &=& \te \big( -\frac{1}{2} C_{ab}\eta^{ab} + \frac{1}{4}\eta_{\beta\gamma} \Tr[C^{\beta\gamma}X]
-\frac{1}{4}\eta_{\beta\gamma}\Tr[(U^\beta -U^\beta{}^T)(C^\gamma -C^\gamma{}^T)] \big) I_0
\cr
&&\te +\frac{1}{8}\Tr[(U^\alpha+U^\alpha{}^T)(\eta_{\alpha\beta}\partial_\gamma - \eta_{\gamma\beta}\partial_\alpha- \eta_{\gamma\alpha}\partial_\beta ) C^{\beta\gamma}] \, I_0\ , \la{r17}
\\
\delta L_2 &=&\te \ \ \frac{1}{4} \eta_{\alpha\beta} {\bar\Psi}^\alpha_a Y_{0F} \Psi^\beta_b\eta^{ab} I_0
+\frac{1}{8}\zeta^{\gamma c} {\bar\Psi}^\alpha_a \Gamma_c (\eta_{\alpha\beta}\partial_\gamma - \eta_{\gamma\beta}\partial_\alpha- \eta_{\gamma\alpha}\partial_\beta ) \Psi^\beta_b\eta^{ab} I_0
\cr
&&\te +\frac{1}{2} {\bar\Psi}_a^{\alpha\beta} I_{\alpha\beta\gamma\delta} (\partial)\Psi_b^{\gamma\delta}
\cr
&&\te -\frac{1}{2} \eta_{\alpha\beta} {\bar\Psi}_a^{\alpha\beta} Y_{0F} \Psi_b \eta^{ab} I_0
+\frac{1}{4} \zeta_c^{\gamma} {\bar\Psi}_a^{\alpha\beta} \Gamma^c (\eta_{\alpha\beta} \partial_\gamma
- \eta_{\alpha\gamma} \partial_\beta - \eta_{\beta\gamma} \partial_\alpha)\Psi_b \eta^{ab} I_0
\cr
&&\te + {\bar\Psi}_a^{\alpha\beta} I_{\alpha\beta\gamma}(\partial) \Psi^\gamma_b \eta^{ab}
+ \frac{1}{2} \zeta_\alpha^c {\bar\Psi}_a^{\alpha} \Gamma_c \Psi_b \eta^{ab} I_0
\ , \la{r18}
\\
\delta L_3 &=& \te \ \
\frac{1}{8} \zeta^\gamma{}^{c}{\bar\Psi}^\alpha_a \Gamma_c \Psi^\beta_b (U^\gamma-U^\gamma{}^T)^{ab}
(\eta_{\alpha\beta}\eta_{\gamma\delta} + \eta_{\alpha\gamma}\eta_{\beta\delta} + \eta_{\alpha\delta}\eta_{\beta\gamma} )I_0
\cr
&&+\te \frac{1}{16} \zeta^{\gamma c}{\bar\Psi}^\alpha_a\Gamma_c\Psi^\beta_b (U^\delta - U^{\delta}{}^T)^{ab}
(\eta_{\alpha\beta}\eta_{\gamma\delta} + \eta_{\alpha\gamma}\eta_{\beta\delta} + \eta_{\alpha\delta}\eta_{\beta\gamma} )I_0
\cr
&&+ \te \frac{1}{2}{\bar\Psi}(x_1)^{\alpha\beta}_a I_{\alpha\beta\gamma\delta}(\partial_1, \partial_2)\Psi(x_2)^{\gamma\delta}_b x^{ab}
\cr
&& \te +\frac{1}{2}{\bar\Psi}(x_1)^{\alpha\beta}_a (I_{\alpha\beta\gamma\delta\rho}(\partial_1, \partial_2) U^{\rho}
- I_{\alpha\delta\gamma\beta\rho}(\partial_2, \partial_1) U^{\rho}{}^T)^{ab}\Psi(x_2)^{\gamma\delta}_b
\cr
&& + \te \frac{1}{8}\zeta^{c \gamma}{\bar\Psi}_a^{\alpha\beta} \Gamma_c \Psi_b (U^\delta-U^\delta{}^T)^{ab}(\eta_{\alpha\beta}\eta_{\gamma\delta} + \eta_{\alpha\gamma}\eta_{\beta\delta} + \eta_{\alpha\delta}\eta_{\beta\gamma} ) I_0
\cr
&&+\te \frac{1}{4} \zeta^{\gamma c} {\bar\Psi}_a^{\alpha\beta} \Gamma_c \Psi_b^\delta x^{ab}
(\eta_{\alpha\beta}\eta_{\gamma\delta} + \eta_{\alpha\gamma}\eta_{\beta\delta} + \eta_{\alpha\delta}\eta_{\beta\gamma} ) I_0
\cr
&&+ {\bar\Psi}(x_1)_a^{\alpha\beta} (I_{\alpha\beta\gamma\rho}(\partial_1,\partial_2)U^\rho-I_{\gamma\beta\alpha\rho}(\partial_2,\partial_1) U^\rho{}^T)^{ab} \Psi(x_2)_b^\gamma \ , \la{r19}
\\
\delta L_4 &=& \te \ \
\frac{1}{2}\zeta^{\gamma c}{\bar\Psi}_a^{\alpha\beta}\Gamma_c \Psi^\delta_b (U^\rho-U^\rho{}^T)^{ad}(U^\xi-U^\xi{}^T)^{db}I_{\alpha\beta\gamma\delta\rho\xi}
\cr
&&\te
+\frac{1}{2}\zeta^{\gamma c}{\bar\Psi}_a^{\alpha\beta}\Gamma_c \Psi^{\delta\rho}_b [(U^\xi-U^\xi{}^T)^{ad}X^{db}
+X^{ad}(U^\rho-U^\rho{}^T)^{db}]I_{\alpha\beta\gamma\delta\rho\xi}
\cr
&&\te +\frac{1}{2} {\bar\Psi}_a^{\alpha\beta}(x_1) I_{\alpha\beta\gamma\delta}(\partial_1,\partial_2,U)^{ab} \Psi^{\delta\rho}_b (x_2)
\cr
&&+\te
\frac{1}{2}\zeta^{\gamma c}{\bar\Psi}_a^{\alpha\beta}\Gamma_c \Psi^{\delta\sigma}_b (U^\rho-U^\rho{}^T)^{ad}(U^\xi-U^\xi{}^T)^{de}(U^\zeta-U^\zeta{}^T)^{eb}I_{\alpha\beta\gamma\delta\rho\xi\sigma\zeta}\ , \la{r20}
\\
\delta L_5 &=&\te \frac{1}{2}\Tr[Y_{0F}Y_{2f}] I_0 \ . \la{r21}
\ee
The standard dimensional regularisation integrals used to derived these expressions are ($d= 2-\epsilon$)
\be
I_0 &=& \te \int \frac{d^d l}{l^2} \ ,
\qquad \qquad \,
I^{(0)}_2 =\te \int d^d l\frac{l_\alpha l_\beta }{l^2 (l+p)^2} = \frac{1}{2}\eta_{\alpha\beta} I_0+\textrm{finite} \ , \no
\\
I^{(1)}_2 &=&\te \int d^d l\frac{l_\alpha l_\beta l_\gamma }{l^2 (l+p)^2} = -\frac{1}{4}(\eta_{\alpha\beta} p_\gamma + \eta_{\alpha\gamma} p_\beta+ \eta_{\beta\gamma} p_\alpha) I_0+\textrm{finite} \ , \no
\\
I^{(0)}_3 &=& \te \int d^d l\frac{l_\alpha l_\beta l_\gamma l_\delta }{l^2 (l+p)^2 (l+q)^2} =
\frac{1}{8} H_{\alpha\beta\gamma\delta}
 I_0+\textrm{finite} \ , \no
\\
I_{\alpha_1\alpha_2\alpha_3 \alpha_4\alpha_5 \alpha_6} &=&\te \int d^d l\frac{l_{\alpha_1} l_{\alpha_2}l_{\alpha_3}l_{\alpha_4}
l_{\alpha_5}l_{\alpha_6}}{(l^2 )^4} = \frac{1}{48}H_{\alpha_1\alpha_2\alpha_3 \alpha_4\alpha_5 \alpha_6} I_0+\textrm{finite} \ , \no
\\
I_{\alpha_1\alpha_2\alpha_3 \alpha_4\alpha_5 \alpha_6\alpha_7 \alpha_8} &=&\te \int d^d l\frac{l_{\alpha_1} l_{\alpha_2}l_{\alpha_3}l_{\alpha_4}
l_{\alpha_5}l_{\alpha_6}l_{\alpha_7}l_{\alpha_8}}{(l^2 )^5} =
\frac{1}{384}H_{\alpha_1\alpha_2\alpha_3 \alpha_4\alpha_5 \alpha_6\alpha_7 \alpha_8} I_0+\textrm{finite} \ . \quad \ \ \la{r277}
\ee
The tensors $H$ are given iteratively by:
\be
H_{\alpha_1\alpha_2\alpha_3 \alpha_4} &=& \eta_{\alpha_1\alpha_2}\eta_{\alpha_3\alpha_4} + \eta_{\alpha_1\alpha_3}\eta_{\alpha_2\alpha_4} 
+ \eta_{\alpha_1\alpha_4}\eta_{\alpha_2\alpha_3}
\no \\
H_{\alpha_1\alpha_2\alpha_3 \alpha_4\alpha_5 \alpha_6} &=& \eta_{\alpha_1\alpha_2} H_{\alpha_3 \alpha_4\alpha_5 \alpha_6} 
+\eta_{\alpha_1\alpha_3} H_{\alpha_2 \alpha_4\alpha_5 \alpha_6} +\eta_{\alpha_1\alpha_4} H_{\alpha_2 \alpha_3\alpha_5 \alpha_6} 
\no\\
&& +\eta_{\alpha_1\alpha_5} H_{\alpha_2 \alpha_3\alpha_4 \alpha_6} +\eta_{\alpha_1\alpha_6} H_{\alpha_2 \alpha_3\alpha_4 \alpha_5} 
\no\\
H_{\alpha_1\alpha_2\alpha_3 \alpha_4\alpha_5 \alpha_6\alpha_7 \alpha_8} &=& \eta_{\alpha_1\alpha_2} H_{\alpha_3 \alpha_4\alpha_5 \alpha_6\alpha_7 \alpha_8} +\eta_{\alpha_1\alpha_3} H_{\alpha_2 \alpha_4\alpha_5 \alpha_6\alpha_7 \alpha_8} +\eta_{\alpha_1\alpha_4} H_{\alpha_2 \alpha_3\alpha_5 \alpha_6\alpha_7 \alpha_8} 
\no\\
&& +\eta_{\alpha_1\alpha_5} H_{\alpha_2 \alpha_3\alpha_4 \alpha_6\alpha_7 \alpha_8} +\eta_{\alpha_1\alpha_6} H_{\alpha_2 \alpha_3\alpha_4 \alpha_5\alpha_7 \alpha_8} +\eta_{\alpha_1\alpha_7} H_{\alpha_2 \alpha_3\alpha_4 \alpha_5\alpha_6 \alpha_8} 
\no\\
&&+\eta_{\alpha_1\alpha_8} H_{\alpha_2 \alpha_3\alpha_4 \alpha_5\alpha_6 \alpha_7} \ .
\la{hhhhhhh}
\ee
Some integrals which lead to derivatives acting on the background-dependent fermion mass term $Y_{0F}$
were left unevaluated:
\be &&\no \te
I_{\alpha\beta\gamma}(\partial) =\int d^d l \; l_\beta (l+i\partial )_\alpha (l+i\partial)_\gamma \ \frac{1}{ l\llap/ + Y_{0F}}\frac{1}{(l+i\partial)^2}\ , \\ &&\te
I_{\alpha\beta\gamma\delta} (\partial) = \int d^d l \; l_\alpha l_\gamma (l+ i\partial)_\beta (l+ i\partial)_\delta\ \frac{1}{l\llap/ + Y_{0F}} \frac{1}{(l+i\partial)^2}\ , \no
\\&&\te
I_{\alpha\beta\gamma\rho} (\partial_1,\del_1, \partial_2) = \int d^d l \; l_\beta (l+ i\partial_1)_\alpha(l+ i\partial_1)_\rho (l+ i\partial_2)_\gamma\ \frac{1}{l\llap/ + Y_{0F}} \frac{1}{(l+i\partial_1)^2}\frac{1}{(l+i\partial_2)^2}\no
\\ &&\te
I_{\alpha\beta\gamma\delta} (\partial_1,\partial_2) = \int d^d l \; l_\beta l_\gamma (l+ i\partial_1)_\alpha (l+ i\partial_2)_\delta\ \frac{1}{l\llap/ + Y_{0F}} \frac{1}{(l+i\partial_1)^2}\ \frac{1}{(l+i\partial_2)^2}\no
\\ &&\te
I^{}_{\alpha\beta\gamma\delta\rho} (\partial_1,\partial_2) = \int d^d l \; l_\beta l_\gamma (l+ i\partial_1)_\beta (l+ i\partial_1)_\rho (l+ i\partial_2)_\delta\ \frac{1}{l\llap/ + Y_{0F}} \frac{1}{(l+i\partial_1)^2}\frac{1}{(l+i\partial_2)^2}\no
\\ &&\te
I_{\alpha\beta\gamma\delta}(\partial_1,\partial_2,U)^{ab} = \int d^d l\; \frac{l_\beta l_\gamma }{l
\llap/
+ Y_{0F}}(l+i\partial_1)_\alpha (l+i\partial_2)_\delta \no
\\ &&\te
(-(l+i\partial_1)_\rho U(x_3)^\rho + (l+i\partial_{1+3})_\rho U(x_3)^\rho)^{ad}
((l+i\partial_{2+3})_\rho U(x_3)^\rho - (l+i\partial_2)_\rho U(x_3)^\rho)^{db}\ .
\qquad \qquad \la{r22} 
\ee
The same applies to integrals that lead to derivatives acting on two of the three vertices.

Using the coefficients \rf{r25} extracted from the expanded Lagrangian \rf{r200}, one finds that
the divergent term $\delta L_1$ in \rf{r17} is given by
\be
\delta L_1 &=&\Big[\te \frac{i}{8} \zeta_\alpha^a \zeta_\beta^b
(\eta^{\alpha\beta} {\bar\Theta} \Gamma_a\nabla^2 \Sigma_e \Gamma_b \Theta
- \epsilon^{\alpha\beta} {\bar\Theta} \Gamma_a\nabla^2 \Sigma_o \Gamma_b \Theta)\no
\\
&& \te
-\frac{i}{8} \zeta_\alpha^d \zeta_\beta^f \big(R_d{}^a\delta_f^b + \delta_d^a R_f{}^b
- R_{d}{}^{ab}{}_f- R_{d}{}^{ba}{}_f \big)(\eta^{\alpha\beta} {\bar\Theta} \Gamma_a \Sigma_e \Gamma_b \Theta
- \epsilon^{\alpha\beta} {\bar\Theta} \Gamma_a \Sigma_o \Gamma_b \Theta)
\nonumber\\
&& \te
-\frac{i}{16} \zeta_\alpha^d \zeta_\beta^f
(\nabla^a H_{df}{}^b + \nabla^b H_{df}{}^a ) (\eta^{\alpha\beta} {\bar\Theta} \Gamma_a \Sigma_o \Gamma_b \Theta
- \epsilon^{\alpha\beta} {\bar\Theta} \Gamma_a \Sigma_e \Gamma_b \Theta)
\nonumber\\
&& \te
-{i\ov 2} \zeta_\alpha^a \zeta_\beta^b H_{a}{}^{df} \,
(\eta^{\alpha\beta} {\bar\Theta} \Gamma_d\nabla_f \Sigma_e \Gamma_b \Theta
- \epsilon^{\alpha\beta} {\bar\Theta} \Gamma_d\nabla_f \Sigma_o \Gamma_b \Theta) \Big] I_0 \ ,
\la{r37}
\ee
where $\Sigma_o$, $\Sigma_e$ were defined in \rf{r7} and $I_0 \sim {1 \ov \epsilon} $ is the UV pole factor.

Comparing \rf{r37} to the corresponding terms in the classical action \rf{r6} one can
read off the contributions to the beta-functions for the R-R couplings.
Projecting onto the
independent set of Dirac matrices $\Gamma_{a_1\dots a_n}$ we
indeed observe the presence of the Hodge-de Rham operator terms as in \rf{scinvF1}--\rf{scinvF5}.
There are also similar terms depending on the
$H_3$ field strength
and its derivatives.
The UV singular terms in $\delta L_2, \delta L_3, \delta L_4$ in \rf{r18}--\rf{r20}
containing a single factor of $\Psi^{\alpha\beta}_a$
will have a similar structure.
The first term in ${\delta L}_4$ contains two factors of $H$ and one of R-R field
and should account for all such terms in eqs. \rf{scinvF1}--\rf{scinvF5}.

There are apparently also other UV singular terms that do not appear in \rf{scinvF1}--\rf{scinvF5}:
terms containing two
$\Psi^{\alpha\beta}_a$ factors are independent of the R-R fields and contain only the $H_3$ strength and factors of the curvature tensor.
We expect such terms to combine into the beta-function of the NS-NS fields entering the couplings \rf{noF} and thus yield the same scale
invariance conditions as in eqs \rf{4},\rf{5}.
Moreover, all terms in $\delta L_2$, $\delta L_3$ and $\delta L_4$ which do not contain $\Psi^{\alpha\beta}_a$
are bilinear in R-R fields and all terms in $\delta L_3$ contain at least one additional factor of either $H_3$ flux or the curvature tensor.
We expect such terms to cancel or to vanish upon use of the NS-NS
scale invariance conditions \rf{4},\rf{5}.

\



\end{document}